\documentclass[12pt,fleqn]{article}
\usepackage{comment}
\usepackage{enumitem}
\usepackage[latin9]{inputenc}
\usepackage{geometry}
\usepackage{float}
\usepackage{amsmath}
\usepackage{amssymb}
\usepackage{graphicx}
\usepackage{setspace}
\usepackage[authoryear]{natbib}
\usepackage[unicode=true,
 bookmarks=false,
 breaklinks=false,pdfborder={0 0 0},pdfborderstyle={},backref=false,colorlinks=false]
 {hyperref}
\hypersetup{
 pdfborderstyle=,pdftex,unicode,bookmarksnumbered=false,hyperfootnotes=true,citecolor=blue}

\makeatletter

\providecommand{\tabularnewline}{\\}

\usepackage{amsfonts}
\usepackage{url}\usepackage{verbatim}\usepackage{ifthen}\usepackage{multirow}\usepackage{amsthm}\usepackage{bm}
\usepackage[bf]{caption}
\usepackage{subcaption}
\usepackage{setspace}

\newcommand{\mappa}{\mu}
\newcommand{\sigphi}{\sigma}

\usepackage{tkz-graph}
\usetikzlibrary{positioning}
\usetikzlibrary{arrows.meta}

\tikzstyle{observed}=[circle, inner sep=0mm, outer sep=0mm, minimum size=2mm, draw=black, fill=black]
\tikzstyle{unobserved}=[circle, inner sep=0mm, outer sep=0mm, minimum size=2mm, draw=black, fill=white]

\tikzstyle{notouch}=[shorten <=5pt, shorten >= 5pt, -{Latex[length=2mm, width=1.5mm]}]

\usepackage[colorinlistoftodos]{todonotes}

\newcommand{\guido}[1]{\todo[inline,color=orange!50]{\textbf{Guido: }#1}}

\newcommand{\hyunseung}[1]{\todo[inline,color=brown!50]{\textbf{Hyunseung: }#1}}

\renewcommand{\mathbf}{\boldsymbol}
\renewcommand{\thepage}{}
\renewcommand{\appendix}{\footnotesize\parindent 0cm\setcounter{equation}{0}
\renewcommand{\theequation}{A.\arabic{equation}}
\setcounter{lemma}{0}\renewcommand{\thelemma}{A.\arabic{lemma}}}

\newcommand{\hmu}{\mu}

\newcommand{\hrho}{\rho}

\newcommand{\hrhoe}{\rho}

\newcommand{\hvarphi}{\varphi}
\newcommand{\hkappa}{\mappa}

\newcommand{\mmv}{\mathbb{V}}

\newcommand{\tick}{\checkmark}
\newcommand{\tock}{?}

\newtheorem{remark}{Remark}

\newcommand{\ppo}{{\rm O}}
\newcommand{\ppe}{{\rm E}}
\newcommand{\been}{\mathbf{1}}

\newcommand{\woin}{W_{i}}%{W_{\ppe,i}}

\newcommand{\yoinn}{Y_{i}(0)}%{Y_{\ppe,i}(0)}
\newcommand{\yoine}{Y_{i}(1)}%{Y_{\ppe,i}(1)}
\newcommand{\yoin}{Y_{i}}%{Y_{\ppe,i}}
\newcommand{\xoin}{X_{i}}%{X_{\ppe,i}}
\newcommand{\soin}{S_{i}}%}{S_{\ppe,i}}
\newcommand{\soinn}{S_{i}(0)}%{S_{\ppe,i}(0)}
\newcommand{\soine}{S_{i}(1)}%{S_{\ppe,i}(1)}

\newcommand{\poi}{P_{i}}
\newcommand{\poin}{P_{i}}

\newcommand{\yoio}{Y_{i}}
\newcommand{\xoio}{X_{i}}
\newcommand{\soio}{S_{i}}
\newcommand{\wojn}{W_{j}}

\newcommand{\xojn}{X_{j}}
\newcommand{\sojn}{S_{j}}

\newcommand{\xojo}{X_{j}}
\newcommand{\sojo}{S_{j}}

\newcommand{\yoipo}{Y_{i'}}
\newcommand{\xoipo}{X_{i'}}
\newcommand{\soipo}{S_{i'}}

\newcommand{\yojpo}{Y_{j'}}
\newcommand{\xojpo}{X_{j'}}
\newcommand{\sojpo}{S_{j'}}

\newcommand{\mmy}{\mathbb{Y}}
\newcommand{\mms}{\mathbb{S}}
\newcommand{\mmx}{\mathbb{X}}
\newcommand{\mmw}{\mathbb{W}}

\newcommand{\pr}{{\rm pr}}

\newcommand{\mme}{\mathbb{E}}

\newcommand{\nex}{N_\ppe}
\newcommand{\nobs}{N_\ppo}

\newtheorem{assumption}{Assumption}\newtheorem{theorem}{Theorem}\newtheorem{lemma}{Lemma}\newtheorem{prop}{Proposition}\newtheorem{definition}{Definition}\newcommand{\indep}{\perp\!\!\!\perp}

\def\monthname{\ifcase\month\or
  January\or February\or March\or April\or May\or June\or July\or
  August\or September\or October\or November\or December\fi}
\renewcommand{\thepage}{[\arabic{page}]}
\numberwithin{equation}{section}

\def\monthname{\ifcase\month\or
January\or February\or March\or April\or May\or June\or
July\or August\or September\or October\or November\or December\fi}
\renewcommand{\appendix}{\small\parindent 0cm\setcounter{equation}{0}
\renewcommand{\theequation}{A.\arabic{equation}}
\setcounter{lemma}{0}\renewcommand{\thelemma}{A.\arabic{lemma}}
\setcounter{theorem}{0}\renewcommand{\thetheorem}{A.\arabic{theorem}}}

\usepackage{titling}
\makeatother

\begin{document}
\title{{\Large{}{}The Surrogate Index: Combining Short-Term Proxies to Estimate
Long-Term Treatment Effects More Rapidly and Precisely}\thanks{{\small{}{}We are grateful for discussions with James Dailey, Lawrence
Katz, David Ritzwoller, Dylan Small, Scott Stern, and Liang Xu and for comments from
numerous seminar participants. We thank Kevin Chen, Yechan Park, Emanuel Schertz and James
Stratton for outstanding research assistance. We are particularly grateful to Kevin Chen and David Ritzwoller for pointing out a mistake in the previous version of this paper. This research was funded
through National Science Foundation Grant DMS-1502437, the Chan-Zuckerberg
Initiative, the Bill \& Melinda Gates Foundation,  the Overdeck
Foundation, ONR grants N00014-17-1-2131 and 
 N00014-19-1-2468 and the Sloan Foundation.}} }
\author{\vspace{0.2cm}
Susan Athey\thanks{{\small{}{}Graduate School of Business, Stanford University, and
NBER, athey@stanford.edu. }} \and Raj Chetty\thanks{{\small{}{}Department of Economics, Harvard University, and NBER,
chetty@fas.harvard.edu. }} \and Guido W. Imbens\thanks{{\small{}{}Graduate School of Business, and Department of Economics,
Stanford University, and NBER, imbens@stanford.edu. }} \and Hyunseung Kang\thanks{{\small{}{}Department of Statistics, University of Wisconsin at Madison,
hyunseung@stat.wisc.edu.}} }
\date{\vspace{0.2cm}
\today}

\maketitle\thispagestyle{empty} 
\vspace{-0.2cm}
\begin{abstract}
\noindent A common challenge in estimating the impact of
interventions ({\it e.g.}, job training programs, educational programs) is that many outcomes of interest
({\it e.g.}, lifetime earnings  or other labor market outcomes) are observed with a long delay. 
In biomedical settings this is often addressed by using short-term outcomes as so-called ``surrogates'' for the outcome of interest, {\it e.g.}, tumor size as a surrogate for mortality in cancer studies.
We build on this literature by combining multiple, possibly qualitatively distinct, short-term outcomes ({\it e.g.}, short-run
earnings and employment indicators) systematically into a ``surrogate index.''  Under the Prentice surrogacy assumption, which requires that the primary
outcome is independent of the treatment conditional on the surrogates, we show that the
average treatment effect on the surrogate index equals the treatment
effect on the long-term outcome. We also relate the surrogacy assumption to a set of structural, causal assumptions.
 We then characterize the bias that arises from violations of
each of the key assumptions,  and we
provide simple methods to validate these assumptions using additional observed
outcomes. We apply
our method to analyze the long-term impacts of a multi-site job training
experiment in California. Rather than waiting a full nine years to directly observe 
the long-term impact, we show that it is possible to use short-term (the first six quarters) outcomes as surrogates.
One could have estimated the program's long-term impacts on mean employment
rates using the employment rates observed in the first six quarters, with a 35\% reduction
in standard errors. 
\end{abstract}

%\textbf{JEL Classification: C14, C21, C52}

\noindent {\small{}Keywords:\ Potential Outcomes, Causality, Surrogate
Outcomes, Surrogate Scpore, Surrogate Index, Mediators, Propensity Score, Principal Stratification, Job
Training}{\small\par}

\baselineskip=22.5pt\newpage\setcounter{page}{1} 
\global\long\def\thepage{[%
\arabic{page}]}%
\global\long\def\theequation{\arabic{section}.%
\arabic{equation}}%

%\listoftodos

\section{Introduction}
\label{section:introduction}

A fundamental challenge for evaluating interventions is that the primary outcomes of interest are often hard to measure.  For example, researchers are often interested in the effect of the policy on some long-term outcome but do not observe that in their study. Instead, they observe a number of short-term outcomes that are all related to this primary outcome of interest.

%\guido{test}

One setting where this type of problem arises involves an educational policy maker evaluating a policy that would change class size. The ultimate goal may be to improve long-term labor market outcomes for the students. However, the decision regarding the class size policy needs to be made at a time when only 
short-term outcomes such as test scores or other educational achievement measures are available. Another setting involves policy makers  considering labor market interventions such as job search assistance or human capital acquisition programs, where they may be primarily interested in the long-term labor market attachment of the participants, but in the short run they may only have access to outcomes such as employment records or earnings over a short period of time. In randomized experiments for medical interventions, the ultimate outcome of interest is often survival or quality-adjusted years of life. Survival rates may be high in the short run, and so typically such trials are evaluated in terms of surrogate measures, such as including tumor size or other measures of the progression of the disease, which can be measured earlier. In all of these types of setting, to make a timely decision, the policy maker needs to assess the programs based on short-term outcomes. 
These challenges also arise in business settings. In the context of experimentation in digital technology companies, a discussion of the most important challenges ranks as the top concern that
 ``While most
experiments in the industry run for 2 weeks or less, we
are really interested in detecting the long-term effect of a
change. How do long-term effects differ from short-term
outcomes? How can we accurately measure those long-
term factors without having to wait a long time in every
case?'' (\citet{gupta2019top}, p. 21). 

In these and many other examples, the researcher is faced with 
 making recommendations regarding the future implementation of the intervention on the basis of measurements of its effect on a variety of sometimes disparate and possibly conflicting outcome measures.
A key question is how to balance these different outcomes when making an overall assessment. 
In practice, researchers often deemphasize short-term outcomes for which they do not find statistically significant effects, instead making perhaps somewhat {\it ad hoc} qualitative assessments regarding the relative importance of the remaining short-term outcomes.

In this paper we lay out a framework for analyzing these issues.  We consider the scenario in which researchers do not measure the primary outcome in the context of data containing information on the intervention. Instead, we assume that the researcher has a second, observational, dataset where the researcher observes the surrogates and the primary outcome but does not observe the treatment. In both samples the researcher may also observe variables not affected by the treatment, such as pre-treatment characteristics of the participants. 

We make four main contributions. First, we articulate three key assumptions under which the average effect of the treatment on the primary outcome is identified from the combination of the experimental and observational samples: $(i)$ a standard assumption that the assignment in the experimental sample is \emph{Unconfounded}; $(ii)$ a \emph{Surrogacy} assumption which requires that the causal path from the treatment to the primary outcome goes through the surrogates \citep{prentice1989surrogate, day1996trial, begg2000, frangakis2002principal}; and $(iii)$ a \emph{Comparability} or external validity assumption, which requires that the observational and experimental samples are comparable in the sense that the outcome distributions conditional on surrogates and pre-treatment variables are identical. 

Under these three assumptions, the average effect of the treatment  on the primary outcome can be estimated as the average effect of the treatment on an aggregate of the surrogates, which we label the surrogate index. This index combines the individual surrogates through their predicted value of the primary outcome.
For example, when studying the impact of class size, the primary outcome might be high school graduation, while the two surrogates might be mathematics and reading scores. For the special case of linear models, the proposal boils down to multiplying the causal effects of the intervention on the two scores (which can be estimated in the experimental sample) by the coefficients from a linear regression of the primary outcome on the two scores in the observational sample. The approach replaces a subjective assessment of the relative importance of the two short-term measures by an objective data-driven criterion, namely the predictive power of the scores for the outcome of interest.

In our second contribution, we derive the efficiency bound and propose various efficient estimators under various scenarios, including scenarios with a single sample or two samples, as well as with and without Surrogacy. This allows us to quantify the information content of the Surrogacy assumption.\footnote{We are grateful to Kevin Chen and David Ritzwoller for pointing out an error in one of our earlier efficiency bound calculations., see \citet{chen2023semiparametric} for more details.}

In our third contribution, we provide bounds on the biases that arise in scenarios where either or both of Surrogacy or Comparability are violated. We show that even if these assumptions fail to hold (but unconfoundedness does hold), the proposed estimators still estimate a well-defined causal effect, by providing a principled way of combining short-term outcomes in a single measure through their predicted effect on the long-term outcome.

In our fourth contribution,
we evaluate these methods in the context of a labor market program where we observe long-term (thirty-six quarters) outcomes in four locations. Following an approach popularized by \citep{lalonde}, we put aside part of the data and investigate whether we could have estimated the long-term effects without having long-term experimental data. Specifically, we take one of the locations, Riverside, and put aside the long-term outcome for individuals from that location. Then we take the other three locations, Alameda, Los Angeles, and San Diego, and put aside the treatment assignment for that sample. We investigate whether these two samples allow us to recover the experimental long-term effects in Riverside using surrogates corresponding to the first $T$ quarters of outcomes (employment, earnings, and aid indicators). We find that combining six quarters of outcome data into a surrogate index suffices to obtain estimates close to the long run effects. Using the additional data that were put aside for the main analysis, we also directly test whether the critical assumptions, Surrogacy and Comparability, hold given various alternative sets of surrogates.

We recognize that the credibility of the Surrogacy assumption may be questioned in any given application, especially when viewed in isolation. Therefore, we view the best path forward as building a ``library'' of surrogate indices in which researchers systematically catalog across several studies the smallest set of surrogates that successfully match long-term outcomes of interest ({\it e.g.},  earnings, mortality, educational attainment). If one establishes, for instance, that six quarters of employment and earnings data are sufficient to predict the impacts of many different job training programs -- as our cross-site comparisons of the GAIN program suggest -- then the long-term impacts of future job training programs could be credibly estimated using the established six-quarter surrogate index. We view the empirical application in this paper as providing one element of such a library and hope future work will expand upon it by identifying surrogate indices that match estimated long-term impacts in other applications.

This study is related to three main bodies of literature, surrogacy, mediation, and missing data.
We extend the literature on surrogacy
(\citealt{prentice1989surrogate, day1996trial, fleming1996surrogate,begg2000,xu2001evaluation,lauritzen2004discussion,d2006surrogate,qu2006quantifying,alonso2006unifying,gilbert2008evaluating,
weir2006statistical}) by formally including the presence of a second observational sample that is used to estimate the relationship between surrogates and the primary outcome and articulating the assumptions that justify doing so.
In doing so we allow for uncertainty in the estimation of this surrogates/outcome relationship, whereas the previous literature took this relation as known. We also consider biases arising from violations of Surrogacy and Comparability.

 In addition, 
this study builds on the literature on mediation (\citealt{baron1986moderator,van2004estimation,imai2010general,zheng2012targeted,tchetgen2011semiparametric,vanderweele2015explanation}), which
considers the decomposition of an average treatment effect into the
direct effect of a treatment on an outcome and indirect effects that
flow through a mediator. In the mediation setup, all three key variables
-- the outcome, the treatment, and the mediator -- are observed
for the same units. The goal in the mediation literature is to determine the relative magnitudes
of the direct and indirect effects. In our surrogacy analysis we focus on the case
in which the direct effect is absent by assumption.

This paper is also related to the classical missing data literature in statistics (\citealt{rubin1976inference, rubin2004multiple,  little2014statistical}). 
Our key assumptions are closely related to the Missing At Random (MAR) assumption. 
Our approach can be viewed as a special case of approaches that
combine data sets, {\it e.g.}, \citet{ridder2007econometrics,chen2008semiparametric}.
In particular \citet{rassler2004data,rassler2012statistical} refers
to our setting, where one variable is missing in one part of a sample
and a second variable missing in the remainder of the sample, as a
``data fusion'' setting. \citet{graham2016efficient} discuss efficient
estimation for a particular set of models defined by moment conditions
in such a data fusion setting, where they allow the treatment  to be a general
random variable, rather than a binary indicator as in our setup. 

The paper is organized as follows. Section \ref{section:setup} sets up the problem and introduces the notation.
Section \ref{assumptions} discusses the critical assumptions and links the setup to the mediation and missing data literature. Section \ref{identification} discusses identification and the efficiency bounds. 
Section \ref{section:misspecification} presents formulas for bias when the surrogacy
assumption fails and derives bounds on the degree of bias. Section \ref{estimation} discusses
estimation. Section \ref{application} presents the empirical application.
Section 8 concludes.

\section{Setup and Notation\label{section:setup}}

We define two samples, an Experimental ($\ppe$) sample
and an Observational ($\ppo$) sample, with $\nex$ and $\nobs$ units or individuals, respectively.
It is convenient to view the data as consisting of a single sample
of size $N=\nex+\nobs$, with $P_{i}\in\{\ppo,\ppe\}$ a binary indicator
denoting the sample to which unit $i$ belongs.

For each unit, there is a
 binary treatment of interest, $\woin\in\{0,1\},$ and a scalar primary
outcome, denoted by $\yoin$.  
This outcome is not observed for individuals in the experimental sample. In addition, there are
intermediate or secondary outcomes, which we  refer to as
surrogates (to be defined precisely in Section \ref{section:surrogacy}),
denoted by $\soin$ for each unit. Typically, the surrogate
outcomes are vector-valued in order to make the properties we define
plausible. Finally, we measure pre-treatment covariates $\xoin$ for
each unit, known not to be affected by the
treatment. 

Following the potential outcomes framework or Rubin Causal Model (\citealt{rubin1974estimating,holland1986statistics,imbens2015causal}),
individuals in this group have two pairs of potential outcomes: $(\yoinn,\yoine)$
and $(\soinn,\soine)$. The realized outcomes are related to their
respective potential outcomes as follows. 
\[
\yoin\equiv \yoin(\woin)=\left\{ \begin{array}{ll}
\yoinn\hskip1cm & {\rm if}\ \woin=0,\\
\yoine\hskip1cm & {\rm if}\ \woin=1,
\end{array}\right.\hskip1cm{\rm and}\ \ \soin\equiv \soin(\woin)=\left\{ \begin{array}{ll}
\soinn\hskip1cm & {\rm if}\ \woin=0,\\
\soine\hskip1cm & {\rm if}\ \woin=1.
\end{array}\right.
\]
Overall, the units are characterized by the values of the septuple
$(\yoinn,\yoine,\soinn,\soine,\xoin,\woin,\poin)$. We do not observe the
full septuple for any units. Rather, for units in the experimental
sample we observe the triple $(\xoin,\woin,\soin)$ with support
$(\mmx, \mmw, \mms)$ where $\mmw=\{0,1\}$. %\subsection{Observational Sample}
In the observational sample, we do not observe to which treatment
each of the $\nobs$ individuals were assigned. We observe the triple $(\xoio,\soio,\yoio)$,
with support $\mmx$, $\mms$, and $\mmy$ respectively. 
To simplify the exposition, we analyze the data as if we have a random sample from a population of units for which we observe the quintuple $(\poin,\xoin,\soin,\mathbf{1}_{\poin=\ppe} \woin,\mathbf{1}_{\poin=\ppo} \yoin)$, where we treat $\poin$ as a random variable taking on the values $\{\ppo,\ppe\}$.
\begin{assumption}\label{sampling}
We have a single random sample of size $N$ drawn from the joint distribution of
$(\poin,\xoin,\soin, \woin, \yoin)$, where we observe for each unit in the sample $(\poin,\xoin,\soin,\mathbf{1}_{\poin=\ppe} \woin,\mathbf{1}_{\poin=\ppo} \yoin)$. 
\end{assumption}
We summarize
this data setup in Table \ref{tabel1}.
The setup differs from those in  \citet{athey2020combining} and  \citet{kallus2020role}, where we would also observe the treatment in the observational sample, but in the experimental sample we would still not observe the primary outcome.

\noindent 
\begin{table}
%\label{tabel1}%[ht]
%\caption{Observation Scheme}
\caption{ \textsc{Observation Scheme: \tick \,is observed, \tock \, is missing}}
\label{tabel1}%[ht]

\begin{centering}
%\textsc{Table 1. Observation Scheme: \tick\tick is observed, }\tock\textsc{ is missing}
\par\end{centering}
\centering{}\label{tabel_sampling}\vskip0.3cm \centering{}%
\begin{tabular}{c|ccccc}
 &  &  & Long-Term  &  & Pretreatment \\
 & Sample &Treatment  & Outcome & Surrogate  &  Variables\tabularnewline
Units & $P_{i}$ & $W_{i}$ & $Y_{i}$ & $S_{i}$ & $X_{i}$\tabularnewline
 &  &  &  &  & \tabularnewline
\hline 
 &  &  &  &  & \tabularnewline
1 to $N_{\ppe}$  & $\ppe$ & \tick & \tock & \tick & \tick\tabularnewline
 &  &  &  &  & \tabularnewline
$N_{\ppe}+1$ to $N_{\ppe}+N_{\ppo}$ & $\ppo$ & \tock & \tick & \tick & \tick\tabularnewline
\end{tabular}
\end{table}
%\noindent \textit{\footnotesize{}Notes: }{\footnotesize{}This table summarizes the setup of the dataset for the problem we analyze. The experimental sample has N\ppeN_{\ppe} observations. It includes information on the treatment indicator, the intermediate outcomes (surrogates), and pretreatment variables; however, critically, it does not include data on the long-term outcome, reflecting the fact that the long-term outcome is observed with a long delay after the experiment. The observational sample has N\ppoobservations.Itincludesinformationontheprimaryoutcome,thesurrogates,andthepretreatmentvariables;however,itmaynotincludedataonthetrein=x,\poin=\ppe)N_{\ppo}observations.Itincludesinformationontheprimaryoutcome,thesurrogates,andthepretreatmentvariables;however,itmaynotincludedataonthetrein=x,\poin=\ppe) \\ 

We are interested in the Average Treatment Effect (ATE) on the primary
outcome in the population from which the experimental sample is drawn:
\begin{equation}\label{tau}
\tau\equiv\mme[\yoine-\yoinn|P_{i}=\ppe].
\end{equation}
The same issues we study in the current paper apply to  other estimands, such as the average  treatment effect for the
treated units, or the average for the observational sample. 

An implicit assumption in our setup is that the two variables that are common to both samples, $\soin$ and $\xoin$, measure the same underlying variables in both samples. In some cases it is possible that in one of the two samples, a coarser version is measured, for example age or education may be measured in multi-year categories rather than in years. In that case, a simple solution is to proceed by using the coarser version of the variables as corresponding to the surrogate of pre-treatment variable, thereby relying on strong assumptions. Another complication arises if the unit of observation differs in two  samples, say individuals versus zipcodes. Again additional assumptions are required to link the variables between samples.

Table \ref{tabel_notation} summarizes key definitions and notation.

\noindent 
\begin{table}
\caption{ \textsc{Notation and Definitions}}
%\label{tabel1}%[ht]
%\label{tabel_notation}%[ht]
%\caption{Observation Scheme}
\begin{centering}
%\textsc{Table 2. Notation and Definitions}
\par\end{centering}
\centering{}\label{tabel_sampling}\vskip0.3cm \centering{}%
\begin{tabular}{c|ccccc}
\hline \\
Sampling Indicator & $\poin\in\{\ppe,\ppo\}$\\
Potential Outcomes for Primary Outcome & $\yoin(0),\yoin(1)$\\
Potential Outcomes for Surrogates & $\soin(0),\soin(1)$\\
Binary Treatment Indicator & $\woin\in\{0,1\}$\\
Realized Value for Outcome & $\yoin=\yoin(\woin)$\\
Realized Value for Surrogate & $\soin=\soin(\woin)$\\ \\
Estimand & $\tau\equiv\mme[\yoin(1)-\yoin(0)|\poin=\ppe]$\\
\\
 & $\mu(s,w,x,p)\equiv\mme[\yoin|\soin=s,\woin=w,\xoin=x,\poin=p]$ & 
\\
Surrogate Index & $\mu(s,x,p)\equiv\mme[\yoin|\soin=s,\xoin=x,\poin=p]$ \\
%& \hmu\ppe,w(x)\equiv\mme[\yoin|\xoin=x,\woin=w,\poin=\ppe]\hmu_{\ppe,w}(x)=\mme[\yoin|\xoin=x,\woin=w,\poin=\ppe] \\
& $\mappa(w,x)\equiv 
\mme[\mu(S_i,X_i,O) \mid W_i=w,X_i=x,P_i=E] $
%& μ(w,x)\equiv\mme[μ(Si,Wi,Xi,O)∣Wi=w,Xi=x,Pi=E]\mu(w,x)=\mme[\mu(S_i,W_i,X_i,O) \mid W_i=w,X_i=x,P_i=E] 
\\ \\
& $\sigma^2(s,w,x,p)\equiv\mmv(\yoin|\soin=s,\woin=w,\xoin=x,\poin=p)$ \\
& $\sigma^2(s,x,p)\equiv\mmv(\yoin|\soin=s,\xoin=x,\poin=p)$ \\
& $\sigphi^2(w,x)\equiv\mmv(\yoin|\woin=w,\xoin=x,\poin=\ppo)$\\
%& $\sigphi^2(w,x)\equiv\mmv(\mu(\soin,\xoin,\ppo)|\woin=w,\xoin=x,\poin=\ppe)$
\\ \\
Surrogate Score & $\rho(s,x)\equiv\pr(\woin=1|\soin=s,\xoin=x,\poin=\ppe)$ & \\
Propensity Score & $\rho(x)\equiv\pr(\woin=1|\xoin=x,\poin=\ppe)$ \\
& $\rho\equiv\pr(\woin=1|\poin=\ppe)$ & \\ \\
Sampling Score & $\hvarphi(s,x)\equiv\pr(\poin=\ppe|\soin=s,\xoin=x)$ & \\
& $\hvarphi(x)\equiv\pr(\poin=\ppe|\xoin=x)$ & \\
& $\hvarphi\equiv\pr(\poin=\ppe)$ & 
\\ \\
Conditional Distribution of Surrogates & $\pi(s|w,x)\equiv f_{S_i|W_i,X_i,P_i}(s|w,x,\ppe)$\\
& $\pi(s|x)\equiv f_{S_i|X_i,P_i}(s|x,\ppe)$
\\ \\
Influence Function & $\psi(y,s,w,x,p)$\\ \\
\end{tabular}
\noindent \textit{\footnotesize{}Notes: }{\footnotesize{}This table summarizes the notation. Conditional expectations and variances of the outcome $\yoin$ are denoted by $\hmu(\cdot)$ and $\sigma^2(\cdot)$ respectively. Conditional probabilities of the treatment are denoted by $\hrho(\cdot)$. Conditional probabilities of the sample are denoted by $\hvarphi(\cdot)$ The arguments of these functions can be both the surrogates $\soin$ and the pre-treatment variables $\xoin$, or just the pre-treatment variables $\xoin$.}{\footnotesize\par}
\label{tabel_notation}
\end{table}

\section{The Critical Assumptions: Unconfoundedness, Surrogacy, and Comparability}\label{assumptions}

\label{section_surrogate}In this section, we discuss the three key assumptions
that together allow us to combine the observational and experimental
samples and estimate the causal effect of the treatment on the primary
outcome, exploiting the presence of the surrogates. The first assumption
is {\it  Unconfoundedness} or {\it Ignorability}, common in the program evaluation
literature  (\citealt{rosenbaum1983central,imbens2015causal}), which ensures that adjusting for pre-treatment variables
leads to valid causal effects in the experimental sample. The second assumption is the {\it Surrogacy
condition} due to \citet{prentice1989surrogate},
that allows us to use the surrogate variables to proxy for the primary
outcome. The third assumption is {\it Comparability}, which formalizes the connection between the two samples. This assumption is rarely stated formally, but plays an important role in our analysis. 

\subsection{Unconfoundedness}

For the individuals in the experimental group,  the propensity
score is the conditional probability of receiving the treatment: $\hrho(x)\equiv \pr(\woin=1|\xoin=x,\poin=\ppe).$
We assume that for individuals in the experimental group, treatment
assignment is unconfounded, and we have overlap in the distribution
of pre-treatment variables between the treatment and control groups
(\citealt{rosenbaum1983central, imbens2015causal}):

\begin{assumption}\label{ass:unconf}\textsc{(Unconfounded Treatment
Assignment / Strong Ignorability)} \\
 $(i)$ \[\woin\ \indep\ \Bigl(\yoinn,\yoine,\soinn,\soine\Bigr)\ \Bigr|\ \xoin,P_{i}=\ppe,\]
 $(ii)$ $0<\hrho(x)<1\ {\rm for\ all}\ x\in\mmx.$
\end{assumption}
This assumption, widely used in the causal inference literature, implies that in  the experimental sample, we can estimate
the average causal effect of the treatment on the surrogates 
by adjusting for pre-treatment variables. 
We would also have been able to estimate the causal effect on the primary outcome had the primary outcome been measured in the experimental sample.
In many applications of surrogacy approaches, the treatment in the experimental sample is assigned completely randomly. In that case this assumption is satisfied by design. However, unconfoundedness is all that is required.

\subsection{Surrogacy}
\label{section:surrogacy}

 Next we discuss the second critical assumption, surrogacy. We also introduce two concepts, the {\it surrogacy score}, similar to the propensity score, and the {\it surrogacy index}, to combine multiple surrogates.

\subsubsection{The Prentice Criterion}

 \citealt{prentice1989surrogate} defines a surrogate as a post-treatment variable where conditioning on it makes the outcome and the treatment independent: \begin{assumption}\label{ass:surro}\textsc{(Surrogacy, Prentice Criterion)}\\
 $(i)$
\[
\woin\ \indep\ \yoin\ \Bigr|\ \soin,\xoin,P_{i}=\ppe.
\]
and
$(ii)$ $0<\rho(s,x)<1, {\rm for\ all}\ s\in\mms,x\in\mmx,\ \textrm{and}\ 0<\pr(\poin=\ppe)<1.$
\end{assumption}
\begin{remark}
If the quadruple $(\yoin,\soin,\woin,\xoin)$ were observed for all units, surrogacy would be a testable condition. With $(\soin,\woin,\xoin)$ observed for units in the experimental sample, and $(\yoin,\soin,\xoin)$ observed for units in the observational sample, this assumption has no testable implications.
\end{remark}
\begin{remark}
Note that Surrogacy is formulated in terms of the realized outcome and surrogate values. In contrast we formulated the ignorability condition 
(Assumption \ref{ass:unconf}) in terms of the potential outcomes. This is partly to connect our discussion to the  surrogacy literature  \citep{prentice1989surrogate, day1996trial}.
\end{remark}

Surrogacy is often debated in empirical applications. \citet{freedman1992statistical} argue that the surrogate may not mediate the full effect of the treatment in many settings. For example, reductions in class size may affect earnings through changes in non-cognitive skills that are not fully captured by standardized test scores (\citealt{heckman2006effects,chetty2011does}).  

\subsubsection{The Surrogacy Index and the Surrogacy Score}

There are two scalar functions of the surrogates that play an important role in the analyses: the surrogate index and surrogate score.

\begin{definition}\textsc{(The Surrogate Index)} The surrogate index
is the conditional expectation of the primary outcome given the surrogate
outcomes and the pre-treatment variables, conditional on the sample:
\[
\hmu(s,x,p)\equiv \mme\left[\left.\yoio\right|\soio=s,\xoio=x,P_{i}=p\right].
\]
\end{definition}\begin{remark}  The surrogate index in the observational sample, $\hmu(s,x,\ppo)$, is identified
because we observe the triple $(\yoin,\soin,\xoin)$ in the observational
sample.
\end{remark}

\begin{definition}\textsc{(The Surrogate Score) }The surrogate score
is the conditional probability of having received the treatment given
the value for the surrogate outcomes and the covariates in the experimental sample: 
\[
\hrho(s,x)\equiv\pr(\woin=1|\soin=s,\xoin=x,P_{i}=E).
\]
\end{definition}
The  
surrogacy score plays is similar to the role the propensity score plays in analyses under unconfoundedness \citep{rosenbaum1983central}. Here if the surrogacy condition holds conditional on $(S_i,X_i)$, it also holds conditional on
 the surrogacy score. 

\begin{prop}\label{prop1}\textsc{(Surrogate Score)} Suppose Surrogacy (Assumption
\ref{ass:surro}) holds. Then: 
\[
\woin\ \indep\ \yoin\ \Bigr|\ \hrho(\soin,\xoin),P_{i}=\ppe.
\]
\end{prop} 
\noindent All proofs are given in the Appendix.

\subsubsection{The Benefits of Multiple Surrogates}
\label{benefits}

One theme of this paper is that having multiple short-term variables can make a surrogacy approach more plausible, the same way multiple pre-treatment variables can make the unconfoundedness assumption more plausible. Here we discuss some illustrative examples.

The first example is illustrated in Figure 1.A. Suppose the treatment is an educational intervention. This treatment affects the outcome of interest, some labor market outcome, e.g., earnings, through a number of different channels corresponding to different skill sets. These channels may include mathematics skills, language skills, and social skills. Using only one of these variables as a surrogate would lead to biased estimates because they would ignore the other causal paths. In this case the set of three short-term variables collectively satisfy Unconfoundedness and Surrogacy.

The second case is illustrated in Figure 1.B. In this setup there is a variable, labeled ``skills', that satisfies the critical assumptions for surrogacy. However, skills is not observed by the researcher. Instead we have two noisy measures of this surrogate, say both a written and an oral exam. Collectively these two variables may still not satisfy Surrogacy, since there may be impacts of skills on earnings not captured by the exams, but the bias from using both would be less than the bias from using only one candidate surrogate.
%%%%%%%%%%%%%%%%%%%%%%%%%%%%%%%%%%%%%%%%%
% figure 1a-1d
%%%%%%%%%%%%%%%%%%%%%%%%%%%%%%%%%%%%%%%%
\begin{figure}[H]
    % Subfigure 1a
    \begin{subfigure}[b]{0.48\textwidth}
        \centering
        \textsf{\textbf{\scriptsize{}Figure 1.a Surrogacy Assumption Satisfied}}{\scriptsize\par}
        \vspace{2.72em}
      \begin{tikzpicture}[>=stealth, node distance=0.8cm]
\node[observed, label=above left:{\scriptsize{}\(\rm Education\)}] (1) {}; 
\node[observed, right=2cm of 1, label=above:{\scriptsize{}\(\rm Language\ Skills\)}] (2) {}; 
\node[observed, above=of 2, label={ {\scriptsize{}\(\rm Math\ Skills\)}} ] (3) {}; 
\node[observed, below=of 2, label=below: { {\scriptsize{}\(\rm Social\ Skills\)}} ] (4) {}; 
\node[observed, right=2cm of 2, label=above right:{\scriptsize{}\(\rm Wage\)}] (5) {}; 
\draw [->, notouch] (1.east) -- (2.west); 
\draw [->, notouch] (2.east) -- (5.west); 
\draw [->, notouch] (1.south east) -- (3.north west); 
\draw [->, notouch] (3.north east) -- (5.south west); 
\draw [->, notouch] (1.south) -- (4.north west); 
\draw [->, notouch] (4.north east) -- (5.south);
\end{tikzpicture}
    \end{subfigure}
    \hfill
    % Subfigure 1b
    \begin{subfigure}[b]{0.48\textwidth}
        \centering
        \textsf{\textbf{\scriptsize{}Figure 1.b Multiple Surrogates}}{\scriptsize\par}
        \vspace{2.72em}
       \begin{tikzpicture}[>=stealth, node distance=0.8cm]
\node[observed, label=above:{\scriptsize{}\(\rm Education\)}] (1) {}; 
\node[unobserved, right=1.3cm of 1, label=above right:{\scriptsize{}\(\rm Skills\)}] (2) {}; 
\node[observed, above=of 2, label=above: {\scriptsize{} {\(\rm Written\ Exam\)}} ] (3) {}; 
\node[observed, below=of 2, label=below:{\scriptsize{} {\(\rm Oral\ Exam\)}} ] (4) {}; 
\node[observed, right=1.3cm of 2, label=above right:{\scriptsize{}\(\rm Wage\)}] (5) {}; 
\draw [->, notouch] (1.east) -- (2.west); 
\draw [->, notouch] (2.east) -- (5.west); 
\draw [->, notouch] (2.north) -- (3.south); 
\draw [->, notouch] (2.south) -- (4.north); 
%\draw [->, notouch] (1.south east) -- (3.north west); 
%\draw [->, notouch] (3.north east) -- (5.south west); 
%\draw [->, notouch] (1.south) -- (4.north west); 
%\draw [->, notouch] (4.north east) -- (5.south);
\end{tikzpicture}
    \end{subfigure}

% Subfigure 1c
\begin{subfigure}[b]{0.40\textwidth}
    \centering
    \textsf{\textbf{\scriptsize{}Figure 1.c: Multiple Surrogates, Scenario 1}}{\scriptsize\par}
    \vspace{2.72em}
   \begin{tikzpicture}[>=stealth, node distance=0.8cm]
\node[observed, label=above left:{\scriptsize{}\(\rm Informative\ Ad\)}] (1) {}; 
\node[unobserved, right=1.3cm of 1, label=below:{\scriptsize{}\(\rm Interested\ in\ item\)}] (2) {}; 
\node[unobserved, right=1.3cm of 2, label=above:{\scriptsize{}\(\rm Engaged\ with\ Website\)}] (3) {}; 
\node[observed, above=of 2, label=above:{\scriptsize{} {\(\rm Click\ on\ Ad\)}} ] (4) {}; 
\node[observed, below=of 3, label=below:{\scriptsize{} {\(\rm Spend\ Time\ on\ Website\)}} ] (5) {}; 
\node[observed, right=1.3cm of 3, label=above right:{\scriptsize{}\(\rm Purchase\ Item\)}] (6) {}; 
\draw [->, notouch] (1.east) -- (2.west); 
\draw [->, notouch] (2.east) -- (3.west); 
\draw [->, notouch] (3.east) -- (6.west); 
\draw [->, notouch] (2.south) -- (4.north); 
\draw [->, notouch] (3.north) -- (5.south); 
%\draw [->, notouch] (1.south east) -- (3.north west); 
%\draw [->, notouch] (3.north east) -- (5.south west); 
%\draw [->, notouch] (1.south) -- (4.north west); 
%\draw [->, notouch] (4.north east) -- (5.south);
\end{tikzpicture}
\end{subfigure}
\hfill
% Subfigure 1d
\begin{subfigure}[b]{0.48\textwidth}
    \centering
    \textsf{\textbf{\scriptsize{}Figure 1.d: Multiple Surrogates, Scenario 2}}{\scriptsize\par}
    \vspace{2.72em}
   \begin{tikzpicture}[>=stealth, node distance=0.8cm]
\node[observed, label=above left:{\scriptsize{}\(\rm Click\ Bait\ Ad\)}] (1) {}; 
\node[unobserved, right=1.3cm of 1, label=below:{\scriptsize{}\(\rm Interested\ in\ item\)}] (2) {}; 
\node[unobserved, right=1.3cm of 2, label=above:{\scriptsize{}\(\rm Engaged\ with\ Website\)}] (3) {}; 
\node[observed, above=of 2, label=above:{ \scriptsize{}{\(\rm Click\ on\ Ad\)}} ] (4) {}; 
\node[observed, below=of 3, label=below:{\scriptsize{} {\(\rm Spend\ Time\ on\ Website\)}} ] (5) {}; 
\node[observed, right=1.3cm of 3, label=above right:{\scriptsize{}\(\rm Purchase\ Item\)}] (6) {}; 
\draw [->, notouch] (1.south east) -- (4.north west); 
\draw [->, notouch] (2.east) -- (3.west); 
\draw [->, notouch] (3.east) -- (6.west); 
\draw [->, notouch] (2.south) -- (4.north); 
\draw [->, notouch] (3.north) -- (5.south); 
%\draw [->, notouch] (1.south east) -- (3.north west); 
%\draw [->, notouch] (3.north east) -- (5.south west); 
%\draw [->, notouch] (1.south) -- (4.north west); 
%\draw [->, notouch] (4.north east) -- (5.south);
\end{tikzpicture}
\end{subfigure}

    \label{fig:part1}
\end{figure}

The third case is illustrated in Figure 1.C. Here there is a pathway from the treatment, an informative advertisement about an item, to the outcome, an indicator for the individual purchasing the advertised item, going through two variables that on their own could each serve as surrogates. These two variables are whether someone has interested in the item, and whether the individual engaged with the website where the item was sold. However, we only measure noisy versions of these surrogates. For the first surrogate we observe whether an individual clicked on the advertisement for the item, and for the second surrogate we observe the time spent on the website. Neither of these two observed variables is a valid surrogate, but the combination of the two generally removes more of the bias than a single one.

Figure 1.D illustrates further the concerns with only using a single surrogate, the possibility of focusing on treatments that improve the surrogate variable but not the primary outcome. Suppose that a researcher uses the single variable ``click on ad'' as a surrogate for the effect of the ad on purchases. If the observational sample was based on informative advertisements, there is likely a positive correlation between clicking on the advertisement and purchases. However, if the new treatment is uninformative, {\it e.g.,} clickbait advertisement, with no effect on the actual interest in the item, the surrogacy analysis using click behavior as the surrogate will be ineffective. Using both click behavior and time spent on the website as surrogates will likely reduce the bias. The same argument implies that using multiple tests as surrogates can reduce problems with ``teaching to the test,'' where the long-run impact of an intervention is not well captured by scores on a test.

\subsection{Comparability}

 Surrogacy and Unconfoundedness by themselves are not sufficient for
consistent estimation of $\tau$ because they do not place
restrictions on how the relationship between $\yoin$ and $\soin$ in the
observational sample compares to that in the experimental sample. 
As far as we know, such restrictions were not previously articulated in the surrogacy literature because the setup is typically one with just the separate experimental sample. However, a comparability assumption is implicit in the way the postulated relationship between the surrogate and the primary outcome is used in that literature.
Related assumptions about the possibility of using causal estimates in one location to predict causal effects in a second location on the basis of distributions of pre-treatment variables are discussed in
\cite{hotz2005predicting} and the literature on transportability, \cite{pearl2014external}.

\subsubsection{The Comparability Assumption}

 Let $\hvarphi\equiv \pr(\poin=\ppe)$  be the probability of a unit being part of the experimental sample. We introduce the \emph{Sampling Score,} the propensity to be in the experimental sample:
\begin{definition}\textsc{(Sampling Score)} \\
The sampling score is $\hvarphi(s,x)\equiv \pr(\poin=\ppe|\soin=s,\xoin=x).$
\end{definition} 

The third key assumption we make is that the conditional distribution
of $\yoin$ given $(\soin,\xoin)$ in the observational sample is
the same as the conditional distribution of $\yoio$ given $(\soio,\xoio)$
in the experimental sample, and that the support of $(\soin,\xoin)$ in the experimental sample is a subset of that in the observational sample. Formally,
\begin{assumption}\label{ass:comp}\textsc{(Comparability
of Samples)} 
\\ $(i) \hspace{.25cm}\poin\ \indep\ \yoin\ \Bigr|\ \soin,\xoin,$
\\$(ii)  \hspace{.25cm}\hvarphi(s,x)<1\ \ {\rm for\ all}\ s\in\mms\ \ {\rm and\ }x\in\mmx.$
\end{assumption}

Similar to Unconfoundedness and Surrogacy this is a strong assumption, but unlike those assumptions it is rarely discussed explicitly. As we show in Section \ref{section:misspecification}, by making it explicit we can discuss the biases arising from violations and improve the intuition when this assumption may be of concern. If the observational and experimental samples are substantially different in terms of the distribution of pre-treatment variables and surrogates, it would likely be more controversial to assume that conditional on those variables the outcome distributions are identical.

\subsubsection{The Surrogate Index and the Sampling Score}

We let $\hmu(s,w,x,p)$ denote the conditional expectation of the primary  outcome
given pre-treatment variables, surrogates, treatment, and sample: 
\begin{equation}
\hmu(s,w,x,p)\equiv \mme\left[\left.\yoin\right|\soin=s,\xoin=x,\woin=w,P_{i}=p\right].\label{een}
\end{equation}
Comparability and Surrogacy together allow us to
impute the missing primary outcomes in the experimental sample, as
shown by the following proposition.

\begin{prop}\textsc{(Surrogate Index)} $(i)$ Suppose Assumption
\ref{ass:surro} (Surrogacy) holds. Then: 
\[
\hmu(s,w,x,\ppe)=\hmu(s,x,\ppe),\hskip1cm{\rm for\ all}\ s\in\mms,\ x\in\mmx,\ \ {\rm and}\ w\in\mmw.
\]
$(ii)$ Suppose Assumption \ref{ass:comp} (Comparability) holds. Then: 
\[
\hmu(s,x,\ppe)=\hmu(s,x,\ppo)\ \ \ {\rm for\ all}\ s\in\mms, \ {\rm and}\ x\in\mmx.
\]
$(iii)$ Suppose Assumptions \ref{ass:surro} (Surrogacy)  and \ref{ass:comp} (Comparability) 
hold. Then: 
\[
\hmu(s,w,x,\ppe)=\hmu(s,x,\ppo)\ \ \ {\rm for\ all}\ s\in\mms,x\in\mmx,\ {\rm and}\ w\in\mmw.
\]
\label{prop2} \end{prop} 
Because we can estimate $\mu(s,x,\ppo)=\mme[\yoin|\soin=s,\xoin=x,\poin=\ppo]$, we can impute the missing $\yoin$ in the experimental sample as $\mu(\soin,\xoin,\ppo)$.

\subsection{Surrogacy, Mediation,  Instrumental Variables, Directed Acyclical Graphs, and Missing Data}

To provide context for the setup here and the key assumptions, it is useful to make a link to three related  literatures,  on mediation, instrumental variables, and missing data  respectively. We describe the causal structures for surrogacy, mediation, and instrumental variables using a directed acyclical graph (DAG) \citep{pearl}. The interpretations provided in this subsection are note essential to the main results in the next section.

\subsubsection{Directed Acyclical Graph Representations}

The surrogacy, mediation, and instrumental variables  literatures all study causal structures involving a causally linked sequence of three (sets) of variables. They differ in three key aspects: $(i)$ the assumptions they make on the causal structure, $(ii)$ the estimands that are the  primary focus of the analysis, and $(iii)$ the data available for the analyses. The literatures also differ in the labels typically used for the three variables. 
In Table \ref{tabel_three_lit} we list the labels, estimands, and some of the assumptions.
\noindent 
\begin{table}%[ht]
\caption{\textsc{ Surrogacy versus Mediation versus Instrumental Variables}}
\label{tabel_three_lit}
\begin{centering}
%\textsc{Table 3. Surrogacy versus Mediation versus Instrumental Variables}
\par\end{centering}
\centering{}\label{tabel_sampling}\vskip0.3cm \centering{}%
\begin{tabular}{c|ccccc}
\hline \\
& Surrogacy & Mediation & Instrumental Var\\ \\
Left Variable (L) & Treatment ($W$) &  Treatment ($W$) & Instrument ($Z$) \\
Middle Variable (M) & Surrogate ($S$) & Mediator ($M$)  & Treatment ($W$)\\
Right Variable (R) & Outcome ($Y$) & Outcome ($Y$)  & Outcome ($Y$)\\ \\
Estimand & Effect of L on R& Direct  and Indirect & Effect of M on R \\ 
&& Effect of L on R
\\ \\
Direct Effect of L on R & No  & Yes & No \\
Unobs Conf between L and M & No 
& No & No \\Unobs Conf between M and R  & No 
& No & Yes \\
All Variables Observed Together & No & Yes & Yes
\end{tabular}
\end{table}

In Figures 2.A-2.C we show the differences in structures in DAG form in a single sample setting (so that we need not be concerned with the comparability assumption). 
Figure 2.A illustrates the surrogacy setup, with a causal link from the treatment to the surrogate and from the surrogate to the outcome.   There is  no unobserved confounder for the causal relation between treatment and surrogate, which would violate 
Assumption \ref{ass:unconf}
(Unconfounded Treatment
Assignment / Strong Ignorability). 
There is no direct causal link from the treatment to the outcome. 
There are also no
unobserved confounders for the causal relation between  surrogate and the outcome.
These two features of the DAG (no direct link between treatment and outcome and no unobserved confounder for the relation between surrogate and outcome imply Assumption (\ref{ass:surro})  
(Surrogacy).

\iffalse

\guido{proof for this last argument: the graph, even if in the absence of unconfounded treatment assignment, so allowing for an unobserved confounder for the relation between treatmena and surrogate  implies $S=f(W,\varepsilon,\nu)$, $Y=g(S,\eta)$, $W=h(\nu,\xi)$, with $(\eta<\nu,\varepsilon,\xi)$ jointly independent. Then
\[ \eta\indep\ \nu,\xi,\varepsilon\]
which implies
\[ \eta\indep\ \nu,\xi,\varepsilon| f(h(\nu,\xi),\varepsilon,\nu) \]
which implies
\[ g( f(h(\nu,\xi),\varepsilon,\nu)\eta)\indep\ h(\nu,\xi)| f(h(\nu,\xi),\varepsilon,\nu) \]
which is
\[ Y\indep W |S\]
\[ \]
}  

\fi

Figure 2.B shows a mediation example where Assumption \ref{ass:surro}
is violated because there is a direct effect of the treatment on the
outcome that does not pass through the surrogate. In this case $\soin$ is a typically labelled a mediator, rather than  a surrogate. In the mediation case the direct effect of the treatment on the outcome is estimable because all three variables, treatment, mediator and outcome are observed in the same sample.

\vspace{0.5cm}

\begin{figure}[h]
\begin{subfigure}[b]{0.32\textwidth}
\centering
\noindent \begin{centering}
\textsf{\textbf{\scriptsize{}Figure 2.A. Surrogacy Assumption Satisfied}}{\scriptsize\par}
\par\end{centering}
\vspace{2.72em}

\begin{tikzpicture}[
            >=stealth,
            node distance=1.5cm
            ]
            \node[observed, label=above:{\scriptsize{}\({\rm Treatment}\)}] (1) {};
            \node[observed, right=of 1, label=above:{\scriptsize{}\({\rm Surrogate}\)}] (2) {};
            \node[observed, right=of 2, label=above:{\scriptsize{}\({\rm Outcome}\)}] (4) {};
            \draw [->, notouch] (1.east) -- (2.west);
            \draw [->, notouch] (2.east) -- (4.west);
            \draw [->, notouch] (1.east) -- (2.west);
        \end{tikzpicture}
\end{subfigure}
\hfill
\begin{subfigure}[b]{0.32\textwidth}
    \centering
    \noindent \begin{centering}
    \textsf{\textbf{\scriptsize{}Figure 2.B. Violation of Surrogacy due to Direct Effect (Mediation Setup)}}{\scriptsize\par}
    \par\end{centering}
    \vspace{1.35em}
    \begin{tikzpicture}[
            >=stealth,       node distance=1.5cm            ]
            \node[observed, label=below:{\scriptsize{}\({\rm Treatment}\)}] (1) {};
            \node[observed, right=of 1, label=above:{\scriptsize{}\({\rm Surrogate}\)}] (2) {};
            \node[observed, right=of 2, label=above right:{\scriptsize{}\({\rm Outcome}\)}] (4) {};
            \draw [->, notouch] (1.east) -- (2.west);
            \draw [->, notouch] (2.east) -- (4.west);
            \draw [->, notouch] (1.east) -- (2.west);
            \draw [->, notouch] (1.north east) to [out=55, in=125] (4.north west);
        \end{tikzpicture}
\end{subfigure}
\hfill
\begin{subfigure}[b]{0.32\textwidth}
    \centering
    \noindent \begin{centering}
    \textsf{\textbf{\scriptsize{}Figure 2.C. Violation of Surrogacy Assumption due to Unobserved Confounder (IV Setup) }}{\scriptsize\par}
    \par\end{centering}
    \vspace{2.0em}
    \begin{tikzpicture}[
            >=stealth,
            node distance=1.5cm        ]
            \node[observed, label=above:{\scriptsize{}\({\rm Instrument}\)}] (1) {};
            \node[observed, right=of 1, label=below:{\scriptsize{}\({\rm Treatment}\)}] (2) {};
            \node[observed, right=of 2, label=below right:{\scriptsize{}\({\rm Outcome}\)}] (4) {};
            \node[unobserved, above right=of 2, label=above:{\scriptsize{}\({\rm Unobserved \ Confounder}\)}] (5) {};
            \draw [->, notouch] (1.east) -- (2.west);
            \draw [->, notouch] (2.east) -- (4.west);
            \draw [->, notouch] (1.east) -- (2.west);
            \draw [dashed, ->, notouch] (5.south west) -- (2.north east);
            \draw [dashed, ->, notouch] (5.south east) -- (4.north west);
        \end{tikzpicture}
\end{subfigure}
\end{figure}

Figure 2.C shows a DAG representation of the standard instrumental
variables (IV) model familiar to economists.
The first difference from the surrogacy setup in Figure 2.A is that in the instrumental variables setting the interest is in the causal effect of the variable in the middle of the three variable chain (the surrogate $S$ in the surrogacy setting, and the treatment $W$ in the instrumental variables setting), on the outcome, whereas in the surrogacy setting the primary interest is in the effect of the first variable in the chain (the treatment $W$ in the surrogacy setting and the instrument $Z$ in the instrumental variables setting) on the outcome.
In the instrumental variables case the surrogacy estimand is immediately identified as the intention-to-treat effect of the instrument, since the instrument and the surrogate are observed in the same sample. Under the assumptions of the surrogacy setup, the target for an instrumental variables analysis, the effect of the surrogate on the primary outcome, is immediately identified.
The instrumental variables settings is characterized by the presence of an unobserved confounder that affects both the treatment of interest and the outcome. The presence of that unobserved confounder violates Surrogacy, even if the treatment has no direct effect on the long-term outcome (\citealt{frangakis2002principal,rosenbaum1984consequences,joffe2009related,vanderweele2015explanation}). 

The presence of this unobserved confounder also violates the comparability assumption if the marginal distribution of the treatment $W$ differs between the observational and experimental samples, as will typically be the case.
In both the surrogacy and the instrumental variables cases, we assume the absence of a direct effect of the first variable in the causal chain (the treatment $W$ in the surrogacy case and the instrument in the instrumental variables case) on the primary outcome. In the surrogacy setting, this assumption is part of the Surrogacy assumption, while in the instrumental variables setting this is typically referred to as the exclusion restriction 
\citep{angrist1996identification}.

\subsubsection{A Missing Data Representation}

In the Online Appendix we also discuss a missing data interpretation of the surrogacy approach. Eessentially we show that the following joint conditional independence assumption,
\begin{equation}\label{missing}
P_{i}\ \indep\ Y_{i}\ \indep\ W_{i}\ \Bigr|\ S_{i},X_{i},
\end{equation}
implies both surrogacy and comparability.

This missing data characterization is useful because it  allows one to use insights from the missing data literature, both for the current problems and for generalizations.
Given (\ref{missing}) we can use the conditional distribution of $Y_i$ given $(\soin,\xoin)$ in the observational sample with $P_i=\ppo$ to impute the missing outcomes in the experimental sample with $P_i=\ppe$, and we can use the conditional distribution of $\woin$ given $(\soin,\xoin)$ in the experimental sample wth $P_i=\ppe$ to impute the missing treatments in the observational sample with $P_i=\ppo$. 

This observation directly extends to more general imputation problems. Suppose we have two samples where in one sample, indicated by $\poin=\ppe$ we observe one set of variables, $(Z_{i1},Z_{i2})$ and in the second sample, indicated by $\poin=\ppo$ we observe a partially overlapping set of variables, $(Z_{i2},Z_{i3})$. Then the analogous assumption that allows the imputation of all missing variables is
$\poin\indep Z_{i1}\indep Z_{i3}|Z_{i2}.$

\section{Identification and Semiparametric Efficiency Bounds}\label{identification}

\subsection{Three Identification Results}

We now present our central identification result. We analyze three
different representations of the average treatment effect that lead
to three estimation strategies, somewhat similar to inverse propensity score weighting, regression, and influence function estimators for average treatment effects under unconfoundedness \citep{imbens2004}. The motivation for developing the
different representations is that estimators corresponding to those
different representations can have different properties in finite
samples, just like they do in the unconfoundedness setting. Estimators based on the first representation require estimation of the surrogate
index, but not the surrogate score. Estimators based on the second representation instead
require estimation of the surrogate score, but not the surrogate
index. Estimators based on the third representation require estimation of both, but have attractive double robustness properties.

%−\been\poin=\ppo1−q(t(\soio,\xoio)1−t(\soio,\xoio)1−qq)(\yoin−h(\soio,\xoio))(1−\hrho(\soio,\xoio))1−e(\xoio)\quad{}- \frac{\been_{\poin=\ppo}}{1 - q} \left(\frac{t(\soio,\xoio)}{1 - t(\soio,\xoio)} \frac{1 - q}{q}\right) \frac{(\yoin - h(\soio,\xoio)) (1 - \hrho(\soio,\xoio))}{1 - e(\xoio)} 
We define the following four objects, all functionals of distributions
that are directly estimable from the data. First define the statistical estimand, the average difference in the surrogate index between treated and control, adjusted for pretreatment variablles, in the experimental sample:
\begin{equation}\label{estimand}
\tau^*\equiv \mme\left[\Bigl\{
\mme\Bigl[
 \mme\left[\left.
\yoin
\right|\soin,\xoin,\poin=\ppo\right]
\Bigr| \woin=1,\xoin,\poin=\ppe
\Bigr]
\right.
\end{equation}
\[
\left.\left.
-
\mme\Bigl[
 \mme\left[\left.
\yoin
\right|\soin,\xoin,\poin=\ppo\right]
\Bigr| \woin=0,\xoin,\poin=\ppe
\Bigr]
\Bigr\}\right| \poin=\ppe\right].
\]
Next,  with a surrogate
index representation: 
\begin{equation}
\tau^{\ppe}\equiv \mme\left[\left.\hmu(\soin,\xoin,\ppo)\cdot\frac{\woin}{\hrho(\xoin)}-\hmu(\soin,\xoin,\ppo)\cdot\frac{1-\woin}{1-\hrho(\xoin)}\right|P_{i}=\ppe\right],\label{estimand1}
\end{equation}
then a surrogate score representation, 
\begin{equation}
\tau^{\ppo}\equiv \mme\left[\yoio\cdot\frac{\hrho(\soio,\xoio)\cdot \hvarphi(\soio,\xoio)\cdot(1-\hvarphi)}{\hrho(\xoio)\cdot(1-\hvarphi(\soio,\xoio))\cdot \hvarphi}\right.\label{estimand2}
\end{equation}
\[
\hskip2cm\left.\left.-\yoio\cdot\frac{(1-\hrho(\soio,\xoio))\cdot \hvarphi(\soio,\xoio)\cdot(1-\hvarphi)}{(1-\hrho(\xoio))\cdot(1-\hvarphi(\soio,\xoio))\cdot \hvarphi}\right|P_{i}=\ppo\right].
\]
The third representation is  based on the influence function. We first 
define
\[ \hkappa(w,x)\equiv \mme[\hmu(\soin,\xoin,\ppo)|\woin=w,\xoin=x,\poin=\ppe].\]
Then the influence function is
\begin{equation}\label{effscore}
\psi(y,s,w,x,p)=\frac{\been_{p=\ppe}}{\hvarphi}\left(\frac{w \cdot (\hmu(s,x,\ppo)-\hkappa(1,x))}{{\hrho}(x)}-\frac{(1-w)\cdot ({\hmu}(s,x,\ppo)- \hkappa(0,x) )}{1-{\hrho}(x)}\right) 
\end{equation}
\[\hskip2cm+ \frac{\been_{p=\ppe}}{\hvarphi} \Bigl(\hkappa(1,x)  - \hkappa(0,x)-\tau   \Bigr)
\]
\[\hskip2cm+\frac{\been_{p=\ppo}}{\hvarphi}
\frac{\hvarphi(s,x)}{1-\hvarphi(s,x)}\frac{(y-{\hmu}(s,x,\ppo))\left({\hrho}(s,x)-{\hrho}(x)\right)}{{\hrho}(x)(1-{\hrho}(x))} \]
with the estimand
\begin{align}
\tau^{\ppo,\ppe} & =\mme\left[\psi(\yoio,\soio,\woin,\xoin,P_{i})+\tau\right].
\end{align}
\begin{remark}
An earlier version of the paper had a mistake in the representation of the influence function. We are grateful to Kevin Chen and David Ritzwoller for pointing this out. See 
\citet{chen2023semiparametric} for details. 
\end{remark}
\begin{theorem}\textsc{(Identification)}
$(i)$
Suppose that Assumption \ref{sampling} holds. Then, assuming all expectations are finite,
\[\tau^*= \tau^{\ppe}=\tau^{\ppo}=\tau^{\ppo,\ppe}.\]
$(ii)$
 Suppose that  Assumptions \ref{sampling}--\ref{ass:comp}
hold. Then the average treatment effect is equal to the following
three estimable functions of the data: 
\[
\tau\equiv\mme[\yoine-\yoinn|P_{i}=\ppe]=
\tau^*=\tau^{\ppe}=\tau^{\ppo}=\tau^{\ppo,\ppe},
\]
$(iii)$ Jointly Assumptions \ref{sampling}, \ref{ass:unconf}$(i)$, \ref{ass:surro}$(i)$ and \ref{ass:comp}$(i)$ have no testable implications.
\label{theorem1} \end{theorem}
\begin{remark}
The first part of the theorem implies that the four functionals of the joint distribution of $(\mathbf{1}_{\poin=\ppo}\yoin,\soin,\mathbf{1}_{\poin=\ppe} \woin,\xoin, \poin )$ are identical, irrespective of the Unconfoundedness, Surrogacy, and Comparability assumptions.
\end{remark}
\begin{remark}
Just like in the unconfoundedness case  \citep{newey1994asymptotic,chernozhukov2016double}, the influence function representation  is doubly robust. 
\citet{chen2023semiparametric} show that if the functions in the influence function that represent conditional expectations of the outcome,
$\hmu(s,x,\ppo)$ and $\hkappa(w,x)$, are correctly specified, then the influence function has expectation zero irrespective of the functions used for the various propensity score, $\hrho(s,x)$,  $\hrho(x),$ $\rho$, and the sampling score
 $\hvarphi(s,x)$. Similarly, if the various propensity score, $\hrho(s,x)$, $\hrho(x),$ $\rho$,
and the sampling score $\hvarphi(s,x)$ are correct, the influence function has expectation zero, irrespective of the functions used for the conditional outcome expectations $\hmu(s,x,\ppo)$ and $\hkappa(w,x)$.

\end{remark}

\subsection{Semiparametric Efficiency Bounds}

In this subsection we present two pairs of semiparametric efficiency bound results \citep{bickel1993efficient,newey1990semiparametric} for two different data configurations. The first directly refers to the main setup in this paper with the experimental and observational sample.
This result  is essentially shown in \cite{chen2023semiparametric} which corrects a mistake in an earlier version of the current paper. 
\begin{theorem}
\label{var_bound}    
 Suppose Assumptions   \ref{sampling}--\ref{ass:comp} hold.
 Then\\
 $(i)$   the semiparametric efficiency bound, normalized by the  square root of the sample size  $N$, is
 
\[
\mmv=\mme[\psi(\yoio,\soio,\woin,\xoin,P_{i})^2]\]
\[\qquad=\mme\bigg[ \frac{1-\hvarphi(\soin,\xoin)}{\hvarphi^2} \left( \left( \frac{\hvarphi(\soin,\xoin)}{1 -\hvarphi(\soin,\xoin)} \frac{\hrho(\soin,\xoin)- \hrho(\xoin)}{\hrho(\xoin)(1-\hrho(\xoin))} \right)^2 \sigma^2(\soin,\xoin,\ppo)  \right)  \]
% \[ + \frac{\hvarphi(\soin,\xoin)}{\hvarphi^2}  \left(\hkappa(1,\xoin) - \hkappa(0,\xoin) - \tau \right)^2   \]
\[+\frac{\hvarphi(\xoin)}{\hvarphi^2} \Bigl(\hkappa(1,\xoin)  - \hkappa(0,\xoin)  -\tau \Bigr)^2  \]
\[\left. + \frac{\hvarphi(\soin,\xoin)}{\hvarphi^2} \left(  \frac{ 
(1-\rho(\soin,\xoin))(\hmu(\soin,\xoin,\ppo) - \hkappa(0,\xoin))^2}{(1 - \hrho(\xoin))^2} + \frac{\rho(\soin,\xoin)(\hmu(\soin,\xoin,\ppo) - \hkappa(1,\xoin))^2}{\hrho(\xoin)^2} \right) \right].\]
% \[
% \mmv=\mme[\psi(\yoio,\soio,\woin,\xoin,P_{i})^2]\]
% \[\qquad=\mme\left[ \frac{1-\hvarphi(\xoin)}{\hvarphi^2} \left( \left( \frac{\hvarphi(\soin,\xoin)}{1 -\hvarphi(\soin,\xoin)} \frac{\hrho(\soin,\xoin)- \hrho(\xoin)}{\hrho(\xoin)(1-\hrho(\xoin))} \right)^2 \sigma^2(\soin,\xoin,\ppo)  \right) \right. \]
% \[\left. + \frac{\hvarphi(\xoin)}{\hvarphi^2} \left( \left(\hkappa(1,\xoin) - \hkappa(0,\xoin) - \tau \right)^2 + \frac{ (\hmu(\soin,\xoin,\ppo) - \hkappa(0,\xoin))^2}{1 - \hrho(\xoin)} + \frac{(\hmu(\soin,\xoin,\ppo) - \hkappa(1,\xoin))^2}{\hrho(\xoin)} \right) \right].\]
$(ii)$
If in addition the observational sample is large relative to the experimental sample, and $\sup_{s,x}\hvarphi(s,x)\rightarrow 0$, then the efficiency bound, now normalized by the expected sample size of the experimental sample, $\mme[N_\ppe]=\hvarphi N$ simplifies to
\[\mme\left[ \left.  \left( \hkappa(1,\xoin) - \hkappa(0,\xoin) - \tau \right)^2 + \frac{ (1-\woin)(\hmu(\soin,\xoin,\ppo) - \hkappa(0,\xoin))^2}{(1 - \hrho(\xoin))^2} + \frac{\woin(\hmu(\soin,\xoin,\ppo) - \hkappa(1,\xoin))^2}{\hrho(\xoin)^2} \right| \poin=\ppe \right].\]
\end{theorem}
\begin{remark}
This variance in  part $(ii)$ of Theorem \ref{var_bound}  is smaller than the effiency bound we would obtain in a randomized experiment where we do observe the primary outcome and did not observe the surrogate. 
The bound in that case is well known since \citet{hahn1998role}, 
\[\mme\left[ \left.  \left( \hkappa(1,\xoin) - \hkappa(0,\xoin) - \tau \right)^2 + \frac{(1-\woin) (\yoin - \hkappa(0,\xoin))^2}{(1 - \hrho(\xoin))^2} + \frac{\woin(\yoin - \hkappa(1,\xoin))^2}{\hrho(\xoin)^2} \right| \poin=\ppe \right].\]
This advantage in terms of asymptotic precision of using the (true) predicted outcome $\mu(S_i,X_i,\ppo)$ rather than the actual outcome $Y_i$ has been noted previously in \citep{day1996trial} in a setting with binary outcomes.
In the general case this gain is equal to
\[\mme\left[ \left.  \frac{ (1-\woin)(\yoin-\hmu(\soin,\xoin,\ppo))^2}{(1 - \hrho(\xoin))^2} + \frac{\woin(\yoin-\hmu(\soin,\xoin,\ppo))^2}{\hrho(\xoin)^2} \right| \poin=\ppe \right].\] %Intuitively, the surrogacy assumption brings additional information to bear on the problem -- namely that any variation in $\yoin$ conditional on $\soin$ is orthogonal to $\woin$ and hence is simplynoise that increases the residual variance of the outcome and reduces precision.  For example, if a program affects individuals' labor market trajectories entirely by changing their first job placement, then any subsequent changes in employment outcomes simply add noise. Hence, one can estimate the program's long-term effect on earnings more precisely by focusing solely on the portion of lifetime earnings that projects onto characteristics of an individual's first job.
\end{remark}

%To formalize this result, let  $\sigma_{w}^{2}(x)=\mmv(\yoin|\xoin=x,\woin=w)$. 
Next we consider the case where in a single sample we observe the treatment, primary outcome, surrogates and pre-treatment variables.
In this single sample case we do not need the fifth variable, $P_i\in\{\ppe,\ppo\}$. To maintain consistency with the other parts of the discussion and to avoid ambiguity, we keep the notation as before. In this case we can think of  $\poin$ always taking the value $\poin=\ppe$.  We calculate the efficiency bound both without the assumption that surrogacy holds and with the assumption that surrogacy holds. We do so for a data generating process where surrogacy does hold, to see the information gain from that assumption.

\begin{theorem}\label{theorem:effbound}  
Suppose Assumptions \ref{ass:unconf} and \ref{ass:surro} 
hold. 
$(i)$ The variance bound
without assuming surrogacy is
\[
\mmv_{{\rm ns}}=\mme\biggl[\sigma^{2}(\soin,\xoin,\ppe)\cdot\left(\frac{\hrho(\soin,\xoin)}{\rho(\xoin)^{2}}+\frac{1-\hrho(\soin,\xoin)}{(1-\rho(\xoin))^{2}}\right)
+\left(\mu(1,X_i)-\mu(0,X_i)-\tau\right)^2
\]
\[
\hskip2cm+\frac{\hrho(\soin,\xoin)}{\rho(\xoin)^{2}}\cdot\left(\mu(\soin,\xoin,\ppe)-\mu(1,\xoin)\right)^{2}+\frac{1-\hrho(\soin,\xoin)}{(1-\rho(\xoin))^{2}}\cdot\left(\mu(\soin,\xoin,\ppe)-\mu(0,\xoin)\right)^{2}
\biggr].\]
%\guido{end new}
$(ii)$ The efficiency gain from assuming surrogacy is
\[
\Delta=\mmv_{{\rm ns}} - \mmv_{{\rm s}} = \mathbb{E}\left[\sigma^2\left(S_i, X_i, E\right) \frac{\rho\left(S_i, X_i\right)\left(1-\rho\left(S_i, X_i\right)\right)}{\rho\left(X_i\right)^2\left(1-\rho\left(X_i\right)\right)^2}\right]  \geq 0 ,
\]
where $\mmv_{\rm s}$ is the variance bound for the case with surrogacy,
\[
\mmv_{{\rm s}}=\mme\left[\sigma^2(\soin,\xoin,\ppe)  \left( \frac{\hrhoe(\soin,\xoin)- \hrho(\xoin)}{\hrho(\xoin)(1-\hrho(\xoin))} \right)^2+ \left( \hkappa(1,\xoin) - \hkappa(0,\xoin) - \tau \right)^2  \right.\]
\[\left. + \frac{\hrhoe(\soin,\xoin)}{\hrho(\xoin)^2} (\hmu(\soin,\xoin,\ppe) -\hkappa(1,\xoin))^2
+ \frac{1-\hrhoe(\soin,\xoin)}{ (1 - \hrho(\xoin))^2} (\hmu(\soin,\xoin,\ppe) - \hkappa(0,\xoin))^2
 \right]
\]

% \[
% \mmv_{{\rm s}}=\mme\left[\sigma^2(\soin,\xoin,\ppe)  \left( \frac{\hrhoe(\soin,\xoin)- \hrho(\xoin)}{\hrho(\xoin)(1-\hrho(\xoin))} \right)^2+ \left( \hkappa(1,\xoin) - \hkappa(0,\xoin) - \tau \right)^2  \right.\]
% \[\left. + \frac{1}{\hrho(\xoin)} (\hmu(\soin,\xoin,\ppe) -\hkappa(1,\xoin))^2
% + \frac{1}{1 - \hrho(\xoin)} (\hmu(\soin,\xoin,\ppe) - \hkappa(0,\xoin))^2
%  \right]
% \]
\end{theorem}
\begin{remark} The expression for $\mmv_{\rm ns}$ is equivalent to the efficiency bound in \citep{hahn1998role}. It is written here in terms of the surrogates to facilitate the comparison to the efficiency bound exploiting surrogacy.
\end{remark}
\begin{comment}
\begin{remark}
The difference between the two bounds, $\mmv_{{\rm ns}}-\mmv_{{\rm s}}$,
is the efficiency gain from the Surrogacy assumption. 
\end{remark}
    
\end{comment}

\begin{remark}
Note that the variance bound in part $(ii)$ of Theorem \ref{theorem:effbound} differs from that in Theorem \ref{var_bound}$(ii)$  which was derived under the same surrogacy assumption, but assuming that the observational sample was infinitely large, so the relation between the surrogates and the primary outcome was known  without error. The result in $(ii)$ captures just the value of the surrogacy assumption.
\end{remark}
%\begin{remark}$\Delta$ stems from the ratio $(\hrhoe(\soin,\xoin)-\hrho(\xoin))^2/(\hrho(\xoin)(1-\hrho(\xoin))$. The expectation of the numerator, conditional on $\xoin$, is equal to the variance of the conditional expectation of $\mme[\woin|\soin,\xoin]$ conditional on $\xoin$, whereas the denominator is equal to the variance of $\woin$ conditional on $\xoin$, so the ratio is less than one in expectation, capturing the gain from knowledge that surrogacy holds.\end{remark}

\section{Violations of the Surrogacy and Comparability Assumptions:  Biases and Bounds}
\label{section:misspecification}

The three critical assumptions, Unconfoundedness, Surrogacy, and Comparability, are strong. 
There is a large literature studying the sensitivity to unconfoundedness conditions 
\citep{rosenbaumrubin_sensitivity, imbens2003, cinelli2020making} or bounds \citep{manski_bounds}.
Multiple studies have also raised concerns that in practice Surrogacy may not be satisfied \citep{begg2000, freedman1992statistical, frangakis2002principal,rosenbaum1984consequences,joffe2009related,vanderweele2015explanation}, although we are not aware of formal sensitivity or bounds analyses.
Violations of Comparability have not been explored because this assumption has not been previously formalized.
In this section we examine the biases
that arise from violations of Surrogacy and Comparability. We first characterize
these biases and then derive estimable bounds on the magnitude of the biases
that can arise from such violations. 
\subsection{Biases}

We begin by characterizing the probability limit of estimators based
on the representations of the estimand, $\tau^{\ppe}$, $\tau^{\ppo}$, and $\tau^{\ppo,\ppe}$, in Theorem \ref{theorem1}
when the Surrogacy and Comparability assumptions are violated, as
well as in cases where the surrogate index is misspecified. Throughout
the section, we maintain Unconfoundedness in the experimental sample (Assumption \ref{ass:unconf}), and the random sampling assumption (Assumption \ref{sampling}).
We denote the probability limit of the estimators by $\underline{\tau}$ to differentiate it from the average treatment effect $\tau=\mme[Y_i(1)-Y_i(0)|\poin=\ppe].$

%Recall the definition $\mu_{\ppe}(s,x,w)\equiv \mme\left[\left.\yoin\right|\soin=s,\xoin=x,\woin=w,P_{i}=\ppe\right]$. If surrogacy holds, then by Proposition \ref{prop2}, $\mu_{\ppe}(s,x,w)=\hmu(s,x,\ppe)$ for all $(s,x,w)$. If comparability holds, then $\hmu(s,x,\ppe)=\hmu(s,x,\ppo)$.

\begin{theorem} $(i)$ Suppose Assumption \ref{ass:unconf} (Unconfoundedness)
 holds, but Assumptions \ref{ass:surro}
(Surrogacy) and \ref{ass:comp} (Comparability) do not necessarily
hold. Then 
\[
\underline{\tau}\equiv \tau^{\ppo}=\tau^{\ppe}=\tau^{\ppe,\ppo}=\mme\left[\left.\hmu(\soine,\xoin,\ppo)-\hmu(\soinn,\xoin,\ppo)\right|\poin=\ppe\right].
\]
$(ii)$ Suppose Assumptions \ref{ass:unconf} (Unconfoundedness) and \ref{ass:comp} (Comparability)  hold, but Assumption \ref{ass:surro} (Surrogacy) does not necessarily hold. Then the difference between the average causal effect and the estimand is \[\textrm{\rm (surrogacy-bias)}\quad \tau-\underline{\tau}=
\mme\left[\left.\Bigl\{\hmu(\soin,1,\xoin,\ppe)-\hmu(\soin,0,\xoin,\ppe)\Bigr\}\cdot\frac{\hrhoe(\soin,\xoin)\cdot(1-\hrhoe(\soin,\xoin))}{\hrho(\xoin)\cdot(1-\hrho(\xoin))}\right|\poin=\ppe\right].
\]
$(iii)$. 
Suppose Assumptions \ref{ass:unconf} (Unconfoundedness) and
\ref{ass:surro} (Surrogacy)  hold,
but Assumption \ref{ass:comp} (Comparability) does not necessarily
hold. Then the difference between the average causal effect and the
estimand is 
\[\textrm{\rm (comparability-bias)}\quad
\tau-\underline{\tau}=\mme\left[\left.\Bigl\{ \hmu(\soin,\xoin,\ppe)-\hmu(\soin,\xoin,\ppo)\Bigr\}\cdot\frac{\hrhoe(\soin,\xoin)-\hrho(\xoin)}{\hrho(\xoin)\cdot(1-\hrho(\xoin))}\right|\poin=\ppe\right].
\]
$(iv)$. 
Suppose Assumption \ref{ass:unconf} (Unconfoundedness) holds, but Assumptions
\ref{ass:surro} (Surrogacy)  and
 \ref{ass:comp} (Comparability) do not necessarily
hold. Then the difference between the average causal effect and the
estimand is 
\begin{align*}
\textrm{\rm (total\ bias)}\quad
 & \tau-\underline{\tau}=\mme\left[\left(\mu(S_{i},1,X_{i},\ppe)-\mu(S_{i},0,X_{i},\ppe)\right)\cdot\frac{(1-\hrho(S_{i},X_{i}))\cdot \hrho(S_{i},X_{i})}{(1-\hrhoe(X_{i}))\cdot \hrhoe(X_{i})}\mid P_{i}=\ppe\right]\\
 & \quad{}+\mme\left[\left(\hmu(\soin,\xoin,\ppe)-\hmu(S_{i},X_{i},\ppo)\right)\cdot\frac{\hrho(S_{i},X_{i})-\hrhoe(X_{i})}{(1-\hrhoe(X_{i}))\cdot \hrhoe(X_{i})}\mid P_{i}=\ppe\right].
\end{align*}

\label{theorem:bias} \end{theorem}
\begin{remark}
Theorem \ref{theorem:bias}$(i)$ shows that even without Surrogacy
and Comparability, we estimate a valid average causal effect as long as unconfoundedness
holds. The treatment effect we estimate is the average effect of the
treatment on the surrogate index -- a principled aggregate of intermediate
outcomes -- rather than the average effect on the primary outcome.
This result also shows that the interpretation does not change with the choice of estimator (using the surrogate score approach,
the surrogate index approach, or the influence function). Theorem \ref{theorem:bias}$(ii-iv)$
show how violations of Comparability or Surrogacy affect the difference
between what is being estimated and the average treatment effect on
the primary outcome.
\end{remark}

\begin{remark}
The bias from violations of Surrogacy (Theorem \ref{theorem:bias}$(ii)$)
 consists of two factors.
 The first factor is small if the treatment does not explain much of the variation in $\yoin$ and therefore $\hmu(s,1,x,\ppe)$ and
 $\hmu(s,0,x,\ppe)$ 
are close. The second factor is small if the surrogate explains a large share of the variation in $\woin$, so that the surrogate score is close to zero or one and therefore $\mme[\hrhoe(\soin,\xoin)\cdot(1-\hrhoe(\soin,\xoin))]$ is close to zero.
\end{remark}

\begin{remark}
The bias from violations of Comparability (Theorem \ref{theorem:bias}$(iii)$)
 also consists of two factors. The first is the difference between the surrogacy index $\hmu(s,x,\ppo)$ and its counterpart in the experimental sample, $\hmu(s,x,\ppe)$.  The second factor depends on the deviation between the surrogacy score and the propensity score, $\rho(\soin,\xoin)-\rho(\xoin)$. If the treatment does not have much effect on the surrogates, violations of Comparability do not generate much bias, because the bias that comes  from a combination of the effect of the treatment on the surrogates and the effect of the surrogates on the outcome, will be small in that case.
\end{remark}

\subsection{Bounds on the  Bias}
\label{subsec:bounds_bias}
In this subsection we explore bounds on the parameter of interest. We show that in general these bounds are uninformative. However, if outcomes themselves are bounded, for example, if the outcomes are binary, informative bounds can be derived. Moreover, we present bounds given assumptions on the range of violations of the Surrogacy and Comparability assumptions.

\begin{lemma}
Suppose Assumptions \ref{ass:unconf} (Unconfoundedness) and
\ref{ass:comp} (Comparability) 
hold, but Assumption \ref{ass:surro} (Surrogacy) does not necessarily
hold. Then:\\
$(i)$
If the outcome can take on values on the whole real line, then there is no value for the average treatment effect $\tau$ that can be ruled out.\\
$(ii)$ if the outcome is binary, 
then the average treatment effect $\tau$ is inside the interval
\[
\biggl\{
\mme\left[\left.\hmu(\soine,\xoin,\ppo)-\hmu(\soinn,\xoin,\ppo)\right|\poin=\ppe\right]+ \mme\left[\left.
\Delta^L_S(\soin,\xoin)
\frac{\hrhoe(\soin,\xoin)(1-\hrhoe(\soin,\xoin))}{\hrho(\xoin)(1-\hrho(\xoin))}\right|\poin=\ppe\right],
\]
\[
\mme\left[\left.\hmu(\soine,\xoin,\ppo)-\hmu(\soinn,\xoin,\ppo)\right|\poin=\ppe\right]+\mme\left[\left.
\Delta_S^U(\soin,\poin)
\frac{\hrhoe(\soin,\xoin)(1-\hrhoe(\soin,\xoin))}{\hrho(\xoin)(1-\hrho(\xoin))}\right|\poin=\ppe\right]
\biggr\},
\]
where
\[ \Delta^L_S(s,x)=-
\min\left(\frac{1-\hmu(s,x,\ppo)}{\hrhoe(s,x)},\frac{\hmu(s,x,\ppo)}{1-\hrhoe(s,x)}\right)
%\max\left(-1,\frac{\hmu(s,x,\ppo)-1}{\hrho(s,x)}\right)
\qquad
\Delta_S^U(s,x)=
\min\left(\frac{\hmu(s,x,\ppo)}{\hrhoe(s,x)},\frac{1-\hmu(s,x,\ppo)}{1-\hrhoe(s,x)}\right),
%\min\left(1,\frac{\hmu(s,x,\ppo)}{\hrho(s,x)}\right).
\]
and these bounds on the bias are sharp.\\
$(iii)$ if the direct effect of the treatment on the outcome $\mu(\soin,1,\xoin,\ppe)-\mu(\soin,0,\xoin,\ppe)$ is bounded in absolute value by $c$, then the average treatment effect $\tau$ is inside the interval
\[
\biggl\{
\mme\left[\left.\hmu(\soine,\xoin,\ppo)-\hmu(\soinn,\xoin,\ppo)\right|\poin=\ppe\right]-c\cdot \mme\left[\left.
\frac{\hrhoe(\soin,\xoin)(1-\hrhoe(\soin,\xoin))}{\hrho(\xoin)(1-\hrho(\xoin))}\right|\poin=\ppe\right],
\]
\[
\mme\left[\left.\hmu(\soine,\xoin,\ppo)-\hmu(\soinn,\xoin,\ppo)\right|\poin=\ppe\right]+c\cdot \mme\left[\left.
\frac{\hrhoe(\soin,\xoin)(1-\hrhoe(\soin,\xoin))}{\hrho(\xoin)(1-\hrho(\xoin))}\right|\poin=\ppe\right]
\biggr\},
\]
and this bound is sharp.
\label{lemma1}
\end{lemma}
\begin{remark}
To provide some intuition for the sharpness of the bounds, consider the surrogacy bias in Theorem \ref{theorem:bias}. The bias has two factors, with the second estimable from the data. The first factor is the difference $\hmu(\soin,1,\xoin,\ppe)-\hmu(\soin,0,\xoin,\ppe)$. The data are not directly informative about this difference beyond the fact that the weighted average
$\rho(\soin,\xoin)\hmu(\soin,1,\xoin,\ppe)+(1-\rho(\soin,\xoin))\hmu(\soin,0,\xoin,\ppe)$ is equal to the estimable quantity $\hmu(\soin,\xoin,\ppo)$. In the absence of any restrictions on the outcome this implies there are no restrictions on $\hmu(\soin,w,\xoin,\ppe)$ or on the difference $\hmu(\soin,1,\xoin,\ppe)-\hmu(\soin,0,\xoin,\ppe)$, and thus not on  the bias or the average treatment effect. Given restrictions on the range of the outcome this representation directly leads to upper and lower bounds on the bias and the average treatment effect.
\end{remark}
\begin{lemma}
 Suppose Assumptions \ref{ass:unconf} (Unconfoundedness) and
\ref{ass:surro} (surrogacy) hold,
but Assumption \ref{ass:comp} (Comparability) does not necessarily
hold. Then:\\
%\[\mme\left[\left.\Bigl\{ \hmu(\soin,\xoin,\ppe)-\hmu(\soin,\xoin,\ppo)\Bigr\}\cdot\frac{\hrho(\soin,\xoin)-e(\xoin)}{e(\xoin)\cdot(1-e(\xoin))}\right|P_{i}=\ppe\right].\]\noindent
$(i)$
If the outcome can take on value on the whole real line, then there is no value for the  average treatment effect $\tau$ that can be ruled out.\\
$(ii)$ if the outcome is binary, then the average treatment effect $\tau$ is inside the interval
\[\left\{
\mme\left[\left.\hmu(\soine,\xoin,\ppo)-\hmu(\soinn,\xoin,\ppo)\right|\poin=\ppe\right]+
\mme\left[\left.\Bigl\{ \mathbf{1}_{\hrhoe(\soin,\xoin)<\hrho(\xoin)}-\hmu(\soin,\xoin,\ppo)\Bigr\}\frac{\hrhoe(\soin,\xoin)-\hrho(\xoin)}{\hrho(\xoin)(1-\hrho(\xoin))}\right|\poin=\ppe\right],\right.
\]
\[\left.
\mme\left[\left.\hmu(\soine,\xoin,\ppo)-\hmu(\soinn,\xoin,\ppo)\right|\poin=\ppe\right]+
\mme\left[\left.\Bigl\{ \mathbf{1}_{\hrhoe(\soin,\xoin)>\hrho(\xoin)}-\hmu(\soin,\xoin,\ppo)\Bigr\}\frac{\hrhoe(\soin,\xoin)-\hrho(\xoin)}{\hrho(\xoin)(1-\hrho(\xoin))}\right|\poin=\ppe\right]\right\},
\]
with width
\[2 \mme\left[\left. \mathbf{1}_{\hrhoe(\soin,\xoin)>\hrho(\xoin)}\cdot\frac{\hrhoe(\soin,\xoin)-\hrho(\xoin)}{\hrho(\xoin)\cdot(1-\hrho(\xoin))}\right|\poin=\ppe\right]
,
\]
and these bounds on the bias are sharp.\\
$(iii)$ if  $\mu(\soin,\xoin,\ppe)-\hmu(\soin,\xoin,\ppo)$ is bounded in absolute value by $c$, then the average treatment effect $\tau$ is inside the interval
\[\left\{
\mme\left[\left.\hmu(\soine,\xoin,\ppo)-\hmu(\soinn,\xoin,\ppo)\right|\poin=\ppe\right]-c\cdot \mme\left[\left.\frac{|\hrhoe(\soin,\xoin)-\hrho(\xoin)|}{\hrho(\xoin)\cdot(1-\hrho(\xoin))}\right|\poin=\ppe\right],\right.
\]
\[\left.
\mme\left[\left.\hmu(\soine,\xoin,\ppo)-\hmu(\soinn,\xoin,\ppo)\right|\poin=\ppe\right]+c\cdot \mme\left[\left.\frac{|\hrhoe(\soin,\xoin)-\hrho(\xoin)|}{\hrho(\xoin)\cdot(1-\hrho(\xoin))}\right|\poin=\ppe\right]\right\},
\]
and these bound are sharp.
\label{lemma2}
\end{lemma}

\section{Estimation}\label{estimation}

In this section, we first present four estimators for the average treatment effect. The first, the surrogate index estimator, is related to previously proposed estimators with the difference that in the earlier literature the surrogate index was implicitly assumed to be known. We
then discuss three new alternative estimators. The last of these new estimators is a matching
estimator. Although matching estimators are generally not efficient
in settings with unconfoundedness (\citealt{rubin2006matched,abadie2006,abadie2016matching}),
they are widely applied, and it is instructive to see
how a matching strategy can be used here.
\subsection{Surrogate Index}

Suppose we estimate the surrogate index as $\hat{\hmu}(s,x,\ppo)$ and the propensity score as $\hat{\hrho}_\ppe(x)$.
We take an average of the surrogate index in the experimental sample
for the treatment and control groups, after adjusting for the propensity
score. A natural estimator, corresponding to (\ref{estimand1}), is
the following difference of the two averages over the experimental
sample: 
\begin{equation}
\hat{\tau}^{\ppe}=\frac{1}{\sum_{i=1}^{N_{\ppe}}\woin/\hat{\hrho}(\xoin)}\sum_{i=1}^{N_{\ppe}}\hat{\hmu}(\soin,\xoin,\ppo)\cdot\frac{\woin}{\hat{\hrho}(\xoin)}\label{est_drie}
\end{equation}
\[
\hskip3cm-\frac{1}{\sum_{i=1}^{N_{\ppe}}(1-\woin)/(1-\hat{\hrho}(\xoin))}\sum_{i=1}^{N_{\ppe}}\hat{\hmu}(\soin,\xoin,\ppo)\cdot\frac{1-\woin}{1-\hat{\hrho}(\xoin)}.
\]
We refer to this as the surrogate index estimator. Note that compared
to the representation in  Theorem \ref{theorem1}, we normalize the weights so
that the weights sum up to one. This tends to improve the finite sample
properties of related estimators in other settings substantially (\citealt{hirano2003efficient,busso2014new}).

In the case where the estimator for the surrogate index ${\mu}(s,x,\ppo)$
was based on a linear specification for the regression of the primary
outcome on the intermediate outcome, $\hmu(s,x,\ppo)=\gamma_{0}+\gamma_{S}'s+\gamma_{X}'x$,
this leads to 
\[
\hat{\tau}^{\ppe}=\hat{\gamma}_{S}'\hat{\tau}_{S},
\]
where $\hat{\tau}_{S}$ is an estimator for the average effect of the treatment on the surrogates, $\mme\left[\soine-\soinn\right].$
In the simplest case without pre-treatment variables and where the
experimental sample is randomized, $\hat{\tau}_{S}=\overline{S}_{1}-\overline{S}_{0}$,
where $\overline{S}_{1}$ and $\overline{S}_{0}$ are the average
values of the surrogate outcomes. Here, the estimator simplifies to
the difference in the estimated surrogate index in the treatment group
and the control group: 
$
\hat{\tau}^{\ppe}=\hat{\gamma}_{S}'(\overline{S}_{1}-\overline{S}_{0})$. This expression is also familiar from the mediation literature (\textit{e.g.},
\citealt{baron1986moderator}) and the surrogacy literature \citep{day1996trial}. However, we emphasize that in general,
there may be interactions between the surrogates and pre-treatment
variables, and in that case the linear specification need not be not adequate.

\subsection{Surrogate Score Estimator}

We now use the second representation
for $\tau$ in the main theorem to derive an alternative estimator.
Let $\hat{\rho}(x)$, $\hat{\rho}(s,x),\hat{\varphi}(s,x),\hat{\varphi}(x)$, and $\hat{\varphi}$, be estimators
for $\rho(x)$, $\hrho(s,x),\varphi(s,x),\varphi(x)$, and $\varphi$ respectively.

The surrogate score estimator is based on averaging the following
expression over the observational sample: 
\begin{equation}
\hat{\tau}^{\ppo}=\frac{1}{\sum_{i|\poin=\ppo}\omega_{1,i}}\sum_{i|\poin=\ppo}\yoio\cdot\omega_{1,i}
-\frac{1}{\sum_{i|\poin=\ppo}\omega_{0,i}}\sum_{i|\poin=\ppo}\yoio\cdot\omega_{0,i}
,\label{est_een}
%\hat{\tau}^{\ppo}=\frac{1}{\sum_{i=1}^{N_{\ppo}}\omega_{1,\hat{\hrho}(s,x),\hat{\hrho}(x),\hat{\hrho},\hat{\hvarphi}(s,x),\hat{\hvarphi}(x),\hat{\hvarphi}}}\sum_{i=1}^{N_{\ppo}}\yoio\cdot\omega_{1,\hat{\hrho}(s,x),\hat{\hrho}(x),\hat{\hrho},\hat{\hvarphi}(s,x),\hat{\hvarphi}(x),\hat{\hvarphi}}    -\frac{1}{\sum_{i=1}^{N_{\ppo}}\omega_{0,\hat{\hrho}(s,x),\hat{\hrho}(x),\hat{\hrho},\hat{\hvarphi}(s,x),\hat{\hvarphi}(x),\hat{\hvarphi}}}\sum_{i=1}^{N_{\ppo}}\yoio\cdot\omega_{0,\hat{\hrho}(s,x),\hat{\hrho}(x),\hat{\hrho},\hat{\hvarphi}(s,x),\hat{\hvarphi}(x),\hat{\hvarphi}},\label{est_een}
\end{equation}
where for $w=0,1$ the weights are 
%\[\omega_{w,\hat{\hrho}(s,x),\hat{\hrho}(x),\hat{\hrho},\hat{\hvarphi}(s,x),\hat{\hvarphi}(x),\hat{\hvarphi}}=\frac{\hat{\hrho}(\soio,\xoio)^{w}\cdot(1-\hat{\hrho}(\soio,\xoio))^{1-w}\cdot\hat{\hvarphi}(\soio,\xoio)\cdot(1-\hat{\hvarphi})}{\hat{\hrho}(\xoio)^{w}\cdot(1-\hat{\hrho}(\xoio))^{1-w}\cdot(1-\hat{\hvarphi}(\soio,\xoio))\cdot \hat{\hvarphi}}.\]
\begin{equation}\label{weights1}\omega_{w,i}=\frac{\hat{\hrho}(\soio,\xoio)^{w}\cdot(1-\hat{\hrho}(\soio,\xoio))^{1-w}\cdot\hat{\hvarphi}(\soio,\xoio)\cdot(1-\hat{\hvarphi})}{\hat{\hrho}(\xoio)^{w}\cdot(1-\hat{\hrho}(\xoio))^{1-w}\cdot(1-\hat{\hvarphi}(\soio,\xoio))\cdot \hat{\hvarphi}}.\end{equation}

\subsection{Influence Function Estimator}

We can also base estimation on
the efficient score given in (\ref{effscore}). Given estimators
for the propensity score, the surrogate score, and the sampling score,
we can estimate the average treatment effect as 
\begin{equation}\label{if_estimator}\hat{\tau}^{E,O}=\sum_{i=1}^{N}\Biggl\{
\frac{\been_{\poin=\ppe}}{\hat{\hvarphi}}\left(\frac{\woin\cdot \hat{\hmu}(\soin,\xoin,\ppo)}{\hat{\hrho}(\xoin)}-\frac{(1-\woin)\cdot \hat{\hmu} (\soin,\xoin,\ppo)}{1-\hat{\hrho}(\xoin)}\right) 
\end{equation}
\[+ \frac{\been_{\poin=\ppe}}{\hat{\hvarphi}} \Biggl(\hat{\hkappa}(1,\xoin)  \left( 1 - \frac{\woin}{\hat{\hrho}(\xoin)} \right) - \hat{\hkappa}(0,\xoin)  \left( 1 - \frac{1 - \woin}{1 - \hat{\hrho}(\xoin)} \right) \Biggr)
\]
\[
\hskip2cm+\frac{\been_{\poin=\ppo}}{1-\hat{\hvarphi}}\left(\frac{\hat{\hvarphi}(\soin,\xoin)}{1-\hat{\hvarphi}(\soin,\xoin)}
\frac{1-\hat{\hvarphi}}{\hat{\hvarphi}}\right)
\frac{(\yoin-\hat{\hmu}(\soin,\xoin,\ppo))\left(\hat{\hrho}(\soin,\xoin)-\hat{\hrho}(\xoin)\right)}{\hat{\hrho}(\xoin)(1-\hat{\hrho}(\xoin))} \Biggr\}.
\]
Based on the results in \citet{newey1994asymptotic}, it follows that
under standard conditions the two estimators above and the surrogate
index estimator all reach the semi-parametric efficiency bound, and are first-order equivalent.

The recent literature on double robust estimation of average treatment effects under unconfoundedness \citep{chernozhukov2016double} suggests that this estimator may have superior properties in small samples.

\subsection{Double Matching Estimator}

% Finally, we consider a matching
% estimator. Although matching estimators are generally not efficient
% in settings with unconfoundedness (\citealt{rubin2006matched,abadie2006,abadie2016matching}),
% they are intuitive and widely applied, and it is instructive to see
% how a matching strategy can be used here.

Consider unit $i$ in the experimental sample with $\xoin=x$ and
$\soin=s$, and suppose this is a treated unit with $\woin=1$. We
need to find three matches for this unit. First, we need to find a
unit with the opposite treatment in the same (experimental) sample.
Specifically, we need to find the closest unit in the experimental
sample, in terms of pre-treatment variables, among the units with
$\woin=0$. Suppose this unit is unit $j$, with $\wojn=0$, and the
value of the pre-treatment variables for this unit are $\xojn=x'$,
and the surrogate outcomes are $\sojn=s'$. As a result of the matching
we should have $x\approx x'$, but potentially $s$ could be quite
different from $s'$. Next, we need to find for each of the two units
$i$ and $j$ a match in the observational sample. Find the unit in
the observational sample closest to unit $i$, in terms of both pre-treatment
variables and surrogates. Let $i'$ be the index for this unit,
and let the value of the outcome for this unit be $\yoipo$, and the
values of the pre-treatment variables and surrogates $\xoipo$ and
$\soipo$. Now as a result of the matching $\xoio\approx\xoipo$ and
$\soio\approx\soipo$. Finally, find the unit in the observational
sample closest to unit $j$, in terms of both pre-treatment variables
and surrogates. Let the value of the outcome for this unit be $\yojpo$,
and the values of the pre-treatment variables and surrogates $\xojpo$
and $\sojpo$, with $\xojo\approx\xojpo$ and $\sojo\approx\sojpo$.

Then we combine these matches to estimate the causal effect for unit
$i$, $\yoine-\yoinn$, as the difference in average outcomes for
the two matches from the observational sample: 
\begin{equation}
\widehat{\yoine-\yoinn}=\yoipo-\yojpo.\label{eq:three-match-equation}
\end{equation}
The matching estimator for $\tau$ would then be the average value
of (\ref{eq:three-match-equation}) over the experimental sample.
The double matching estimator is then
\[ \hat\tau^{\rm match}=\frac{1}{N^\ppe}\sum_{i:\poin\ppe}\left\{ W_i \left( Y_{i'}-Y_{j'}\right)
+(1-W_i) \left( Y_{j'}-Y_{i'}\right)\right\}.
\]

\section{Application: Impacts of Job Training on Employment}\label{application}

In this section, we apply our method to estimate the causal effect
of the Greater Avenues to Independence (GAIN) job training program
on long-term labor market outcomes.
GAIN was a job assistance program implemented in California in the
 1980s to help welfare recipients find work (\citealt{riccio1989gain,friedlander1995evaluating, hotz2006evaluating}). MDRC conducted a
randomized trial to evaluate the GAIN program's employment impacts
in six counties in California in the late 1980s. We focus primarily
on the GAIN trial in Riverside, which was widely heralded as the program
that had the largest treatment effects on earnings. The Riverside
program emphasized a ``jobs first'' approach to re-entry into the
labor force, encouraging unemployed workers to take any job they find;
in contrast, other sites focused more heavily on developing human
capital through training programs (\citealt{hotz2006evaluating}).

We have available long-term outcomes for the four GAIN sites, including employment, earnings, and receipt of aid over the first thirty-six quarters after random assignment. We take the average of the thirty-six employment indicators and earnings in Riverside as our primary outcomes. We then investigate whether we could have predicted the long-term impact on these outcomes using only the first $T$ quarters of all outcomes (including  employment, earnings, and aid) as surrogates, as well as using pre-treatment variables (characteristics of the individuals as well as lagged employment, earnings and aid outcomes). The Riverside data on the treatment, surrogates and pre-treatment variables play the role of of our experimental sample. We use the data from the combination of the other three locations (Alameda, Los Angeles, and San Diego) as our observational sample. For the observational sample we only use the information on the surrogates, pre-treatment variables, and outcome, but not the treatment assignment, nor the indicator for the location.

We begin by presenting a brief summary of the samples.
We then describe how we construct our surrogate index. Next we illustrate our theoretical
results by evaluating the magnitude of the gains from using surrogate
indices in terms of time and precision relative to existing experimental
estimates of the program's long-term impacts in Riverside. We also
show how one can validate the surrogacy assumption using intermediate
outcomes and bound the degree of bias arising from potential violations
of surrogacy.

\subsection{The GAIN Program}

The GAIN treatment was randomly assigned to welfare (Aid for Families
with Dependent Children) recipients,  a very low-income population.
The treatment group consisted of $N_{\ppe,T}=4405$ participants, which
the control group consisted of $N_{\ppe,C}=1040$ participants who were
not eligible for the additional services in the GAIN program. 
The data we use come from the
\citet{hotz2006evaluating} which followed study participants for nine years
after assignment of the treatment, measuring quarterly employment
rates and earnings\footnote{ All income variables were converted to 1999 dollars using cost-of-living deflators; see footnote 21 of \citet{hotz2000long} for more information.} from the Unemployment Insurance database. They
found that the treatment effects of the Riverside GAIN program on
employment rates and earnings were initially large, but declined over
time, as shown in Figure 3A, which plots employment rates  by quarter for individuals in
the experimental (Riverside) treatment and control groups, and in Figure 3B, which shows the correspond results for quarterly earnings.

\begin{comment}
\begin{figure}[htbp]
    \centering
    \includegraphics[width=0.7\textwidth]{Figures/Figure3A_Employ.png} 
    \label{fig:Figure3A_Covs_filt}
\end{figure}

\begin{figure}[htbp]
    \centering
    \includegraphics[width=0.7\textwidth]{Figures/Figure3B_Earn.png} 
    \label{fig:Figure3A_Covs_filt}
\end{figure}
\end{comment}
\begin{figure}[htbp]
\centering
\begin{subfigure}[b]{0.49\textwidth}
\centering
\includegraphics[width=\textwidth]{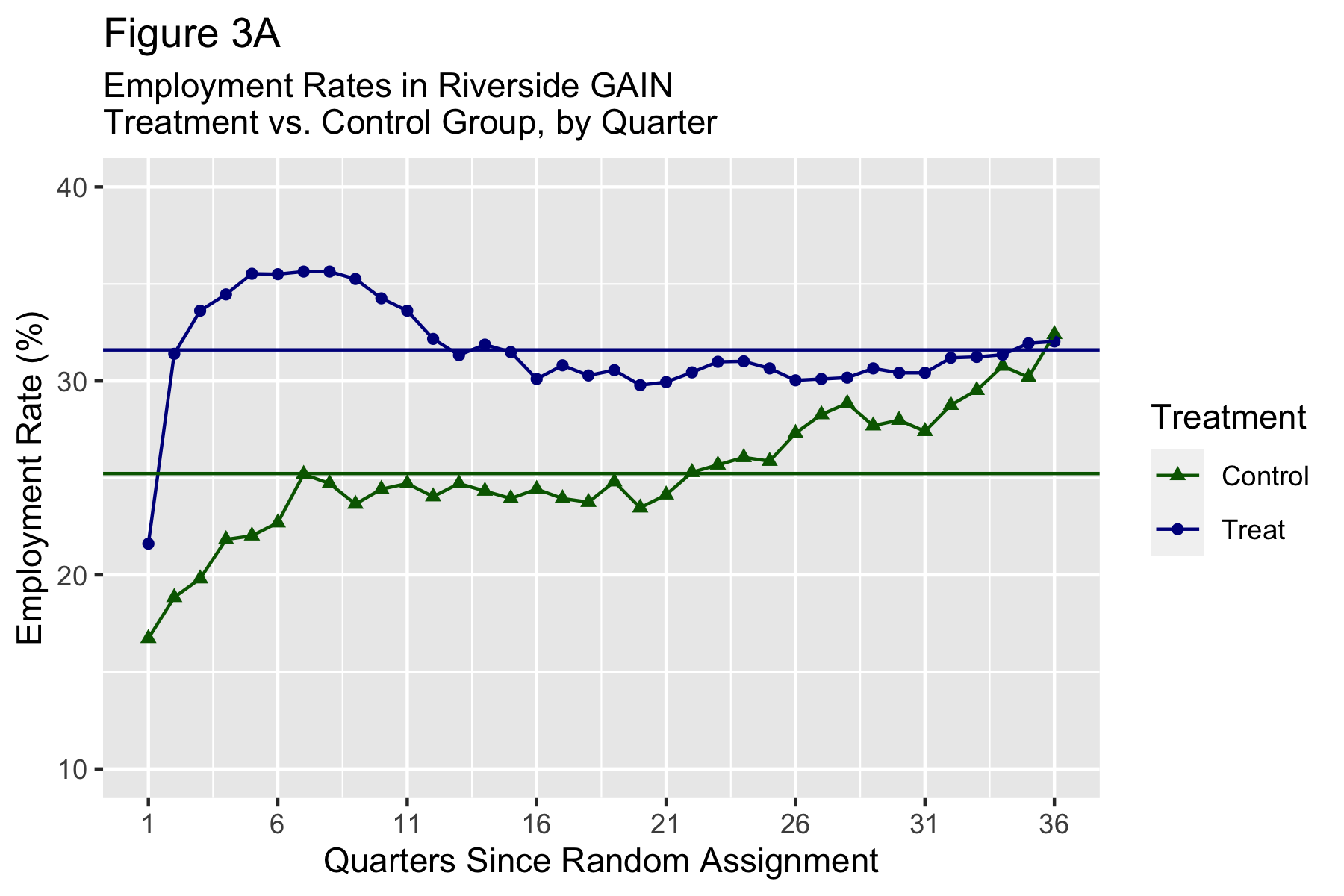}
\caption{Employment}
\label{fig:Figure3A_Employ}
\end{subfigure}
%\hfill
\begin{subfigure}[b]{0.50\textwidth}
\centering
\includegraphics[width=\textwidth]{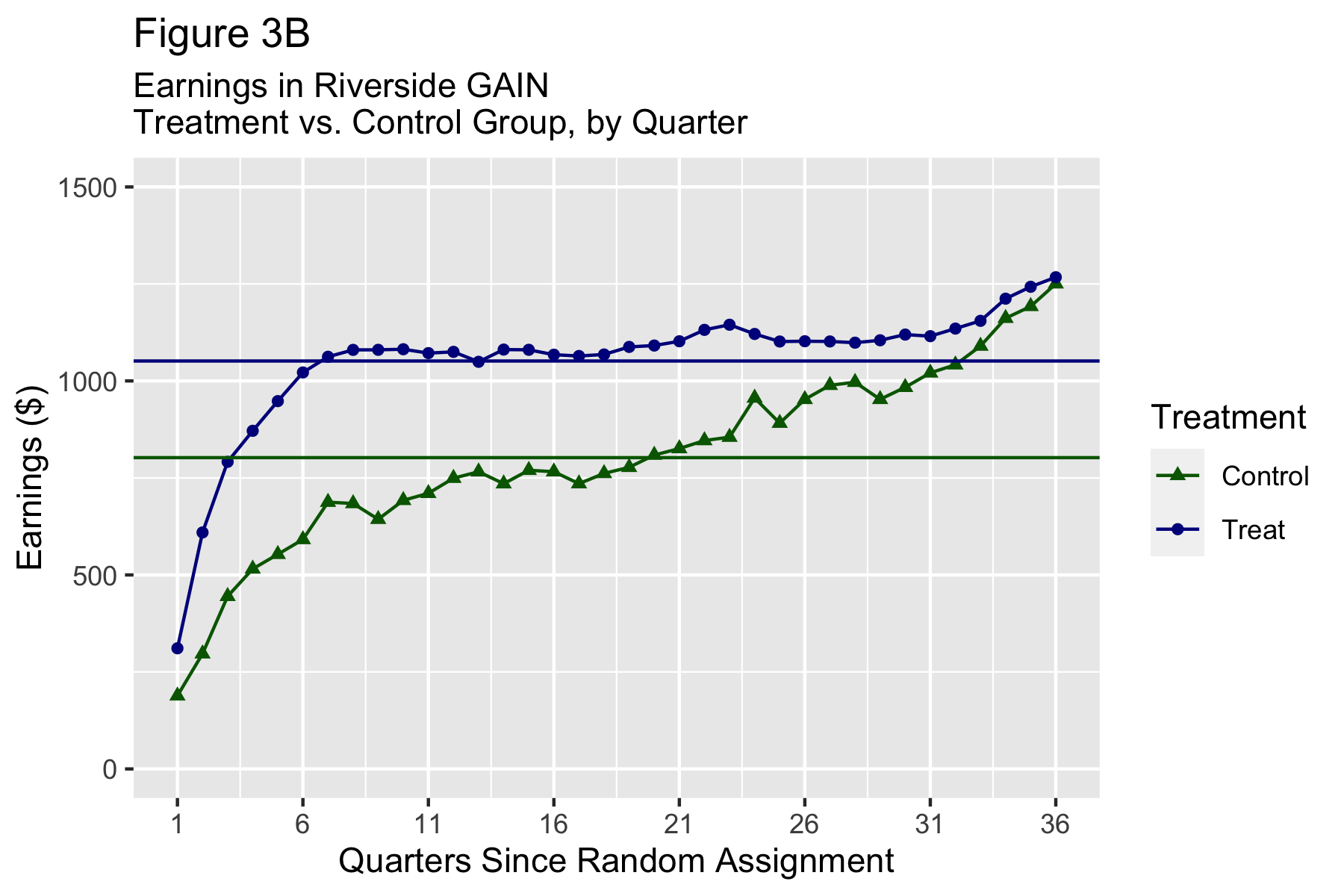}
\caption{Earnings}
\label{fig:Figure3B_Earn}
\end{subfigure}
%\caption{Employment and Earnings}
\label{fig:Figure3}
\end{figure}

In Riverside, the estimated causal effects on the primary outcomes were a 6.4 (s.e. = 1.2) percentage point (pp) increase
in average quarterly employment rates, and an \$249  (s.e. \$84) increase in average quarterly earnings, in both cases averaged over the 36 quarter post-treatment. Our question is whether these
impacts could have been estimated more quickly by using short-term
employment, earnings and aid receipt as surrogates. 

The observational sample includes the other three locations, Alameda, Los Angeles and San Diego, for a total of $N_\ppo= 13,725$ individuals.

In the online appendix Table \ref{tab:Normalized_Diff} presents information on the pre-treatment variables. Clearly the two samples, Riverside and the combination of the other three locations, are substantially different prior to the intervention in terms of permanent characteristics such as ethnicity, as well as in pre-treatment outcomes.

\subsection{Three Estimators}

We discuss here the estimators for the average effect of the program
We wish to consider different set of surrogates, indexed by the number of periods $t$ we want to use as surrogates.  To capture this we index the surrogate for individual $i$, $S_i^t$, by the superscript $t$. $S_i^t$ contains the employment indicators, earnings outcomes and aid receipt indicators for the $t$ quarters after the intervention.

\subsubsection{Surrogate Index Estimator}

To construct the surrogacy index we estimate a
linear regression model using least squares, for the individuals in the observational sample 
\begin{equation}
Y_{i}=\beta_{0}+\beta_{S}^\top S_{i}^t+\beta_X^\top X_i+\varepsilon_{i}.\label{eq:surr_index_estimation}
\end{equation}
The predicted value from this regression, which we denote by $\hat{Y}_{i}$,
is our surrogate index for mean employment
based on surrogates up to quarter $t$. We then
compute this surrogate index for each of the individuals in the experimental
sample and estimate the treatment effect based on the
surrogate index as 
\begin{equation}
\hat{\tau}^\ppo=\frac{1}{N_{\ppe,T}}\sum_{i=1}^{N^\ppe}\hat{Y}_{i}W_{i}-\frac{1}{N_{\ppe,C}}\sum_{i=1}^{N_\ppe}\hat{Y}_{i}(1-W_{i}).\label{eq:treat_effect_estimation}
\end{equation}
If we use the all 36 quarters  of employment indicators are surrogates, then the regression of $Y_i$ on the set of surrogates will fit perfectly, $\hat{Y}_i$ will be equal to $Y_i$, and the estimated effect will be identical to the original experimental estimate. The question is whether using a much more limited set of surrogates will get us close to the experimental benchmark.

\subsubsection{Surrogate Score Estimator}

For the surrogate score estimator we first estimate a logistic regression of the treatment indicator on the pretreatment variables and the surrogates. We specify
\[ 
 \ln\left(
\frac{\rho(S_i^t,X_i)}{1-\rho(S_i^t,X_i)}
\right)\equiv \ln\left(
\frac{\pr(W_i=1|S_i^t,X_i,\poin=\ppe)}{1-\pr(W_i=1|S_i^t,X_i,\poin=\ppe)}
\right)=
\alpha_0+\alpha_{S}^\top S_{i}^t+\alpha_X^\top X_i,\]
and estimate this on the experimental (Riverside) sample.

Next we estimate the propensity score, also as a logistic regression,
\[ 
 \ln\left(
\frac{\rho(X_i)}{1-\rho(X_i)}
\right)\equiv \ln\left(
\frac{\pr(W_i=1|X_i,\poin=\ppe)}{1-\pr(W_i=1|X_i,\poin=\ppe)}
\right)=
\delta_0+\delta_X^\top X_i,\]
and estimate this again on the experimental (Riverside) sample. 
In principle the random assignment implies that the $\delta_X$ should be close to zero in this case.

Finally we estimate the comparability score
\[ 
 \ln\left(
\frac{\varphi(S_i^t,X_i)}{1-\varphi(S_i^t,X_i)}
\right)\equiv \ln\left(
\frac{\pr(\poin=\ppe|X_i,S_i^t)}{1-\pr(\poin=\ppe|X_i,S_i^t)}
\right)=
\gamma_0+\gamma_{S}^\top S_{i}^t+\gamma_X^\top X_i,\]
and estimate this on the combined observational and experimental samples.

The surrogate score estimator is based on averaging the following
expression over the observational sample: 
\begin{equation}
\hat{\tau}^{\ppo}=\frac{1}{\sum_{i|\poin=\ppo}\omega_{1,i}}\sum_{i|\poin=\ppo}\yoio\cdot\omega_{1,i}
-\frac{1}{\sum_{i|\poin=\ppo}\omega_{0,i}}\sum_{i|\poin=\ppo}\yoio\cdot\omega_{0,i},
\end{equation}
where  the weights are as before in Equation (\ref{weights1}).
%\[ \omega_{w,\hat{\hrho}(s,x),\hat{\hrho}(x),\hat{\hrho},\hat{\hvarphi}(s,x),\hat{\hvarphi}(x),\hat{\hvarphi}}=\frac{\hat{\hrho}(\soio,\xoio)^{w}\cdot(1-\hat{\hrho}(\soio,\xoio))^{1-w}\cdot\hat{\hvarphi}(\soio,\xoio)\cdot(1-\hat{\hvarphi})}{\hat{\hrho}(\xoio)^{w}\cdot(1-\hat{\hrho}(\xoio))^{1-w}\cdot(1-\hat{\hvarphi}(\soio,\xoio))\cdot \hat{\hvarphi}}. \]

\subsubsection{Influence Function Estimator}

For the influence function estimator we first estimate the surrogacy index, the surrogacy score, the propensity score, and the comparability score as before. We then plug those into the estimator in Equation (\ref{if_estimator}).
%\[\hat{\tau}^{E,O}=\sum_{i=1}^{N}\Biggl\{ \frac{\been_{\poin=\ppe}}{\hat{\hvarphi}}\left(\frac{\woin\cdot \hat{\hmu}(\soin,\xoin,\ppo)}{\hat{\hrho}(\xoin)}-\frac{(1-\woin)\cdot \hat{\hmu} (\soin,\xoin,\ppo)}{1-\hat{\hrho}(\xoin)}\right)  \] \[+ \frac{\been_{p=\ppe}}{\hat{\hvarphi}} \Biggl(\hat{\hkappa}(1,\xoin)  \left( 1 - \frac{\woin}{\hat{\hrho}(\xoin)} \right) - \hat{\hkappa}(0,\xoin)  \left( 1 - \frac{1 - \woin}{1 - \hat{\hrho}(\xoin)} \right) \Biggr) \] \[ \hskip2cm+\frac{\been_{p=\ppo}}{1-\hat{\hvarphi}}\left(\frac{\hat{\hvarphi}(\soin,\xoin)}{1-\hat{\hvarphi}(\soin,\xoin)} \frac{1-\hat{\hvarphi}}{\hat{\hvarphi}}\right) \frac{(\yoin-\hat{\hmu}(\soin,\xoin,\ppo))\left(\hat{\hrho}(\soin,\xoin)-\hat{\hrho}(\xoin)\right)}{\hat{\hrho}(\xoin)(1-\hat{\hrho}(\xoin))} \Biggr\}. \]

\subsection{Results}

Here we discuss two sets of results. First the estimates for the average effect of the intervention on the two primary outcomes under various assumptions about the surrogates. Second, we test the Surrogacy and Comparability assumptions directly.

\subsubsection{Estimation Results}

\begin{figure}[htbp]
    \centering
    \includegraphics[width=0.7\textwidth]{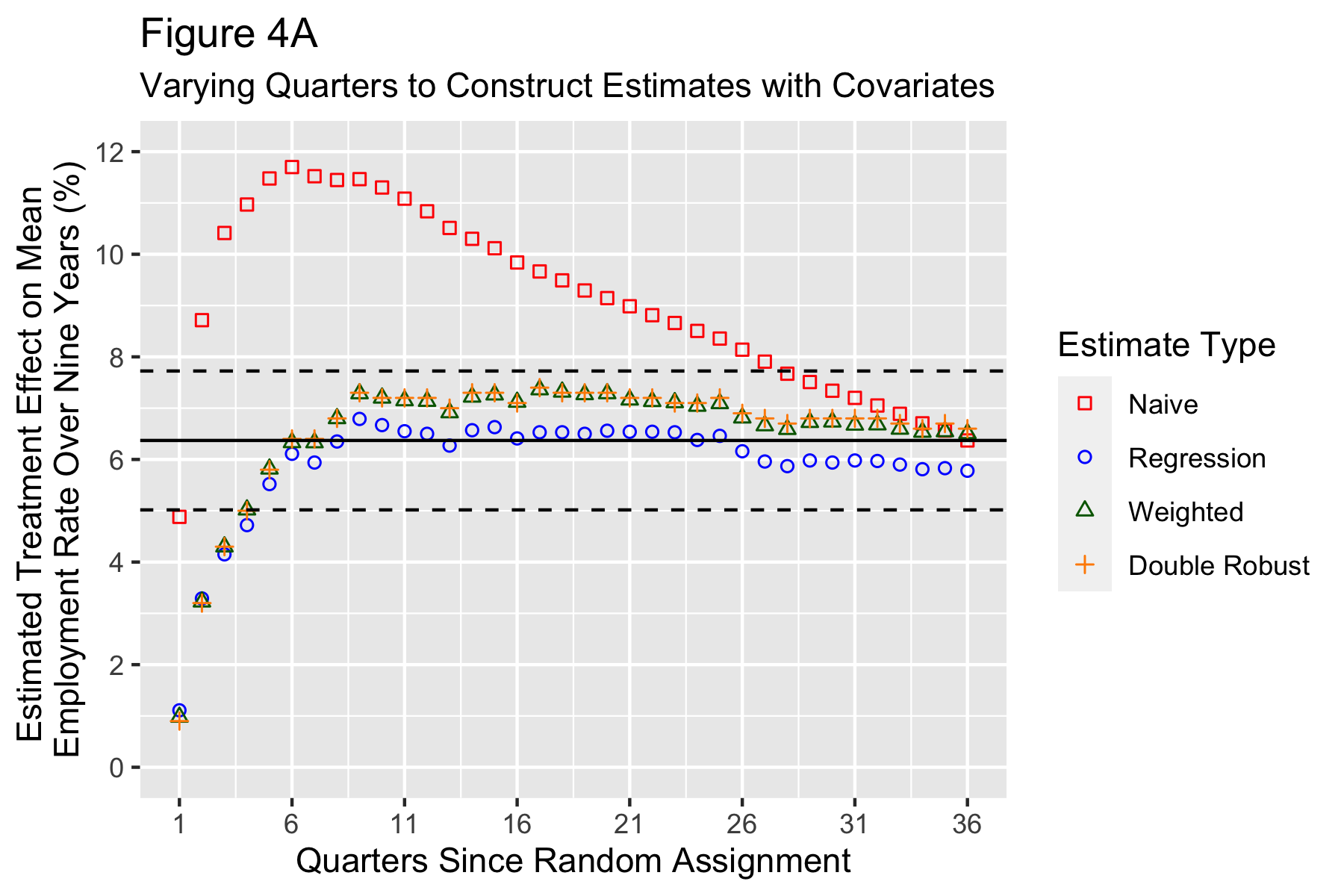} 
    \label{fig:Figure3A_Covs_filt}
\end{figure}

\begin{figure}[htbp]
    \centering
    \includegraphics[width=0.7\textwidth]{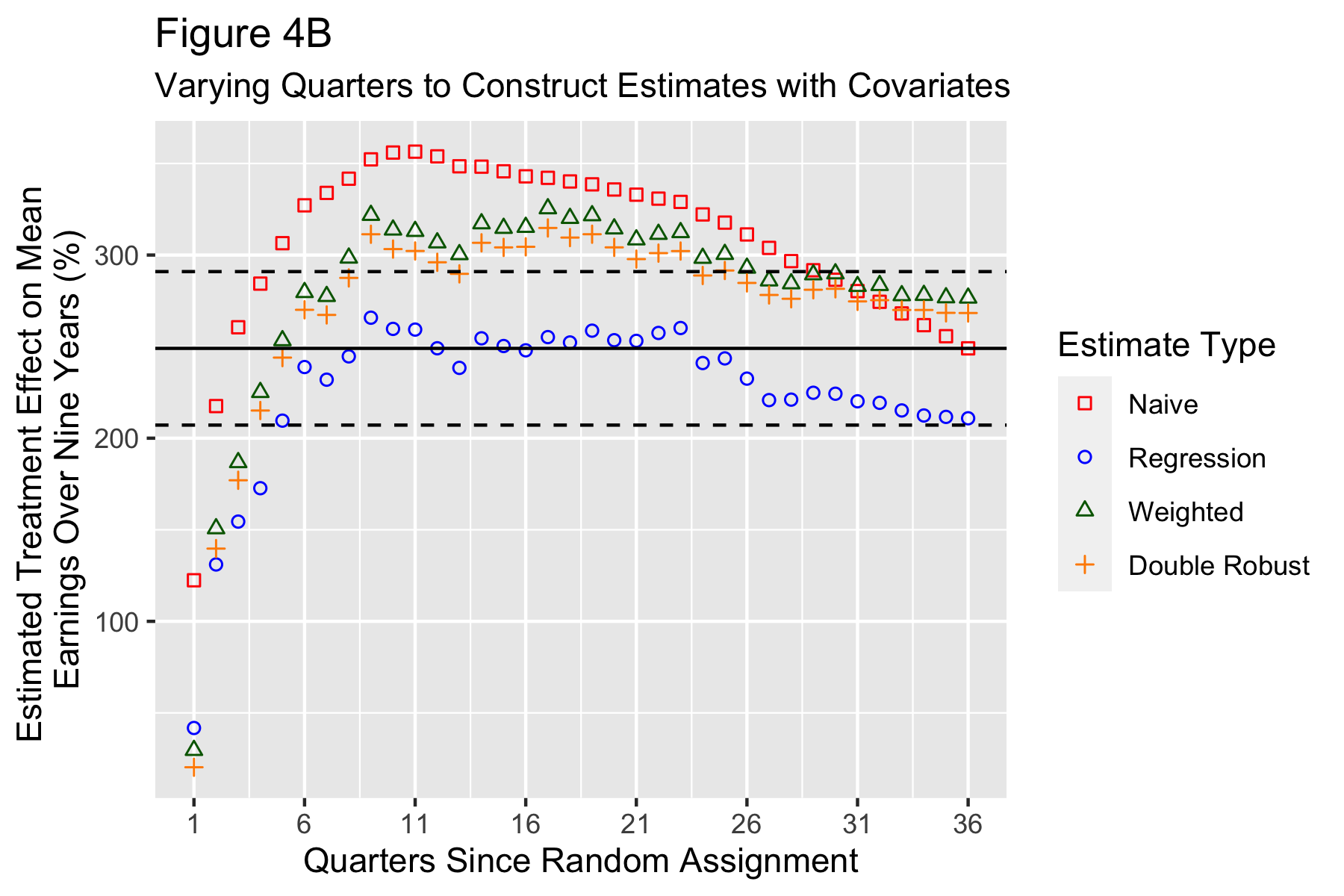} 
    \label{fig:Figure3A_Covs_filt}
\end{figure}
As the discussion after the surrogate index estimator shows, we  recover the experimental estimates if we use all 36 quarters of employment indicators as surrogates. The question is whether we can do approximately as well with fewer than 36 quarters of surrogates. In Figures 4A and 4B we compare the experimental estimates of the effect on the primary outcomes (0.064 for the employment outcome, and \$249 for the earnings outcome) to the three sets of surrogate estimates, as a function of how many periods of surrogates we use, ranging from 1 quarter to 36 quarters. To put this in perspective we also include in these two figures what we label the ``naive'' estimator where we estimate the effect on the long-term outcome as the effect on the first $t$ quarters of the outcome.
In Tables \ref{tab:Fig3A_Results} and \ref{tab:Fig3B_Results} we report a subset of the numbers underlying these estimates with the corresponding standard errors.

\begin{comment}
    
\begin{figure}[htbp]
    % Subfigure A
    \begin{subfigure}[b]{0.6\textwidth}
        \centering
        \includegraphics[width=\textwidth]{Figures/Figure4A_Employ.png} 
        \caption{Figure 4A: Employment}
        \label{fig:Figure4A_Employ}
    \end{subfigure}
    \hfill
    % Subfigure B
    \begin{subfigure}[b]{0.6\textwidth}
        \centering
        \includegraphics[width=\textwidth]{Figures/Figure4B_Earn.png} 
        \caption{Figure 4B: Earnings}
        \label{fig:Figure4B_Earn}
    \end{subfigure}
    \caption{Figure 4: Employment and Earnings}
    \label{fig:Figure4}
\end{figure}
\end{comment}

\begin{table}[!htbp] \centering 
  \caption{\textsc{Estimates for Effect on Employment. Experimental Benchmark: 0.064 (s.e. 0.012)}} 
  \label{tab:Fig3A_Results} 
\begin{tabular}{@{\extracolsep{5pt}} ccccccccc} 
\\[-1.8ex]\hline 
\hline \\[-1.8ex] 
t & \multicolumn{2}{c}{Naive} & \multicolumn{2}{c}{Surrogacy Index}  & \multicolumn{2}{c}{Surrogacy Score} & \multicolumn{2}{c}{Influence Function}  \\ 
& est & (s.e.)& est & (s.e.)& est & (s.e.)& est & (s.e.)\\
\hline \\[-1.8ex] 
1& 0.049 &(0.013) &0.011 &(0.003)& 0.010&(0.002) & 0.009 &(0.003)\\ 
2& 0.087&(0.012) & 0.033&(0.003) & 0.032 &(0.003)& 0.032 &(0.004)\\ 
3& 0.104&(0.011)& 0.042 &(0.004)& 0.043 &(0.004)& 0.043 &(0.004)\\ 
4& 0.110&(0.011) & 0.047&(0.005) & 0.050&(0.005) & 0.050&(0.005) \\ 
5& 0.115&(0.011) & 0.055&(0.005) & 0.058&(0.005) & 0.058 &(0.005)\\ 
6& 0.117&(0.010) & 0.061&(0.006) & 0.063 &(0.006)& 0.064&(0.006) \\ 
12& 0.108 &(0.010)& 0.065 &(0.007)& 0.071&(0.008) & 0.072 &(0.008)\\ 
18& 0.095 &(0.010)& 0.065&(0.008) & 0.073&(0.009) & 0.073 &(0.009)\\ 
24& 0.085 &(0.010)& 0.064&(0.009) & 0.070&(0.010) & 0.071 &(0.010)\\ 
30& 0.073 &(0.010)& 0.059 &(0.009)& 0.067&(0.010) & 0.068 &(0.010)\\ 
36& 0.064 &(0.010)& 0.058&(0.009) & 0.065 &(0.010)& 0.066&(0.010)\\ 
\hline \\[-1.8ex] 
\end{tabular} 
\end{table}

\begin{table}[!htbp] \centering 
  \caption{\textsc{Estimates for Effect on Earnings. Experimental Benchmark: \$249 (s.e. \$83)}} 
  \label{tab:Fig3B_Results} 
\begin{tabular}{@{\extracolsep{5pt}} ccccccccc} 
\\[-1.8ex]\hline 
\hline \\[-1.8ex] 
t & \multicolumn{2}{c}{Naive} & \multicolumn{2}{c}{Surrogacy Index}  & \multicolumn{2}{c}{Surrogacy Score} & \multicolumn{2}{c}{Influence Function}  \\ 
& est & (s.e.)& est & (s.e.)& est & (s.e.)& est & (s.e.)\\
\hline \\[-1.8ex] 
1 & 122.426 & (28.936) & 41.765 & (13.448) & 29.639 & (9.450) & 20.329 & (12.572)\\
2 & 217.535 & (30.086) & 131.052 & (18.305) & 150.678 & (18.934) & 139.714 & (20.316)\\ 
3 & 260.571 & (31.786) & 154.463 & (23.414) & 186.779 & (24.128) & 177.021 & (24.925)\\ 
4 & 284.399 & (33.461) & 172.690 & (26.975) & 225.073 & (27.993) & 215.102 & (28.273) \\ 
5 & 306.479 & (35.162) & 209.587 & (29.501) & 253.358 & (30.269) & 244.004 & (30.706)\\ 
6 & 327.145 & (36.572) & 238.845 & (31.520) & 279.654 & (32.487) & 270.099 & (32.875) \\ 
12 & 353.884 & (41.254) & 249.100 & (39.417) & 306.777 & (42.873) & 296.049 & (43.640)\\ 
18 & 340.180 & (43.882) & 252.269 & (44.282) & 320.129 & (45.656) & 309.546 & (46.495)\\ 
24 & 322.179 & (46.487) & 241.037 & (49.798) & 298.344 & (49.705) & 288.837 & (50.831)\\ 
30 & 286.512 & (48.461) & 224.286 & (50.288) & 289.866 & (50.538) & 281.567 & (51.553)\\ 
36 & 249.054 & (49.960) & 210.872 & (50.231) & 276.622 & (51.378) & 268.352 & (52.523)\\ 
\hline \\[-1.8ex] 
\end{tabular} 
\end{table}

We see that the naive estimator does very poorly. It takes more than 25 quarters before the naive estimator is within two standard errors of the experimental estimate. In contrast all three surrogate-based estimators are all within two standard errors when the surrogates include 5 quarters of outcomes, for both outcomes.

\subsubsection{Validation Results and Other Supplementary Analyses}

Given the data available we can also test whether using $t$ quarters of surrogates is sufficient to satisfy Surrogacy and Comparability. 
To test Surrogacy we regress the primary outcome on the pre-treatment variables, the surrogates up to quarter $t$, and the indicator for the treatment; a finding that the treatment has an impact indicates a violation of Surrogacy. We estimate this regression using a logistic regression model, using only the data from the experimental (Riverside) sample. We report in Table \ref{tab:Surrogate_Test_employment} and \ref{tab:Surrogate_Test_earnings} the results from these regressions for a number of different values for $t$, for the employment outcome and the earnings outcome. We report the point estimate, standard error and t-statistic. We see that point estimates for $t\leq 3$ are large and highly statistically significant. After that most of the t-statistics are less than 2, although there are some where the t-statistics are a little above 2, but the coefficient estimates are small.

We do a similar exercise for Comparability. We combine the experimental and observational samples and regress the final outcome on the surrogates, the pretreatment variables, and an indicator for the experimental sample, again using surrogates up to period $t$. We report the estimates on the indicator for the experimental sample, and the corresponding standard error. Here the the point estimates become smaller after $t=12$, but the t-statistics remain large even with a substantial number of surrogate periods, indicating a violation of Surrogacy.

\begin{table}[!htbp] \centering 
  \caption{\textsc{Surrogacy and Comparability  Assumption Tests for Employment Outcome}} 
  \label{tab:Surrogate_Test_employment} 
\begin{tabular}{@{\extracolsep{5pt}} ccccccc} 
\\[-1.8ex]\hline 
\hline \\[-1.8ex] 
& \multicolumn{3}{c}{Surrogacy Assumption}& \multicolumn{3}{c}{Comparability Assumption}\\
t & est & (s.e) & T-Stat& est & (s.e) & T-Stat \\ 
\hline \\[-1.8ex] 
1&  0.052  & (0.010) &  5.4  &  0.008  & (0.005) &  1.7 \\ 
2&  0.034  & (0.009) &  3.7  &  -0.004  & (0.004) &  -0.9 \\ 
3&  0.024  & (0.009) &  2.6  &  -0.006  & (0.004) &  -1.5 \\ 
4&  0.018  & (0.009) &  2.0  &  -0.007  & (0.004) &  -1.8 \\ 
5&  0.010  & (0.008) &  1.2  &  -0.010  & (0.004) &  -2.5 \\ 
6&  0.004  & (0.008) &  0.5  &  -0.011  & (0.004) &  -3.0 \\ 
12 &  -0.004  & (0.006) &  -0.7  &  -0.015  & (0.003) &  -5.0 \\ 
18 &  -0.007  & (0.004) &  -1.6  &  -0.009  & (0.002) &  -4.1 \\ 
24 &  -0.005  & (0.003) &  -1.8  &  -0.005  & (0.001) &  -3.6 \\ 
30 &  -0.002  & (0.001) &  -1.3  &  -0.001  & (0.001) &  -1.9 \\ 
35 &  0.000  & (0.000) &  -2.0  &  0.000  & (0.000) &  -0.8 \\ 
%36 &  0.000  & (0.000) & -- &  0.000  & (0.000) & --\\
\hline \\[-1.8ex] 
\end{tabular} 
\end{table}

\begin{table}[!htbp] \centering 
  \caption{\textsc{Surrogacy and Comparability  Assumption Tests for Earnings Outcome}} 
  \label{tab:Surrogate_Test_earnings} 
\begin{tabular}{@{\extracolsep{5pt}} ccccccc} 
\\[-1.8ex]\hline 
\hline \\[-1.8ex] 
& \multicolumn{3}{c}{Surrogacy Assumption}& \multicolumn{3}{c}{Comparability Assumption}\\
t & est & (s.e) & T-Stat& est & (s.e) & T-Stat \\ 
\hline \\[-1.8ex] 
1&  185.6  & (50.8) & 3.7 &  -35.6  & (25.3) & -1.4\\ 
2&  129.7 & (49.9) & 2.6 &  -65.0  & (24.5) & -2.7\\
3&  94.3  & (48.1) & 2.0 &  -72.8  & (23.5) & -3.1\\
4&  66.3  & (46.3) & 1.4 &  -67.5  & (22.4) & -3.0\\
5&  42.2  & (44.5) & 1.0 &  -71.0  & (21.5) & -3.3\\
6&  12.6  & (42.0) & 0.3 &  -73.9  & (20.5) & -3.6\\
12 &  -19.1  & (31.3) & -0.6 &  -65.2  & (15.3) & -4.2\\
18 &  -41.8  & (22.0) & -1.9 &  -31.2  & (10.8) & -2.9\\
24 &  -20.2  & (13.6) & -1.5 &  -27.4  & (6.7) & -4.1\\
30 &  -10.8  & (5.9) & -1.8 &  -4.7  & (2.8) & -1.6\\
35 &  -0.5  & (1.0) & -0.5 &  -0.4  & (0.5) & -0.8\\
%36 &  0.0  & (0.0) & -1.4 &  0.0 & (0.0) & -2.6\\
\hline \\[-1.8ex] 
\end{tabular} 
\end{table} 

If we are unwilling to make the surrogacy assumption we can still calculate bounds for the effect on employment, using the fact that this outcome is binary. For the case where the first six quarters of post-treatment data are used as surrogates, the lower and upper bound are estimated as -0.186 and 0.124. These are not very informative, because the data now do not allow us to estimate the indirect effect of the treatment on the outcome. 

Similarly, we can calculate bounds for the average effect without assuming Comparability. With six quarters of surrogates the bounds are again wide at  -0.076 and 0.194 respectively. Here the fact that the treatment effect on the surrogates is strong leads to substantial sensitivity to the comparability assumption as formalized in Lemma \ref{lemma2}.

Using the data for Riverside we can also assess the value of the Surrogacy assumption. Using the six quarters of data as surrogates, we find that the  gain from knowledge of Surrogacy (the $\Delta$ in Theorem \ref{theorem:effbound}) is quite large.
The standard error given Surrogacy, $\sqrt{\mmv}_{\rm s}$, is 0.33 times the standard error without knowledge that Surrogacy holds, $\sqrt{\mmv}_{\rm ns}$.

\section{Conclusion}

\label{section:conclusion} We develop new methods
for combining intermediate outcomes to estimate the long-term impacts
of treatments more rapidly and precisely. Our method requires estimating
a ``surrogate index'' -- the conditional expectation of the long-term
outcome given intermediate outcomes -- and then estimating the treatment
effect on the surrogate index. The surrogate index can be estimated
using parametric or nonparametric regression methods. We formalize conditions
under which this method yields unbiased estimates, derive bounds for
the degree of bias when those assumptions fail, and propose a simple
out-of-sample validation approach using ``hold out'' intermediate
outcomes. We show that surrogates can also greatly improve the precision
of estimates even in settings where the treatment effect on the long-term
outcome can be estimated directly, particularly when that outcome
is rare or noisy.

Applying the method to analyze the impacts of the GAIN job training
program in California, we find that using short-term earnings and
employment rates to construct surrogate indices expedite the detection
of long-term treatment effects on employment and earnings by several
years and also substantially increases precision. Furthermore, a single
surrogate index accurately predicts heterogeneity in the long-term
treatment effects of different types of job training programs across
sites, showing that surrogate indices estimated in a given setting
may be generalizable to other settings. The success of the surrogate
index in this application validates the use of short-term employment
outcomes as surrogates for detecting longer-term impacts of job training
programs, an empirical result that can be applied when analyzing ongoing
programs.

Building on this application, it would be useful to systematically
establish surrogate indices that match the long-term treatment effects
estimated in other experiments and quasi-experiments. Over time, this
would allow researchers to collectively build a public library of
surrogate indices for long-term outcomes that could be used to expedite
the analysis of future interventions.

%\end{document}
\vskip0.5cm %\newpage

\vskip0.5cm

{\small{}{}{}}{\small\par}

\begin{comment}
    {\small{}{}\pagebreak{} \bibliographystyle{aea_referencing}
\bibliography{references}
}{\small\par}
\end{comment}
{\small{}{} \bibliographystyle{aea_referencing}
\bibliography{references}
}{\small\par}

\newpage

\begin{center}
\textbf{\large{}{}ONLINE APPENDICES}{\large\par}
\par\end{center}

\noindent \textbf{\large{}{A}.\ Additional Table}{\large\par}

\begin{table}[!htbp] \centering 
  \caption{\textsc{Summary Statistics of Covariates by Location}} 
  \label{tab:Normalized_Diff}
\small 
\begin{tabular}{@{\extracolsep{3pt}} cccccc} 
\\[-1.8ex]\hline 
\hline \\[-1.8ex] 
 & \multicolumn{2}{c}{Riverside $(N_\ppe=5,445)$} & \multicolumn{2}{c}{Other Locations ($N_\ppo= 13,725$)}\\ 
 &  Mean & (Std. Dev.) &  Mean & (Std. Dev.) & t-statistic\\
\hline \\[-1.8ex] 
%Growth in Earnings & -0.009 & (0.012) &\\ 

%Growth in Employment & $$-$0.007$ & $0.013$ &  \\ 
Female & 0.88 & (0.33) & 0.88 & (0.33) & 0.0 \\ 
Highschool Diploma & 0.523 & (0.5) & 0.499& (0.5) & 3.1 \\ 
Children \textless 5 & 0.164& (0.371) & 0.137 &(0.344) & 4.6 \\ 
Single & 0.866& (0.341) & 0.863 &(0.343) & 0.4 \\ 
Grade 17 to 20 & 0.001& (0.033) & 0.005 &0.071) & -5.3 \\ 
Grade 16 & 0.007 &(0.082) & 0.015 &(0.122) &  -5.4 \\ 
Grade 13 to 15 & 0.107& (0.309) & 0.125& (0.33) &  -3.5 \\ 
Grade 12 & 0.358 &(0.479) & 0.327& (0.469) & 4 \\ 
Grade 9 to 11 & 0.395& (0.489) & 0.337& (0.473) & 7.5 \\ 
White & 0.519 &(0.5) & 0.305& (0.46) & $27.4$ \\ 
Hispanic & 0.273 &(0.446) & 0.26 &(0.438) & $1.9$ \\ 
Black & 0.158& (0.365) & 0.341& (0.474) & -28.6 \\ 
Age & 33.6 &(8.2) & 35.4 &(8.8) & -13.1 \\

Lagged Aid for t = 1 Quarter & 0.774 & (0.419) & 0.837 &(0.369) & -9.8\\
Lagged Aid for t = 2 Quarter & 0.651 & (0.477) & 0.769 & (0.421) & -16 \\ 
Lagged Aid for t = 3 Quarter & 0.639 & (0.48) & 0.761 &(0.426) & -16.3 \\ 
Lagged Aid for t = 4 Quarter & 0.634 & (0.482) & 0.751 &(0.433) & -15.6 \\ 
Lagged Earnings for t = 1 Quarter & 452& (1405)  &437 &(1283) & 0.7 \\ 
Lagged Earnings for t = 2 Quarter & 574 &(1553) & 510 & (1433) & 2.6 \\ 
Lagged Earnings for t = 3 Quarter & 598 &(1600) & 543 & (1491) & 2.2 \\ 
Lagged Earnings for t = 4 Quarter & 613 & (1601) & 570 &(1582) & 1.7 \\ 
Lagged Earnings for t = 5 Quarter & 665 &(1701) & 579 & (1619) & 3.2 \\ 
Lagged Earnings for t = 6 Quarter & 698 & (1761) & 580 & (1586) & 4.3 \\ 
Lagged Earnings for t = 7 Quarter & 709 & (1788) & 579 & (1630) & 4.6 \\ 
Lagged Earnings for t = 8 Quarter & 726 & (1839) & 567 & (1631) & 5.6 \\ 
Lagged Earnings for t = 9 Quarter & 719 & (1828) & 570 & (1655) & 5.2 \\ 
Lagged Earnings for t = 10 Quarter & 729 & (1815) & 573 & (1663) & 5.5 \\ 
%\fi
\hline \\[-1.8ex] 
\end{tabular} 
\end{table}

\noindent \textbf{\large{}{B}.\ Related Literature}{\large\par}

\subsubsection*{Critical Assumptions in the Mediation Literature and their Relation to Surrogacy}

In the mediation literature
(\textit{e.g.,} \citealt{baron1986moderator,vanderweele2015explanation}),
the intermediate outcome that we refer to here as the surrogate $S_{i}$
is called a mediator. To emphasize its role as a causal variable in
the mediation literature, we expand the notation and consider potential outcomes $Y_{i}(w,s)$
that are indexed by the treatment and the surrogate. (In terms of
these potential outcomes the original potential outcomes defined in the previous section, $Y_{i}(w)$, indexed only by the treatment $\woin$,
equals $Y_{i}(w)=Y_{i}(w,S_{i}(w))$, for $w\in\mmw$.)
In the setting considered in the mediation literature, we observe
the quadruple $(Y_{i},S_{i},W_{i},X_{i},\poi)$ for all units in the sample
and so there is not necessarily a distinction between the experimental sample and
the observational sample. To capture that we focus in this section on the case where we only have the experimental sample, $\poin=\ppe$, and where we observe the primary outcome $\yoin$ for this sample.

The focus of the mediation literature is on decomposing the causal
effect of the treatment on the outcome into a direct effect that involves
comparing potential outcomes where the surrogate remains fixed, and
an indirect effect that passes through the mediator/surrogate. Three
key estimands are the average \textit{total effect,} 
\[
\tau^{{\rm total}}\equiv\mme\left[Y_{i}(1,S_{i}(1))-Y_{i}(0,S_{i}(0))\right],
\]
the average \textit{natural indirect effect,} where we fix the treatment
at $w=1$, but change the surrogate from $S_{i}(0)$ to $S_{i}(1)$,
\[
\tau^{{\rm nie}}\equiv\mme\left[Y_{i}(1,S_{i}(1))-Y_{i}(1,S_{i}(0))\right],
\]
and the average \textit{natural direct effect}, where we fix the surrogate
at $S_{i}(0)$ and change the treatment from $W_{i}=0$ to $W_{i}=1$:
\[
\tau^{{\rm nde}}\equiv\mme\left[Y_{i}(1,S_{i}(0))-Y_{i}(0,S_{i}(0))\right],
\]
with the latter two adding up to the first: $\tau^{{\rm total}}=\tau^{{\rm nie}}+\tau^{{\rm nde}}$.

These effects are identified in the mediation literature using assumptions
similar to  Assumptions \ref{ass:unconf} and \ref{ass:surro}. The
first assumption in the mediation framework is a reformulation of
the unconfoundedness assumption, Assumption \ref{ass:unconf}. It
rules out the presence of unmeasured confounders between the treatment
and the surrogate, and between the treatment and the outcome.\begin{assumption}\label{ass:unconf_med1}
\textsc{(Unconfounded Treatment
Assignment / Strong Ignorability)} \\
$(i)$ 
$\woin\ \indep\ \Bigl(\soinn,\soine,Y_{i}(0,S_{i}(0)),Y_{i}(1,S_{i}(1))\Bigr)\ \Bigr|\ \xoin,\poi=\ppe,
$\\
 $(ii)$ $0<\hrho(x)<1\ {\rm for\ all}\ x\in\mmx.$
\end{assumption} The second assumption typically made in the mediation literature is another unconfoundedness
assumption that rules out the presence of unobserved confounders between
the surrogate and the outcome, conditional on the treatment. \begin{assumption}\label{ass:unconf_med2}
\[
\soin\ \indep\ \Bigl(Y_{i}(\woin,s)_{s\in\mms}\Bigr)\ \Bigr|\ \woin,\xoin,\poi.
\]
\end{assumption}
This assumption implies that comparisons of primary outcomes for units with different values for the surrogates but identical values for the treatment and pre-treatment variables can be given a causal interpretation.

To make the link to the surrogacy literature we need to add one key assumption that is not commonly made in the
mediation literature. This assumption rules out any direct effect of the treatment
on the outcome, allowing only for an indirect effect through the surrogate.\begin{assumption}\label{ass:excl}
For all $i$, $w,w'\in\mmw,s\in\mms$, 
\[
Y_{i}(w,s)=Y_{i}(w',s).
\]
\end{assumption}
This assumption is similar to the exclusion restriction in instrumental
variables settings, \textit{e.g.,} \citet{imbens1994,angrist1996identification}. In combination with the previous assumption this implies that we can give comparisons in the primary outcome between units with different values for the surrogates but the same values for pre-treatment variables a causal interpretation, without knowing the treatment status.

The following proposition links the  surrogacy and mediation assumptions. \begin{prop}\label{prop3}
Suppose Assumptions \ref{ass:unconf_med1}-\ref{ass:excl} hold. Then
Assumptions \ref{ass:unconf} and \ref{ass:surro} hold. \end{prop}

This connection highlights that at the heart of the surrogacy assumption is a causal relation between the surrogate and the primary outcome that mediates the causal effect of the treatment on the outcome.

\subsubsection*{Surrogacy and Comparability from a Missing Data Perspective}

From a missing data perspective,
Surrogacy and Comparability have parallels
to the missingness at random (MAR) assumption common in the missing
data literature (\citealt{rubin1976inference,little2019statistical}),
and specifically the literature on combining samples with different
sets of variables, (\citealt{ridder2007econometrics,gelman1998not,rassler2004data, graham2016efficient}).
In particular \citep{rassler2012statistical} focuses on a missing data structure closely related to ours.

In our two sample setting, we can think of the complete data as the
quintuple $(Y_{i},S_{i},W_{i},X_{i},P_{i})$. Here, we view the sample
as randomly drawn from a large population, so that we view $P_{i}$
as a stochastic missing data indicator. For the units in the sample
we observe the incomplete data $(\mathbf{1}_{P_{i}=\ppo}Y_{i},S_{i},X_{i},{\mathbf{1}}_{P_{i}=\ppe}W_{i},P_{i})$,
where for units with $P_{i}=\ppo$ the treatment indicator $W_{i}$
is missing, and for units with $P_{i}=\ppe$ the outcome $Y_{i}$
is missing. Now consider the following assumption. \begin{assumption}\label{ass:MAR}\textsc{(Augmented Missing At Random
 Assumption)}\label{ass:missing}\\
 Conditional on $(S_{i},X_{i})$, the three variables $P_{i}$, $Y_{i}$
and $W_{i}$ are jointly independent: 
\[
P_{i}\ \indep\ Y_{i}\ \indep\ W_{i}\ \Bigr|\ S_{i},X_{i}.
\]
\end{assumption} This is slightly different from a standard MAR assumption in \citep{rubin1976inference} where one would assume $P_{i}\indep Y_{i}|S_{i},X_{i}$
and/or $P_{i}\indep W_{i}|S_{i},X_{i}$. We need the stronger assumption
to incorporate surrogacy, as the following proposition shows. \begin{prop}\label{prop4}\textsc{(Missing
Data Model)}\\
 $(i)$ Assumption \ref{ass:missing} implies Assumption \ref{ass:surro} (Surrogacy) 
\[
Y_{i}\ \indep\ W_{i}\ \Bigr|\ S_{i},X_{i},
\]
and  Assumption \ref{ass:comp} (Comparability) 
\[
P_{i}\ \indep\ Y_{i}\ \Bigr|\ S_{i},X_{i}.
\]
$(ii)$ Assumption \ref{ass:MAR} has no testable implications. \end{prop}
%Comparability corresponds to $\yoin$ being independent of $\poin$ given $(\soin,\xoin)$, and surrogacy corresponds to $\woin$ being independent of $\yoin$ given $(\soin,\xoin)$ and given $\poin=E$.
%Assumption  \ref{ass:MAR} is in fact stronger than the combination of these two,
%  Assumption \ref{ass:surro} and \ref{ass:comp}
 %because it also assumes that conditional on $\poin=\ppo$,  $\woin$ is independent of $\yoin$, and it assumes that $\woin$ is independent of $\poin$. Neither are required for our main results, but because we do not need the $\woin$ in the observational sample and because these restrictions  do not imply testable restrictions there is no loss of generality.
Note that even after we have dealt with the missing $Y_{i}$ and missing
$W_{i}$ problems, we still have the missing potential outcomes, which
is why we also need the unconfoundedness assumption.

\noindent \textbf{\large{}{}C. Proofs}{\large\par}

\textit{Proof of Proposition \ref{prop1}:}\textsc{ 
\[
\pr\left(W_{i}=1|Y_{i}=y,\hrho(\soin,\xoin)=r,P_{i}=\ppe\right)=\mme\left[\left.W_{i}\right|Y_{i}=y,\hrho(\soin,\xoin)=r,P_{i}=\ppe\right]
\]
\[
\hskip1cm=\mme\left[\left.\mme\left[\left.W_{i}\right|Y_{i}=y,\soin,\xoin,\hrho(\soin,\xoin)=r,P_{i}=\ppe\right]\right|Y_{i}=y,\hrho(\soin,\xoin)=r,P_{i}=\ppe\right]
\]
\[
\hskip1cm=\mme\left[\left.\mme\left[\left.W_{i}\right|Y_{i}=y,\soin,\xoin,P_{i}=\ppe\right]\right|Y_{i}=y,\hrho(\soin,\xoin)=r,P_{i}=\ppe\right]
\]
\[
\hskip1cm=\mme\left[\left.\mme\left[\left.W_{i}\right|\soin,\xoin,P_{i}=\ppe\right]\right|Y_{i}=y,\hrho(\soin,\xoin)=r,P_{i}=\ppe\right]
\]
\[
\hskip1cm=\mme\left[\left.\hrho(\soin,\xoin)\right|Y_{i}=y,\hrho(\soin,\xoin)=r,P_{i}=\ppe\right]=\hrho(\soin,\xoin),
\]
}which proves the result\textsc{. $\square$}

\vskip0.5cm

%\end{document}

\textit{Proof of Proposition \ref{prop2}:} Part $(i)$ follows directly
from the definitions of $\mu(\cdot,\ppe)$
and Assumption \ref{ass:surro}. Part $(ii)$ follows directly from
the definitions of $\hmu(\cdot,\ppe)$ and $\hmu(\cdot,\ppo)$ and Assumption
\ref{ass:comp}. Part $(iii)$ follows from parts $(i)$ and $(ii)$.
$\square$

\vskip0.5cm

\textit{Proof of Proposition \ref{prop3}:} We wish to show that the
three conditions 
\begin{equation}
\woin\ \indep\ \Bigl(\soinn,\soine,Y_{i}(0,S_{i}(0)),Y_{i}(1,S_{i}(1))\Bigr)\ \Bigr|\ \xoin\label{a}
\end{equation}
\begin{equation}
\soin\ \indep\ \Bigl(Y_{i}(W_{i},s)_{s\in\mms}\Bigr)\ \Bigr|\ \xoin,\woin\label{b}
\end{equation}
and 
\begin{equation}
Y_{i}(w,s)=Y_{i}(w',s)\hskip1cm\forall\ i,w,w'\in\mmw,s\in\mms,\label{c}
\end{equation}
imply 
\begin{equation}
\woin\ \indep\ \Bigl(\yoinn,\yoine,\soinn,\soine\Bigr)\ \Bigr|\ \xoin,\label{een-1}
\end{equation}
\begin{equation}
\woin\ \indep\ \yoin\ \Bigr|\ \soin,\xoin.\label{twee}
\end{equation}
Note that we leave out the conditioning in $P_{i}=\ppe$ in the last
two conditions because we are focused here on the one-sample case.
Condition (\ref{een-1}) follows directly from (\ref{a}) because
$Y_{i}(w)=Y_{i}(w,S_{i}(w))$.

Condition (\ref{c}) implies that we can write $Y_{i}(s)$ without
ambiguity, and by (\ref{a}), we have 
$
\woin\ \indep\ Y_{i}(s)\ \Bigr|\ \xoin.
$
By (\ref{b}) we have 
$\soin\ \indep\ Y_{i}(s)\ \Bigr|\ \xoin,\woin.
$
Combining these implies 
$
\Bigl(\soin,\woin\Bigr)\ \indep\ Y_{i}(s)\ \Bigr|\ \xoin.
$
This in turn implies 
$
W_{i}\ \indep\ Y_{i}(s)\ \Bigr|\ \soin,\xoin,
$
which in turn implies 
$
W_{i}\ \indep\ Y_{i}(\soin)\ \Bigr|\ \soin,\xoin.
$
This is equivalent to the condition we set out to prove, 
$
W_{i}\ \indep\ Y_{i}\ \Bigr|\ \soin,\xoin.
$
$\square$

\vskip0.5cm

\textit{Proof of Proposition \ref{prop4}:} The first part of the
Proposition is immediate. For the second part, note that we can identify
from the data the distributions 
\[
f_{Y_i|S_i,X_i,P_i}(y|s,x,\ppo),\hskip1cmf_{W_i|S_i,X_i,P_i}(w|s,x,\ppe),\hskip1cm{\rm and}\ \ f_{P_i,S_i,X_i}(p,s,x),
\]
but no other distributions. That implies that the joint distribution
of $(Y_i,S_i,W_i,X_i,P_i)$ implied by 
$
f_{Y_i|S_i,W_i,X_i,P_i}(y|s,w,x,p)=f_{Y_i|S_i,X_i,P_i}(y|s,x,\ppo),
$
and 
$
f_{W_i|S_i,X_i,P_i}(w|s,x,\ppo)=f_{W_i|S_i,X_i,P_i}(w|s,x,\ppe),
$
for all $(y,s,s,w,x,p)$ is consistent with the data, and it also satisfies
Assumption \ref{ass:MAR}. $\square$

\vskip0.5cm

\textit{Proof of Theorem \ref{theorem1}:} We prove the case for $\mme[\yoine|P_{i}=\ppe]$,
specifically 
\begin{align}
\mme[\yoine|P_{i}=\ppe] & =\mme\left[\left.\hmu(\soin,\xoin,\ppo)\cdot\frac{\woin}{\hrhoe(\xoin)}\right|P_{i}=\ppe\right]\label{pvier}\\
 & =\mme\left[\left.\yoio\cdot\frac{\hrho(\soio,\xoio)\cdot \varphi(\soio,\xoio)\cdot(1-\varphi)}{\hrhoe(\xoio)\cdot(1-\varphi(\soio,\xoio))\cdot \varphi}\right|P_{i}=\ppo\right]\label{peen}\\
 & =\mme\left[\left.\hmu(S_{i},X_{i},\ppo)\cdot\frac{\hrho(\soio,\xoio)\cdot \varphi(\soio,\xoio)\cdot(1-\varphi)}{\hrhoe(\xoio)\cdot(1-\varphi(\soio,\xoio))\cdot \varphi}\right|P_{i}=\ppo\right]\label{peh}
\end{align}
The proof of $\mme[\yoinn|P_{i}=\ppe]$ is similar. The score function
representation is immediate from these equalities. We note that equality
\eqref{pvier} uses Assumptions \ref{ass:unconf}--\ref{ass:comp}
and equalities \eqref{peen} and \eqref{peh} only use  the overlap condition, Assumption
\ref{ass:comp}$(ii)$.

Consider (\ref{pvier}). By Assumption \ref{ass:unconf} (unconfoundedness),
it follows that 
\[
\mme[\yoine|P_{i}=\ppe]=\mme\left[\left.\yoin\cdot\frac{\woin}{\hrhoe(\xoin)}\right|P_{i}=\ppe\right].
\]
Using the law of iterated expectations, we can first condition on
$\soin$ and $\xoin$ to get 
\[
\mme\left[\left.\yoin\cdot\frac{\woin}{\hrhoe(\xoin)}\right|P_{i}=\ppe\right]=\mme\left[\left.\mme\left[\left.\yoin\cdot\frac{\woin}{\hrhoe(\xoin)}\right|\soin,\xoin,P_{i}=\ppe\right]\right|P_{i}=\ppe\right].
\]
By Assumption \ref{ass:surro} (surrogacy), we have 
\[
\mme\left[\left.\mme\left[\left.\yoin\cdot\frac{\woin}{\hrhoe(\xoin)}\right|\soin,\xoin,P_{i}=\ppe\right]\right|P_{i}=\ppe\right]=\mme\left[\left.\mme\left[\yoin|\soin,\xoin,P_{i}=\ppe\right]\cdot\frac{\mme\left[\woin|\soin,\xoin,P_{i}=\ppe\right]}{\hrhoe(\xoin)}\right|P_{i}=\ppe\right]
\]
By Assumption \ref{ass:comp} (Comparability), $\hmu(s,x,\ppe)=\hmu(s,x,\ppo)$
so that this is equal to 
\[
\mme\left[\left.\hmu(\soin,\xoin,\ppo)\cdot\frac{\mme\left[\woin|\soin,\xoin,P_{i}=\ppe\right]}{\hrhoe(\xoin)}\right|P_{i}=\ppe\right]=\mme\left[\left.\hmu(\soin,\xoin,\ppo)\cdot\frac{\hrho(\soin,(\xoin)}{\hrhoe\xoin)}\right|P_{i}=\ppe\right]
\]
Undoing the law of iterated expectations gives us the desired equality.

Consider (\ref{peen}). By the definition of $\varphi(s,x)$, we have 
\[
\frac{\varphi(s,x)}{(1-\varphi(s,x))}\cdot\frac{1-\varphi}{\varphi}=\frac{\pr\left(\left.\soio=s,\xoio=x\right|P_{i}=\ppe\right)}{\pr\left(\left.\soio=s,\xoio=x\right|P_{i}=\ppo\right)}
\]
where the common support condition assures $1-\varphi(s,x)$ is not
zero. This leads to 
\[
\mme\left[\left.\yoio\cdot\frac{\hrho(\soio,\xoio)\cdot t(\soio,\xoio)\cdot(1-\varphi)}{\hrhoe (\xoio)\cdot(1-t(\soio,\xoio))\cdot \varphi}\right|P_{i}=\ppo\right]=\mme\left[\left.\yoio\cdot\frac{\hrho(\soio,\xoio)}{\hrhoe(\xoio)}\cdot\frac{\pr\left(\left.\soio,\xoio\right|P_{i}=\ppe\right)}{\pr\left(\left.\soio,\xoio\right|P_{i}=\ppo\right)}\right|P_{i}=\ppo\right]
\]
Again, by the law of iterated expectations, conditioning on $\soio$
and $\xoio$ leads to 
\[
\mme\left[\left.\yoio\cdot\frac{\hrho(\soio,\xoio)}{\hrhoe(\xoio)}\cdot\frac{\pr\left(\left.\soio,\xoio\right|P_{i}=\ppe\right)}{\pr\left(\left.\soio,\xoio\right|P_{i}=\ppo\right)}\right|P_{i}=\ppo\right]=\mme\left[\left.\hmu(\soio,\xoio,\ppo)\frac{\hrho(\soio,\xoio)}{\hrhoe(\xoio)}\cdot\frac{\pr\left(\left.\soio,\xoio\right|P_{i}=\ppe\right)}{\pr\left(\left.\soio,\xoio\right|P_{i}=\ppo\right)}\right|P_{i}=\ppo\right]
\]
Using the definition of conditional expectations, we obtain 
\begin{align*}
 & \mme\left[\left.\hmu(\soio,\xoio,\ppo)\frac{\hrho(\soio,\xoio)}{\hrhoe(\xoio)}\cdot\frac{\pr\left(\left.\soio,\xoio\right|P_{i}=\ppe\right)}{\pr\left(\left.\soio,\xoio\right|P_{i}=\ppo\right)}\right|P_{i}=\ppo\right]\\
= & \int \hmu(s,x,\ppo)\frac{\hrho(s,x)}{\hrhoe(x)}\cdot\frac{\pr\left(\left.\soio=s,\xoio=x\right|P_{i}=\ppe\right)}{\pr\left(\left.\soio=s,\xoio=x\right|P_{i}=\ppo\right)}\cdot\pr\left(\left.\soio=s,\xoio=x\right|P_{i}=\ppo\right)dsdx\\
= & \int \hmu(s,x,\ppo)\frac{\hrho(s,x)}{\hrhoe(x)}\pr\left(\left.\soio=s,\xoio=x\right|P_{i}=\ppe\right)dsdx\\
= & \mme\left[\left.\hmu(\soio,\xoio,\ppo)\frac{\hrho(\soio,\xoio)}{\hrhoe(\xoio)}\right|P_{i}=\ppe\right]
\end{align*}
Consider (\ref{peh}). By the law of iterated expectations conditional
on $\soio$ and $\xoio$, we obtain 
\[
\mme\left[\left.\yoio\cdot\frac{\hrho(\soio,\xoio)\cdot \varphi(\soio,\xoio)\cdot(1-\varphi)}{\hrhoe(\xoio)\cdot(1-\varphi(\soio,\xoio))\cdot \varphi}\right|P_{i}=\ppo\right]=\mme\left[\left.\hmu(\soio,\xoio,\ppo)\cdot\frac{\hrho(\soio,\xoio)\cdot \varphi(\soio,\xoio)\cdot(1-\varphi)}{\hrhoe(\xoio)\cdot(1-\varphi(\soio,\xoio))\cdot \varphi}\right|P_{i}=\ppo\right]
\]
where the common support condition assures $1-\varphi(s,x)$ is not
zero. 
Part $(iii)$ follows from Proposition \ref{prop4}, which shows that Surrogacy and Comparability have no testable implications. Standard arguments then imply that unconfoundednes does not generate any testable implications.
$\square$

\vskip0.5cm

\textit{Proof of Theorem \ref{var_bound}:}
For Part (i), we need to calculate the variance of the Efficient Influence Function (EIF) to obtain the efficiency bound.
We provide the detailed calculation for completeness.\footnote{ While our influence function representation coincides with \citet{chen2023semiparametric}, the variance calculation resulted in a slightly different expression.}

Given the EIF:

\begin{equation*}
\psi(y,s,w,x,p)=\frac{\been_{p=\ppe}}{\hvarphi}\left(\frac{w \cdot (\hmu(s,x,\ppo)-\hkappa(1,x))}{{\hrho}(x)}-\frac{(1-w)\cdot ({\hmu}(s,x,\ppo)- \hkappa(0,x) )}{1-{\hrho}(x)}\right) 
\end{equation*}
\[+\frac{\been_{p=\ppe}}{\hvarphi} \Bigl(\hkappa(1,x)  - \hkappa(0,x)  -\tau \Bigr)
\]
\[\hskip2cm+\frac{\been_{p=\ppo}}{\hvarphi}
\left(\frac{\hvarphi(s,x)}{1-\hvarphi(s,x)}\frac{(y-{\hmu}(s,x,\ppo))\left({\hrho}(s,x)-{\hrho}(x)\right)}{{\hrho}(x)(1-{\hrho}(x))} \right)\]

\[ 
\mmv = \bigg[\psi(Y_i,S_i,W_i,X_i)^2 \bigg] 
\]

\[
=\mme \bigg[ \left( \frac{\been_{p=\ppe}}{\hvarphi}\left(\frac{\woin \cdot (\hmu(\soin,\xoin,\ppo)-\hkappa(1,x))}{{\hrho}(\xoin)}-\frac{(1-\woin)\cdot ({\hmu}(\soin,\xoin,\ppo)- \hkappa(0,x) )}{1-{\hrho}(\xoin)}\right) \right)^2  
\]
\[
+ \left(  \frac{\been_{p=\ppe}}{\hvarphi} \Bigl(\hkappa(1,x)  - \hkappa(0,x)  -\tau \Bigr) \right)^2 \]
\[
 + \left( \frac{\been_{p=\ppo}}{\hvarphi}
\left(\frac{\hvarphi(\soin,\xoin)}{1-\hvarphi(\soin,\xoin)}\frac{(\yoin-{\hmu}(\soin,\xoin,\ppo))\left({\hrho}(\soin,\xoin)-{\hrho}(\xoin)\right)}{{\hrho}(\xoin)(1-{\hrho}(\xoin))} \right)\right)^2
\bigg]
\]

Focusing on the first block 
\[\left( \frac{\been_{p=\ppe}}{\hvarphi}\left(\frac{w \cdot (\hmu(s,x,\ppo)-\hkappa(1,x))}{{\hrho}(x)}-\frac{(1-w)\cdot ({\hmu}(s,x,\ppo)- \hkappa(0,x) )}{1-{\hrho}(x)}\right) \right)^2,  \]
noting that $w(1-w)=0$ and hence the cross-term disappearing, we only have 
to take the expectation of 
\[
\left(\frac{\been_{p=\ppe}}{\hvarphi}\frac{w \cdot (\hmu(s,x,\ppo)-\hkappa(1,x))}{{\hrho}(x)}\right)^2
\qquad
{\rm and }\quad
\left(
\frac{\been_{p=\ppe}}{\hvarphi}
\frac{ (1-w)\cdot ({\hmu}(s,x,\ppo)- \hkappa(0,x) )}{1-{\hrho}(x)}\right)^2 
\]
Note that 
\[
\mme \bigg[ \left(\frac{\been_{p=\ppe}}{\hvarphi}\frac{\woin \cdot (\hmu(\soin,\xoin,\ppo)-\hkappa(1,\xoin))}{{\hrho}(\xoin)}\right)^2 \bigg]
\]
\[
= \mme \bigg[ \frac{(\hmu(\soin,\xoin,\ppo)-\hkappa(1,\xoin))^2}{ {\hrho}(\xoin)^2\hvarphi^2} \mme \big[ 
\been_{p=\ppe} \woin | \soin,\xoin \big] \bigg]\quad (\because \text{Tower Property} )
\]
\[
= \mme \bigg[ \frac{(\hmu(\soin,\xoin,\ppo)-\hkappa(1,\xoin))^2}{ {\hrho}(\xoin)^2\hvarphi^2} \hvarphi(\soin,\xoin) \mme \big[ 
 \woin | \soin,\xoin ,\poin = \ppe \big] \bigg]
\]
\[
= \mme \bigg[ \frac{(\hmu(\soin,\xoin,\ppo)-\hkappa(1,\xoin))^2}{ {\hrho}(\xoin)^2\hvarphi^2} \hvarphi(\soin,\xoin)  \hrho(\soin,\xoin) \bigg]
\]
\[
= \mme \bigg[ \frac{\hvarphi(\soin,\xoin) \hrho(\soin,\xoin) }{ \hvarphi^2{\hrho}(\xoin)^2}    (\hmu(\soin,\xoin,\ppo)-\hkappa(1,\xoin))^2\bigg]
\]

Likewise, we can derive 
\[ 
 \mme \bigg[\left(
\frac{\been_{p=\ppe}}{\hvarphi}
\frac{ (1-\woin)\cdot ({\hmu}(\soin,\xoin,\ppo)- \hkappa(0,\xoin) )}{1-{\hrho}(\xoin)}\right)^2 \bigg]
\]
\[
= \mme \bigg[ \frac{\hvarphi(\soin,\xoin) (1-\hrho(\soin,\xoin)) }{ \hvarphi^2(1-\hrho(\xoin))^2}    (\hmu(\soin,\xoin,\ppo)-\hkappa(0,\xoin))^2\bigg]
\]
Collectivizing the two term yields the first block:
\[
 \mme \bigg[ \frac{\hvarphi(\soin,\xoin) }{ \hvarphi^2} \bigg( \frac{1-\hrho(\soin,\xoin) }{(1-\hrho(\xoin))^2} (\hmu(\soin,\xoin,\ppo)-\hkappa(0,\xoin))^2  +  \frac{\hrho(\soin,\xoin) }{{\hrho}(\xoin)^2} (\hmu(\soin,\xoin,\ppo)-\hkappa(1,\xoin))^2  \bigg) \bigg]
\]

Next, for the second block 
\[\frac{\been_{p=\ppe}}{\hvarphi} \Bigl(\hkappa(1,x)  - \hkappa(0,x)  -\tau \Bigr)
\]
we can likewise derive by using the Tower Property with respect to $\xoin$ that 
\[ \mme \bigg[ 
 \left(  \frac{\been_{p=\ppe}}{\hvarphi} \Bigl(\hkappa(1,\xoin)  - \hkappa(0,\xoin)  -\tau \Bigr) \right)^2  \bigg]
= \mme \bigg[ 
   \frac{\hvarphi(\xoin)}{\hvarphi^2} \Bigl(\hkappa(1,\xoin)  - \hkappa(0,\xoin)  -\tau \Bigr)^2  \bigg]
\]
Finally, for the third block
\[
  \left( \frac{\been_{p=\ppo}}{\hvarphi}
\left(\frac{\hvarphi(s,x)}{1-\hvarphi(s,x)}\frac{(y-{\hmu}(s,x,\ppo))\left({\hrho}(s,x)-{\hrho}(x)\right)}{{\hrho}(x)(1-{\hrho}(x))} \right)\right)^2,
\]
note that
\[ 
\mme \bigg[ \left( \frac{\been_{\poin=\ppo}}{\hvarphi}
\left(\frac{\hvarphi(\soin,\xoin)}{1-\hvarphi(\soin,\xoin)}\frac{(\yoin-{\hmu}(\soin,\xoin,\ppo))\left({\hrho}(\soin,\xoin)-{\hrho}(\xoin)\right)}{{\hrho}(\xoin)(1-{\hrho}(\xoin))} \right)\right)^2 \bigg]
\]
\[ 
=\mme \bigg[ \frac{(\hvarphi(\soin,\xoin))^2}{\hvarphi^2(1-\hvarphi(\soin,\xoin))^2} \frac{\left({\hrho}(\soin,\xoin)-{\hrho}(\xoin)\right)^2 }{{((\hrho}(\xoin)(1-{\hrho}(\xoin)))^2)^2}  \mme \bigg[  \been_{\poin=\ppo}
(\yoin-{\hmu}(\soin,\xoin,\ppo))^2 | \soin,\xoin \bigg]  \bigg]
\]
\[ 
=\mme \bigg[ \frac{(\hvarphi(\soin,\xoin))^2}{\hvarphi^2(1-\hvarphi(\soin,\xoin))^2} \frac{\left({\hrho}(\soin,\xoin)-{\hrho}(\xoin)\right)^2 }{{(\hrho}(\xoin)(1-{\hrho}(\xoin)))^2}  \mme \bigg[  (1-\hvarphi(\soin,\xoin))
(\yoin-{\hmu}(\soin,\xoin,\ppo))^2 | \soin,\xoin , \poin = \ppo \bigg]  \bigg]
\]
\[ 
=\mme \bigg[ \frac{(\hvarphi(\soin,\xoin))^2}{\hvarphi^2(1-\hvarphi(\soin,\xoin))^2} \frac{\left({\hrho}(\soin,\xoin)-{\hrho}(\xoin)\right)^2 }{{(\hrho}(\xoin)(1-{\hrho}(\xoin)))^2}  (1-\hvarphi(\soin,\xoin)) \sigma^2(\soin,\xoin,\ppo)  \bigg]
\]
\[ 
=\mme \bigg[ \frac{(\hvarphi(\soin,\xoin))^2}{\hvarphi^2(1-\hvarphi(\soin,\xoin))} \frac{\left({\hrho}(\soin,\xoin)-{\hrho}(\xoin)\right)^2 }{{(\hrho}(\xoin)(1-{\hrho}(\xoin)))^2}   \sigma^2(\soin,\xoin,\ppo)  \bigg]
\]
\[
=\mme \bigg[\frac{1-\hvarphi(\soin,\xoin)}{\hvarphi^2} \left( \left( \frac{\hvarphi(\soin,\xoin)}{1 -\hvarphi(\soin,\xoin)} \frac{\hrho(\soin,\xoin)- \hrho(\xoin)}{\hrho(\xoin)(1-\hrho(\xoin))} \right)^2 \sigma^2(\soin,\xoin,\ppo)  \right)\bigg]
\]

Hence, adding up the three blocks (in the order from the third to the first block) yield the desired efficiency bound: 
\[
\mmv=\mme[\psi(\yoio,\soio,\woin,\xoin,P_{i})^2]\]
\[\qquad=\mme\bigg[ \frac{1-\hvarphi(\soin,\xoin)}{\hvarphi^2} \left( \left( \frac{\hvarphi(\soin,\xoin)}{1 -\hvarphi(\soin,\xoin)} \frac{\hrho(\soin,\xoin)- \hrho(\xoin)}{\hrho(\xoin)(1-\hrho(\xoin))} \right)^2 \sigma^2(\soin,\xoin,\ppo)  \right)  \]
\[+\frac{\hvarphi(\xoin)}{\hvarphi^2} \Bigl(\hkappa(1,\xoin)  - \hkappa(0,\xoin)  -\tau \Bigr)^2  \]
\[\left. + \frac{\hvarphi(\soin,\xoin)}{\hvarphi^2} \left(  \frac{ 
(1-\rho(\soin,\xoin))(\hmu(\soin,\xoin,\ppo) - \hkappa(0,\xoin))^2}{(1 - \hrho(\xoin))^2} + \frac{\rho(\soin,\xoin)(\hmu(\soin,\xoin,\ppo) - \hkappa(1,\xoin))^2}{\hrho(\xoin)^2} \right) \right].\]
\[\qquad=\mme\bigg[ \frac{1}{\hvarphi^2}  \frac{\hvarphi(\soin,\xoin)^2}{1 -\hvarphi(\soin,\xoin)}\left(  \frac{\hrho(\soin,\xoin)- \hrho(\xoin)}{\hrho(\xoin)(1-\hrho(\xoin))} \right)^2 \sigma^2(\soin,\xoin,\ppo)    \]
\[+\frac{\hvarphi(\xoin)}{\hvarphi^2} \Bigl(\hkappa(1,\xoin)  - \hkappa(0,\xoin)  -\tau \Bigr)^2  \]
\[\left. + \frac{\hvarphi(\soin,\xoin)}{\hvarphi^2} \left(  \frac{ 
(1-\rho(\soin,\xoin))(\hmu(\soin,\xoin,\ppo) - \hkappa(0,\xoin))^2}{(1 - \hrho(\xoin))^2} + \frac{\rho(\soin,\xoin)(\hmu(\soin,\xoin,\ppo) - \hkappa(1,\xoin))^2}{\hrho(\xoin)^2} \right) \right].\]

For part $(ii)$, first rewrite the variance bound, normalized by the square root of the expected size of the experimental sample, $\varphi N$, instead of normalized by the total sample size $N$, as
\[\tilde \mmv=\mme\bigg[ \frac{1}{\hvarphi}  \frac{\hvarphi(\soin,\xoin)^2}{1 -\hvarphi(\soin,\xoin)}\left(  \frac{\hrho(\soin,\xoin)- \hrho(\xoin)}{\hrho(\xoin)(1-\hrho(\xoin))} \right)^2 \sigma^2(\soin,\xoin,\ppo)    \]
\[+\frac{\hvarphi(\xoin)}{\hvarphi} \Bigl(\hkappa(1,\xoin)  - \hkappa(0,\xoin)  -\tau \Bigr)^2  \]
\[\left. + \frac{\hvarphi(\soin,\xoin)}{\hvarphi} \left(  \frac{ 
(1-\rho(\soin,\xoin))(\hmu(\soin,\xoin,\ppo) - \hkappa(0,\xoin))^2}{(1 - \hrho(\xoin))^2} + \frac{\rho(\soin,\xoin)(\hmu(\soin,\xoin,\ppo) - \hkappa(1,\xoin))^2}{\hrho(\xoin)^2} \right) \right].\]

Next, we re-write the bound in terms of a conditional expectation in the experimental sample, rather than as the unconditional expectation, (this implies multiplying by $\varphi/\varphi(S_i,X_i)$ or $\varphi/\varphi(X_i)$  appropriately) as

\[\tilde \mmv=\mme\bigg[  \frac{\hvarphi(\soin,\xoin)}{1 -\hvarphi(\soin,\xoin)}\left(  \frac{\hrho(\soin,\xoin)- \hrho(\xoin)}{\hrho(\xoin)(1-\hrho(\xoin))} \right)^2 \sigma^2(\soin,\xoin,\ppo)    \]
\[+ \Bigl(\hkappa(1,\xoin)  - \hkappa(0,\xoin)  -\tau \Bigr)^2  \]
\[\left.\left. +  \left(  \frac{ 
(1-\rho(\soin,\xoin))(\hmu(\soin,\xoin,\ppo) - \hkappa(0,\xoin))^2}{(1 - \hrho(\xoin))^2} + \frac{\rho(\soin,\xoin)(\hmu(\soin,\xoin,\ppo) - \hkappa(1,\xoin))^2}{\hrho(\xoin)^2} \right) \right|\poin=\ppe\right].\]

Now we consider a sequence of data generating processes, where the outcome distribution in the observational sample remains fixed, and the propensity and surrogate scores remain fixed, and only the functions $\varphi(s,x)$, $\varphi(x)$ and the scalar $\varphi$ change, in such a way that $\sup_{s,x}\varphi(s,x)\rightarrow  0$. The the first term converges to zero, leaving us with
\[\bar\mmv=\mme\bigg[\Bigl(\hkappa(1,\xoin)  - \hkappa(0,\xoin)  -\tau \Bigr)^2  \]
\[\left.\left. +  \left(  \frac{ 
(1-\rho(\soin,\xoin))(\hmu(\soin,\xoin,\ppo) - \hkappa(0,\xoin))^2}{(1 - \hrho(\xoin))^2} + \frac{\rho(\soin,\xoin)(\hmu(\soin,\xoin,\ppo) - \hkappa(1,\xoin))^2}{\hrho(\xoin)^2} \right) \right|\poin=\ppe\right].\]
The final step is to note that $\rho(\soin,\xoin)=\mme[\woin|\soin,\xoin,\poin=\ppe]$ so we can write $\bar\mmv$ as
\[\bar\mmv=\mme\bigg[\Bigl(\hkappa(1,\xoin)  - \hkappa(0,\xoin)  -\tau \Bigr)^2  \]
\[\left.\left. +  \left(  \frac{ 
(1-\mme[\woin|\soin,\xoin,\poin=\ppe])(\hmu(\soin,\xoin,\ppo) - \hkappa(0,\xoin))^2}{(1 - \hrho(\xoin))^2} + \frac{\mme[\woin|\soin,\xoin,\poin=\ppe](\hmu(\soin,\xoin,\ppo) - \hkappa(1,\xoin))^2}{\hrho(\xoin)^2} \right) \right|\poin=\ppe\right]\]
\[=\mme\bigg[\Bigl(\hkappa(1,\xoin)  - \hkappa(0,\xoin)  -\tau \Bigr)^2  \]
\[\left.\left. +  \left(  \frac{ 
(1-\woin)(\hmu(\soin,\xoin,\ppo) - \hkappa(0,\xoin))^2}{(1 - \hrho(\xoin))^2} + \frac{\woin(\hmu(\soin,\xoin,\ppo) - \hkappa(1,\xoin))^2}{\hrho(\xoin)^2} \right) \right|\poin=\ppe\right].\]

$\square$

%\textcolor{red}{=========================}

\vskip0.5cm

%{\small{}{}{}\vskip0.5cm}{\small\par}

\textit{Proof of Theorem \ref{theorem:effbound}:}%{\small{}{}{}

The first representation of the efficiency bound without surrogacy
in part $(i)$ of the Theorem is essentially rewriting the efficiency bound in \citet{hahn1998role}, and related results in \citet{robins1995semiparametric,robins1995analysis}. 
The standard version of the efficiency bound is
\[ \mathbb{V}=\mme \left[
\frac{\sigma^2(1,X_i)}{\rho(X_i)}+\frac{\sigma^2(0,X_i)}{1-\rho(X_i)}+\left(\mu(1,X_i)-\mu(0,X_i)-\tau\right)^2
\right].\]
The proof consists of showing that this is equal to the expression for $\mmv_{{\rm ns}}$ in Theorem \ref{theorem:effbound}:
\[
\mmv_{{\rm ns}}=\mme\biggl[\sigma^{2}(\soin,\xoin,\ppe)\cdot\left(\frac{\hrho(\soin,\xoin)}{\rho(\xoin)^{2}}+\frac{1-\hrho(\soin,\xoin)}{(1-\rho(\xoin))^{2}}\right)
\]
\[
\hskip2cm+\frac{\hrho(\soin,\xoin)}{\rho(\xoin)^{2}}\cdot\left(\mu(\soin,\xoin,\ppe)-\mu(1,\xoin)\right)^{2}+\frac{1-\hrho(\soin,\xoin)}{(1-\rho(\xoin))^{2}}\cdot\left(\mu(\soin,\xoin,\ppe)-\mu(0,\xoin)\right)^{2}
\]
\[\hskip2cm +\left(\mu(1,X_i)-\mu(0,X_i)-\tau\right)^2
\biggr].\]
 which amounts to showing the equality of
  \begin{equation}\label{aeen} \mme \left[
\frac{\sigma^2(1,X_i)}{\rho(X_i)}+\frac{\sigma^2(0,X_i)}{1-\rho(X_i)}
\right],\end{equation}
and
\begin{equation} \label{aeen2} \mme\biggl[\sigma^{2}(\soin,\xoin,\ppe)\cdot\left(\frac{\hrho(\soin,\xoin)}{\rho(\xoin)^{2}}+\frac{1-\hrho(\soin,\xoin)}{(1-\rho(\xoin))^{2}}\right)
\end{equation}
\[
\hskip2cm+\frac{\hrho(\soin,\xoin)}{\rho(\xoin)^{2}}\cdot\left(\mu(\soin,\xoin,\ppe)-\mu(1,\xoin)\right)^{2}+\frac{1-\hrho(\soin,\xoin)}{(1-\rho(\xoin))^{2}}\cdot\left(\mu(\soin,\xoin,\ppe)-\mu(0,\xoin)\right)^{2}\biggr].
\]

By unconfoundedness
\[ \sigma^2(1,x)\equiv \mmv(Y_i(1)|X_i=x)=\mmv(Y_i|W_i=1,X_i=x),\]
where as mentioned in the main text, we implicitly condition on the sampling indicator and abstract it from the notation when it does not lead to confusion.

By iterated expectations this is equal to
\[ \mme\left[\left.\mmv\left(Y_i|W_i=1,S_i,X_i=x\right)\right|W_i=1,X_i=x\right]+\mmv\left(\left.\mme[Y_i|W_i=1,S_i,X_i]\right|W_i=1,X_i\right).\]
By surrogacy the conditional distribution of $Y_i$ given $W_i$, $S_i$ and $X_i$ does not vary by $W_i$, so this is equal to
\[ \mme\left[\left.\mmv\left(Y_i|S_i,X_i=x\right)\right|W_i=1,X_i=x\right]+\mmv\left(\left.\mme[Y_i|S_i,X_i]\right|W_i=1,X_i\right)\]
\[ \hskip1cm =\mme\left[\left.\sigma^2(S_i,X_i)\right|W_i=1,X_i=x\right]+\mmv\left(\left.\mu(S_i,X_i)\right|W_i=1,X_i\right).\]

For the first term,
\[ \mme\left[\left.\sigma^2(S_i,X_i)\right|W_i=1,X_i=x\right]=\mme\left[\left.\frac{\sigma^2(S_i,X_i)\rho(S_i,X_i)}{\rho(X_i)}\right|X_i=x\right].\]
For the second term, note that
\[ \mme\left[\left.\mu(S_i,X_i)\right|W_i=1,X_i\right]
= \mme\left[\left.\mme[Y_i|S_i,X_i]\right|W_i=1,X_i\right]
\]
is by surrogacy  equal to
$ \mme\left[\left.\mme[Y_i|W_i=1,S_i,X_i]\right|W_i=1,X_i\right] ,$
which in turn by iterated expectations is equal to
$ \mme\left[\left.Y_i\right|W_i=1,X_i\right] =\mu(1,X_i).$
Hence the second term is
\[ \mmv\left(\left.\mu(S_i,X_i)\right|W_i=1,X_i\right)
=
\mme\left[\left.\left(\mu(S_i,X_i)-\mu(1,X_i)\right)^2\right| W_i=1,X_i\right]
\]
\[\hskip1cm = \mme\left[
\left(\mu(S_i,X_i)-\mu(1,X_i)\right)^2
\frac{\rho(S_i,X_i)}{\rho(X_i)}
\right].\]
Combining the two terms and including the denominator $\rho(X_i)$, we have 
 \[\mme \left[
\frac{\sigma^2(1,X_i)}{\rho(X_i)}
\right]=\mme\left[\frac{\sigma^2(S_i,X_i)\rho(S_i,X_i)}{\rho(X_i)^2}\right]+
 \mme\left[
\left(\mu(S_i,X_i)-\mu(1,X_i)\right)^2
\frac{\rho(S_i,X_i)}{\rho(X_i)^2}
\right].
\]
By the same argument
 \[\mme \left[
\frac{\sigma^2(0,X_i)}{1-\rho(X_i)}
\right]=\mme\left[\frac{\sigma^2(S_i,X_i)(1-\rho(S_i,X_i))}{(1-\rho(X_i))^2}\right]+
 \mme\left[
\left(\mu(S_i,X_i)-\mappa(0,X_i)\right)^2
\frac{1-\rho(S_i,X_i)}{(1-\rho(X_i))^2}
\right]
\]
Hence, adding up the two equalities above shows the desired equivalence of (\ref{aeen}) and (\ref{aeen2}).
This finishes the proof of part $(i)$ of the theorem.

Next, for part $(ii)$ of the theorem, we derive the efficiency bound for the case with surrogacy by first deriving the efficient influence function and then deriving its variance.
To derive the efficient influence function, we follow the proof in \citet{chen2023semiparametric} and \citet{newey1990semiparametric}, specifically the following four steps: (1) constructing the tangent space, (2) deriving the pathwise derivative of the target estimand (i.e. the ATE under surrogacy), (3) showing that the conjectured efficient influence function (EIF) lies in the tangent space, and (4) showing that the pathwise derivative of the target estimand and the conjectured EIF satisfies a key condition in \citet{newey1990semiparametric}.

First, to characterize the tangent space, considering the data density where the functions $f$ denote the density of random variables. 
%To simplify notation, we drop the subscript $i$:
\[
f_{Y_i,S_i,W_i,X_i}(y,s,w,x) = f_{Y_i \mid S_i,X_i} (y\mid s,x) f_{S_i \mid W_i,X_i}(s \mid w,x) f_{W_i \mid X_i}(w \mid x) f_{X_i}(x).
\]
We assume the data density satisfies the regularity and smoothness conditions in Definition (A.1) of Newey (1990).

Let $G^\epsilon$ be a parametric submodel parameterized by $\epsilon \in [0,1]$ where $G^{\epsilon =0} = G$ and $G$ is the true data generating model. Let $f^{\epsilon}$ be the corresponding density function for the parametric submodel. Then, the score of $f_{\epsilon}$ is

{\footnotesize
\begin{align*}
\frac{\delta}{\delta \epsilon} \log(f_{Y_i,S_i,W_i,X_i}^{\epsilon}(y,s,w,x)) &= \frac{\delta}{\delta \epsilon} \log (f_{Y_i \mid S_i,X_i}^\epsilon(y \mid s,x)) + \frac{\delta}{\delta \epsilon} \log(f_{S_i \mid W_i,X_i}^\epsilon (s \mid w,x))  + \frac{\delta}{\delta \epsilon} \log(f_{W_i \mid X_i}^{\epsilon} (w \mid x)) + \frac{\delta}{\delta \epsilon} \log(f_{X_i}^{\epsilon} (x)) \\
&= Q_{Y_i \mid S_i,X_i}^\epsilon(y \mid s,x) + Q_{S_i\mid W_i,X_i}^\epsilon(s \mid w,x) + Q_{W_i \mid X_i}^{\epsilon}(w \mid x) + Q_{X_i}^{\epsilon}(x).
\end{align*}}
We use $Q(\cdot)$'s to denote the score function, i.e. $Q(\cdot) = \frac{\delta}{\delta \epsilon} \log(f^\epsilon(\cdot))$. Evaluating the derivative at $\epsilon=0$ leads us to the score of the true model, i.e.,
\[
Q_{Y_i,S_i,W_i,X_i}(y,s,w,x) =  Q_{Y_i\mid S_i,X_i}(y \mid s,x) + Q_{S_i \mid W_i,X_i}(s \mid w,x) + Q_{W_i\mid X_i}(w \mid x) + Q_{X_i}(x).
\]
The tangent space $\mathcal{T}$ is the mean closure of a linear combination of mean-zero, square-integrable functions $\overline{Q}_1,...,\overline{Q}_4$ that satisfy the following conditions:
\begin{align*}
\mathcal{T} &= \biggl\{ \overline{Q}(y,s,w,x) \in \mathbb{R} \mid \overline{Q}(y,s,w,x) = \overline{Q}_1(y,s,x) + \overline{Q}_2(s,x,w) + \overline{Q}_3(w,x) + \overline{Q}_4(x)  \\
&\hspace{4cm} \mme [\overline{Q}_1(Y_i,s,x) \mid S_i=s,X_i=x] = \mme[\overline{Q}_1(Y_i,s,x) \mid S_i=s,W_i=w,X_i=x] = 0  \\
&\hspace{4cm} \mme[\overline{Q}_2(S_i,x,w) \mid X_i=x,W_i=w] = 0, \quad{} \mme[\overline{Q}_3(W_i,x) \mid X_i=x] = 0, \quad{} \mme[\overline{Q}_4(X_i)] = 0 \biggr\}.
\end{align*}

Second, we derive the pathwise derivative of our estimand. With some abuse of the integral notation, our estimand can be written as follows:
\begin{align*}
\tau &= \mme \biggr[ \mme\biggr[\mme [Y_i \mid S_i,X_i] \mid W_i=1,X_i \biggl]\biggl] - \mme \biggr[  \mme \biggr[\mme[Y_i \mid S_i,X_i] \mid W_i=0,X_i\biggl]\biggl] \\
&= \int \int \int y f_{Y_i \mid S_i,X_i}(y \mid s,x) f_{S_i \mid W_i,X_i}(s \mid 1, x) f_{X_i}(x) dy ds dx \\
&\quad{} - \int \int \int y f_{Y_i\mid S_i,X_i}(y \mid s,x) f_{S_i \mid W_i,X_i}(s \mid 0, x) f_{X_i}(x) dy ds dx.
\end{align*}
The pathwise derivative of the estimand $\tau$ is 

{\footnotesize 
\begin{align*}
\frac{\delta}{\delta \epsilon} \tau  =& \int \int \int y \frac{\delta}{\delta \epsilon} \left\{ f_{Y_i\mid S_i,X_i}^\epsilon(y \mid s,x) f_{S_i \mid W_i,X_i}^{\epsilon}(s \mid 1, x) f_{X_i}^{\epsilon}(x) \right\} dy ds dx \\
&\quad{} - \int \int \int y \frac{\delta}{\delta \epsilon} \left\{ f_{Y_i\mid S_i,X_i}^\epsilon(y \mid s,x) f_{S_i \mid W_i,X_i}^{\epsilon}(s \mid 0, x) f_{X_i}^{\epsilon}(x)\right\} dy ds dx \\
=& \int \int \int y \biggl\{ Q_{Y_i \mid S_i,X_i}^{\epsilon}(y \mid s,x) f_{Y_i \mid S_i,X_i}^{\epsilon}(y \mid s,x) f_{S_i \mid W_i,X_i}^{\epsilon}(s \mid 1,x) f_{X_i}^{\epsilon}(x) \\
&\hspace{2cm} +  f_{Y_i\mid S_i,X_i}^{\epsilon}(y \mid s,x) Q_{S_i \mid W_i,X_i}^{\epsilon}(s \mid 1,x) f_{S_i \mid W_i,X_i}^{\epsilon}(s \mid 1,x) f_{X_i}^{\epsilon}(x) \\
&\hspace{2cm} + f_{Y_i\mid S_i,X_i}^{\epsilon}(y \mid s,x) f_{S_i \mid W_i,X_i}^{\epsilon}(s \mid 1,x) Q_{X_i}^{\epsilon}(x) f_{X_i}^{\epsilon}(x) \biggr\} dy ds dx \\
&\quad{} -  \int \int \int y \biggl\{ Q_{Y_i\mid S_i,X_i}^{\epsilon}(y \mid s,x) f_{Y_i \mid S_i,X_i}^{\epsilon}(y \mid s,x) f_{S_i \mid W_i,X_i}^{\epsilon}(s \mid 0,x) f_{X_i}^{\epsilon}(x) \\
&\hspace{2.5cm} +  f_{Y_i\mid S_i,X_i}^{\epsilon}(y \mid s,x) Q_{S_i \mid W_i,X_i}^{\epsilon}(s \mid 0,x) f_{S_i \mid W_i,X_i}^{\epsilon}(s \mid 0,x) f_{X_i}^{\epsilon}(x) \\
&\hspace{2.5cm} + f_{Y_i\mid S_i,X_i}^{\epsilon}(y \mid s,x) f_{S_i \mid W_i,X_i}^{\epsilon}(s \mid 0,x) Q_{X_i}^{\epsilon}(x) f_{X_i}^{\epsilon}(x) \biggr\} dy ds dx \\
=& \int \int \int y Q_{Y_i\mid S_i,X_i}^{\epsilon} f_{Y_i\mid S_i,X_i}^{\epsilon}(y \mid s,x) f_{X_i}^{\epsilon}(x)\biggl\{ f_{S_i \mid W_i,X_i}^{\epsilon}(s \mid 1,x) - f_{S_i \mid W_i,X_i}^{\epsilon}(s \mid 0,x) \biggr\} dy ds dx \\
&\quad{} + \int \int \int y f_{Y_i \mid S_i,X_i}^{\epsilon}(y \mid s,x) f_{X_i}^{\epsilon}(x)\biggl\{ Q_{S_i \mid W_i,X_i}^{\epsilon}(s \mid 1,x)f_{S_i \mid W_i,X_i}^{\epsilon}(s \mid 1,x) \\
&\hspace{5.5cm} - Q_{S_i \mid W_i,X_i}^{\epsilon}(s \mid 0,x)f_{S_i \mid W_i,X_i}^{\epsilon}(s \mid 0,x) \biggr\} dy ds dx \\
&\quad{} + \int \int \int y f_{Y_i \mid S_i,X_i}^{\epsilon}(y\mid s,x) Q_{X_i}^{\epsilon}(x) f_{X_i}^{\epsilon}(x) \biggl\{ f_{S_i \mid W_i,X_i}^{\epsilon}(s \mid 1,x) - f_{S_i \mid W_i,X_i}^{\epsilon}(s \mid 0,x) \biggr\} dy ds dx.
\end{align*}
}
The derivatives above use the chain rule from calculus and the fact that 
\[
\frac{\delta}{\delta \epsilon} f^{\epsilon} = \frac{\delta}{\delta \epsilon} \log(f^{\epsilon})  f^{\epsilon} = Q^{\epsilon} f^{\epsilon}
\]

Let $\tau'$ denote evaluating the above derivative at $\epsilon = 0$, i.e.
{\footnotesize 
\begin{align*}
\tau' &=  \int \int \int y Q_{Y_i\mid S_i,X_i}(y \mid s,x) f_{Y_i\mid S_i,X_i}(y \mid s,x) f_{X_i}(x)\biggl\{ f_{S_i \mid W_i,X_i}(s \mid 1,x) - f_{S_i \mid W_i,X_i}(s \mid 0,x) \biggr\} dy ds dx \\
&\quad{} + \int \int \int y f_{Y_i\mid S_i,X_i}(y \mid s,x) f_{X_i}(x)\biggl\{ Q_{S_i \mid W_i,X_i}(s \mid 1,x)f_{S_i \mid W_i,X_i}(s \mid 1,x) \\
&\hspace{5.5cm} - Q_{S_i \mid W_i,X_i}(s \mid w=0,x)f_{S_i \mid W_i,X_i}(s \mid 0,x) \biggr\} dy ds dx \\
&\ + \int \int \int y f_{Y_i\mid S_i,X_i}(y\mid s,x) Q_{X_i}(x) f_{X_i}(x) \biggl\{ f_{S_i \mid W_i,X_i}(s \mid 1,x) - f_{S_i \mid W_i,X_i}(s \mid 0,x) \biggr\} dy ds dx \\
&=\mme \Biggl[ \mme \biggl[\mme \bigl[Y_iQ_{Y_i\mid S_i,X_i}(Y_i\mid S_i,X_i)  \bigl| S_i,X_i \bigr] \biggl| W_i=1,X_i \biggr] -  \mme \biggl[\mme \bigl[Y_iQ_{Y_i\mid S_i,X_i}(Y_i\mid S_i,X_i)  \bigl| S_i,X_i \bigr] \biggl| W_i=0,X_i \biggr]  \Biggr] \\
&\quad{} + \mme\Biggl[  \mme \biggl[\mu(S_i,X_i) Q_{S_i \mid W_i,X_i}(S_i \mid W_i=1,X_i) \biggl| W_i=1,X_i \biggr] - \mme \biggl[\mu(S_i,X_i) Q_{S_i \mid W_i,X_i}(S_i \mid W_i=0,X_i) \biggl| W_i=0,X_i \biggr] \Biggr] \\
&\quad{} + \mme[Q_{X_i}(X_i) (\mu(1,X_i) - \mu(0,X_i))]
\end{align*}
}
Third, consider the conjectured efficient influence function (EIF).
{\footnotesize \[
\psi(Y_i,S_i,W_i,X_i) = \frac{(Y_i - \mu(S_i,X_i)) (\hrho(S_i,X_i) - \hrho(X_i))}{\hrho(X_i)(1-\hrho(X_i))} + \frac{W_i(\mu(S_i,X_i) - \mu(1,X_i))}{\hrho(X_i)} - \frac{(1-W_i)(\mu(S_i,X_i) - \mu(0,X_i))}{1-\hrho(X_i)} + \mu(1,X_i) - \mu(0,X_i) - \tau
\]}
We show that $\psi(Y_i,S_i,W_i,X_i)$ is an element of the tangent space $\mathcal{T}$ by showing that different parts of $\psi(Y_i,S_i,W_i,X_i)$ satisfies conditions for $\overline{Q}_1, \overline{Q}_2$, and $\overline{Q}_4$.
\begin{enumerate}
\item For $\overline{Q}_1$, we have $\mme\biggl[\frac{(Y_i - \mu(S_i,X_i)) (\hrho(S_i,X_i) - \hrho(X_i))}{\hrhoe(X_i)(1-\hrho(X_i))} \biggl| S_i=s,X_i=x\biggr] = 0$ by definition of $\mu(S_i,X_i)$ and $\mme \biggl[\frac{(Y_i - \mu(S_i,X_i)) (\hrho(S_i,X_i) - \hrho(X_i))}{\hrho(X_i)(1-\hrho(X_i))} \biggl|  S_i=s,W_i=w,X_i=x\biggr] = 0$ by using statistical surrogacy.
\item  For $\overline{Q}_2$, we have  $\mme \biggl[\frac{W_i(\mu(S_i,X_i) - \mu(1,X_i))}{\hrho(X_i)} \biggl|  W_i=w,X_i=x\biggr] = \frac{w}{\hrho(x)} (\mme [\mu(S_i,X_i) \mid W_i=w,X_i=x] -\mu(1,x)) = 0$ for any value of $w$. Similarly,  $\mme \biggl[\frac{(1-W_i)(\mu(S_i,X_i) - \mu(0,X_i))}{1-\hrho(X_i)} \biggl|  W_i=w,X_i=x\biggr] = \frac{1-w}{1-\hrho(x)} \biggl(\mme \biggl[h(S_i,X_i) \biggl| W_i=w,X_i=x\biggr] -\mu(0,x)\biggr) = 0$ for any value of $w$. 
\item For $\overline{Q}_4$, we have $\mme[ \mu(1,X_i) - \mu(0,X_i) - \tau] = 0$.
\end{enumerate}
By setting $\overline{Q}_3 = 0$, we arrive at $\psi(Y_i,S_i,W_i,X_i) \in \mathcal{T}$.

Fourth, we show that $\tau'$ and $\psi(Y_i,S_i,W_i,X_i)$ satisfy the following relationship that all efficient influence functions must satisfy from Theorem 2.2 in \citet{newey1990semiparametric}:
\begin{equation}
\tau' = \mme [\psi(Y_i,S_i,W_i,X_i) \cdot Q(Y_i,S_i,W_i,X_i)].
\end{equation}
We break the proof of this equality into several steps.
\begin{enumerate}[label=(\alph*)]
\item Let us consider the part of the $\psi(Y_i,S_i,W_i,X_i)$ concerning $\frac{(Y_i - \mu(S_i,X_i)) (\hrho(S_i,X_i) - \hrho(X_i))}{\hrho(X_i)(1-\hrho(X_i))}$. We have
{\footnotesize 
\begin{align*}
&\mme\biggl[\frac{(Y_i - h(S_i,X_i)) (\hrho(S_i,X_i) - \hrho(X_i))}{\hrho(X_i)(1-\hrho(X_i))} Q_{Y_i,S_i,W_i,X_i}(Y_i,S_i,W_i,X_i) \biggr] \\
=&\mme\Biggl[ \frac{1}{\hrho(X_i)(1-\hrho(X_i))} \mme \biggl[ (Y_i - \mu(S_i,X_i)) (\hrho(S_i,X_i) - \hrho(X_i)) Q_{Y_i,S_i,W_i,X_i}(Y_i,S_i,W_i,X_i) \biggl| X_i \biggr] \Biggr] \\
=& \mme\Biggl[ \frac{1}{\hrho(X_i)(1-\hrho(X_i))} E \biggl[(Y_i - \mu(S_i,X_i)) (\hrho(S_i,X_i) - \hrho(X_i))  Q_{Y_i\mid,S_i,X_i}(Y_i \mid S_i,X_i) \biggl|  X_i \biggr] \Biggr] \\
&\quad{} + \mme\Biggl[ \frac{1}{\hrho(X_i)(1-\hrho(X_i))} \mme \biggl[(Y_i - \mu(S_i,X_i)) (\hrho(S_i,X_i) - \hrho(X_i)) Q_{S_i \mid W_i,X_i}(S_i \mid W_i,X_i) \biggl|  X_i \biggr] \Biggr] \\
&\quad{} +  \mme\Biggl[ \frac{1}{\hrho(X_i)(1-\hrho(X_i))} \mme \biggl[(Y_i - \mu(S_i,X_i)) (\hrho(S_i,X_i) - \hrho(X_i)) Q_{W_i \mid X_i}(W_i\mid X_i) \biggl|  X_i \biggr] \Biggr] \\
&\quad{} + \mme \Biggl[ \frac{Q_{X_i}(X_i)}{\hrho(X_i)(1-\hrho(X_i))} \mme \biggl[(Y_i - \mu(S_i,X_i)) (\hrho(S_i,X_i) - \hrho(X_i)) \biggl| X_i \biggr] \Biggr] 
\end{align*}
} The first equality uses the law of total expectation. The second equality uses the definition of $Q_{Y_i,S_i,W_i,X_i}$. We consider each term separately, starting from the bottom. 

For the $Q_{X_i}$ term, we have 
\begin{align*}
 &\mme\Biggl[ \frac{Q_{X_i}(X_i)}{\hrho(X_i)(1-\hrho(X_i))} E \biggl[(Y_i - \mu(S_i,X_i)) (\hrho(S_i,X_i) - \hrho(X_i)) \biggl| X_i \biggr] \Biggr] \\
 =& \mme\Biggl[ \frac{Q_{X_i}(X_i)}{\hrho(X_i)(1-\hrho(X_i))} \mme \biggl[ (\hrho(S_i,X_i) - \hrho(X_i)) \mme \bigl[(Y_i - \mu(S_i,X_i)) \bigl| S_i, X_i \biggr] \biggl| X_i \biggr] \Biggr] \\
 =&\mme  \Biggl[ \frac{Q_4(X_i)}{\hrho(X_i)(1-\hrho(X_i))} \mme  \biggl[ (\hrho(S_i,X_i) - \hrho(X_i)) \cdot 0 \biggl| X_i \biggr] \Biggr] \\
 =& 0.
\end{align*}
The first equality uses the law of total expectation. The second equality uses the definition of $\mu(S_i,X_i)$.

For the $Q_{W_i\mid X_i}$ term, we have
{\footnotesize 
\begin{align*}
&\mme \Biggl[ \frac{1}{\hrho(X_i)(1-\hrho(X_i))} \mme \biggl[(Y_i - \mu(S_i,X_i)) (\hrho(S_i,X_i) - \hrho(X_i)) Q_{W_i\mid X_i}(W_i\mid X_i) \biggl| X_i \biggr] \Biggr] \\
=& \mme \Biggl[ \frac{1}{\hrho(X_i)(1-\hrho(X_i))} \mme \biggl[ Q_{W_i\mid X_i}(W_i \mid X_i) \mme \bigl[ (Y_i - h(S_i,X_i)) (\hrho(S_i,X_i) - \hrho(X_i)) \bigl|  W_i,X_i \bigr] \biggl| X_i \biggr] \Biggr] \\
=& \mme \Biggl[ \frac{1}{\hrho(X_i)(1-\hrho(X_i))} \mme \biggl[ Q_{W_i\mid X_i}(W_i \mid X_i) \mme \biggl[  (\hrho(S_i,X_i) - \hrho(X_i)) \mme \bigl[ (Y_i - \mu(S_i,X_i)) \bigl| S_i, W_i,X_i \bigr] \biggl| W_i,X_i \biggr] \biggl| X_i \biggr] \Biggr] \\
=& \mme \Biggl[ \frac{1}{\hrho(X_i)(1-\hrho(X_i))} \mme \biggl[ Q_{W_i\mid X_i}(W_i \mid X_i) \mme \biggl[  (\hrho(S_i,X_i) - \hrho(X_i))  (\mme \bigl[Y_i \mid S_i,W_i,X_i\bigr] - \mu(S_i,X_i)) \biggl| W_i,X_i \biggr] \biggl| X_i \biggr] \Biggr] \\
=& \mme \Biggl[ \frac{1}{\hrho(X_i)(1-\hrho(X_i))} \mme \biggl[ Q_{W_i\mid X_i}(W_i \mid X_i) \mme \biggl[  (\hrho(S_i,X_i) - \hrho(X_i)) \cdot 0 \biggl| W_i,X_i \biggr] \biggl| X_i \biggr] \Biggr] \\
=&  0
\end{align*}} 
The first and second equalities use the law of total expectation. The third equality is algebra. The fourth equality uses statistical surrogacy. 

For the $Q_{S_i \mid W_i,X_i}$ term, we have
%{\footnotesize 
\begin{align*}
 &\mme \Biggl[ \frac{1}{\hrho(X_i)(1-\hrho(X_i))} \mme \biggl[(Y_i - \mu(S_i,X_i)) (\hrho(S_i,X_i) - \hrho(X_i)) Q_{S_i \mid W_i,X_i}(S_i \mid W_i,X_i) \biggl| X_i \biggr] \Biggr] \\
 =&\mme \Biggl[ \frac{1}{\hrho(X_i)(1-\hrho(X_i))} \mme \biggl[ \mme \biggl[ (Y_i - \mu(S_i,X_i)) (\hrho(S_i,X_i) - \hrho(X_i)) Q_{S_i \mid W_i,X_i}(S_i \mid W_i,X_i) \biggl|  X_i,W_i  \biggr] \biggl| X_i \biggr] \Biggr] \\
 =& \mme \Biggl[ \frac{1}{\hrho(X_i)(1-\hrho(X_i))} \mme \biggl[ \mme \biggl[ (\hrho(S_i,X_i) - \hrho(X_i)) Q_{S_i \mid W_i,X_i}(S_i \mid W_i,X_i) \mme \biggl[ Y_i - \mu(S_i,X_i) \biggl| S_i,X_i,W_i \biggr] \biggl|  X_i,W_i  \biggr] \biggl| X_i \biggr] \Biggr] \\
 =& 0
\end{align*}
 The first two equalities use the law of total expectation. The third equality uses statistical surrogacy.

For the $Q_{Y_i\mid S_i,X_i}$ term, we have
{\footnotesize
\begin{align*}
&\mme \Biggl[ \frac{1}{\hrho(X_i)(1-\hrho(X_i))} \mme \biggl[(Y_i - \mu(S_i,X_i)) (\hrho(S_i,X_i) - \hrho(X_i))  Q_{Y_i\mid S_i,X_i}(Y_i \mid S_i,X_i) \biggl|  X_i \biggr] \Biggr] \\
=&\mme \Biggl[ \frac{1}{\hrho(X_i)(1-\hrho(X_i))} \mme \biggl[ Y_i (\hrho(S_i,X_i) - \hrho(X_i))  Q_{Y_i\mid S_i,X_i}(Y_i \mid S_i,X_i) \biggl|  X_i \biggr] \Biggr] \\
&\quad{} - \mme \Biggl[ \frac{1}{\hrho(X_i)(1-\hrho(X_i))} \mme \biggl[ \mu(S_i,X_i) (\hrho(S_i,X_i) - \hrho(X_i))  Q_{Y_i\mid S_i,X_i}(Y_i \mid S_i,X_i) \biggl| X_i \biggr] \Biggr]
\end{align*}}
The first term above is equal to $\mme \biggl[\frac{Y_i(W_i -\hrho(X_i))Q_{Y_i\mid S_i,X_i}(Y_i \mid S_i,X_i) }{\hrho(X_i) (1-\hrho(X_i))} \biggr]$ because

\begin{align*}
&\mme \Biggl[ \frac{Y_i(W_i -\hrho(X_i)) Q_{Y_i\mid S_i,X_i}(Y_i \mid S_i,X_i)}{\hrho(X_i) (1-\hrho(X_i))} \biggr] \\
=& \mme \biggl[ \frac{1}{\hrho(X_i)(1-\hrho(X_i))} \mme \biggl[Y_i(W_i -\hrho(X_i)) Q_{Y_i\mid S_i,X_i}(Y_i \mid S_i,X_i) \mid S_i,X_i \biggr] \Biggr] \\
=& \mme \Biggl[ \frac{1}{\hrho(X_i)(1-\hrho(X_i))} \mme \biggl[Y_i Q_{Y_i\mid S_i,X_i}(Y_i \mid S_i,X_i) \biggl| S_i,X_i \biggr]  \mme \biggl[ W_i - \hrho(X_i) \biggl| S_i,X_i \biggr]  \Biggr] \\
=& \mme \Biggl[ \frac{1}{\hrho(X_i)(1-\hrho(X_i))} \mme \biggl[Y_i Q_{Y_i\mid S_i,X_i}(Y_i \mid S_i,X_i) \biggl|  S_i,X_i \biggr]  (\hrho(S_i,X_i) - \hrho(X_i)) \biggr] \\
=& \mme \Biggl[ \frac{Y_i (\hrho(S_i,X_i) - \hrho(X_i))  Q_{Y_i\mid S_i,X_i}(Y_i \mid S_i,X_i)}{\hrho(X_i)(1-\hrho(X_i))} \Biggr]
\end{align*}
The first equality uses the law of total expectation. The second equality uses statistical surrogacy where $Y_i \perp W_i | S_i,X_i$ implies $Y_i,S_i,X_i \perp W_i,X_i | S_i,X_i$. The third equality is the definition of the surrogate score. The fourth equality uses the law of total expectation.

The second term above simplifies to zero because
{\footnotesize \begin{align*}
&\mme \Biggl[ \frac{1}{\hrho(X_i)(1-\hrho(X_i))} \mme \biggl[ \mu(S_i,X_i) (\hrho(S_i,X_i) - \hrho(X_i))  Q_{Y_i\mid S_i,X_i}(Y_i \mid S_i,X_i) \biggl| X_i \biggr] \Biggr] \\
=& \mme \Biggl[ \frac{1}{\hrho(X_i)(1-\hrho(X_i))} \mme \biggl[ \mu(S_i,X_i) (\hrho(S_i,X_i) - \hrho(X_i)) \mme \biggl[  Q_{Y_i\mid S_i,X_i}(Y_i \mid S_i,X_i) \biggl| S_i,X_i \biggr] \biggl| X_i \biggr] \Biggr] \\
=& \mme \Biggl[ \frac{1}{\hrho(X_i)(1-\hrho(X_i))} \mme \biggl[ \mu(S_i,X_i) (\hrho(S_i,X_i) - \hrho(X_i)) \cdot 0 \biggl| X_i \biggr] \Biggr] \\
=& 0
\end{align*}}
The first equality uses the law of total expectation. The second equality uses the the mean-zero property of the score function $Q_{Y_i\mid S_i,X_i}(Y_i\mid S_i,X_i)$.

Finally, we can rewrite $\frac{Y_i(W_i -\hrho(X_i)) Q_{Y_i\mid S_i,X_i}(Y_i \mid S_i,X_i)}{\hrho(X_i) (1-\hrho(X_i))}$ as 
{\footnotesize
\[
 \frac{Y_i(W_i -\hrho(X_i)) Q_{Y_i\mid S_i,X_i}(Y_i \mid S_i,X_i)}{\hrho(X_i) (1-\hrho(X_i))} = \frac{Y_iW_iQ_{Y_i\mid S_i,X_i}(Y_i\mid S_i,X_i)}{\hrho(X_i)} - \frac{Y_i(1-W_i)Q_{Y_i\mid S_i,X_i}(Y_i\mid S_i,X_i)}{1-\hrho(X_i)}
\]}
Also, in expectation, each term above equals to 
{\scriptsize
\begin{align*}
\mme \biggl[ \frac{W_iY_iQ_{Y_i\mid S_i,X_i}(Y_i\mid S_i,X_i)}{\hrhoe(X_i)}  \biggr] &
%= E\biggl[ \frac{1}{\hrhoe(X_i)} E \biggl[ W_iY_iQ_1(Y_i\mid S_i,X_i) \mid X_i \biggr] \biggr] 
= \mme \biggl[ \frac{1}{\hrhoe(X_i)} \mme \biggl[ Y_iQ_{Y_i\mid S_i,X_i}(Y_i\mid S_i,X_i) \mid W_i = 1, X_i \biggr] \hrhoe(X_i) \biggr] = \mme \biggl[ \mme \biggl[ Y_iQ_{Y_i\mid S_i,X_i}(Y_i\mid S_i,X_i) \mid W_i = 1, X_i \biggr] \biggr], \\
\mme \biggl[ \frac{(1-W_i)Y_iQ_{Y_i\mid S_i,X_i}(Y_i\mid S_i,X_i)}{1-\hrhoe(X_i)}  \biggr] &= %\mme \biggl[ \frac{1}{1-\hrhoe(X_i)} E \biggl[ (1-W_i)Y_iQ_1(Y_i\mid S_i,X_i) \mid X_i \biggr] \biggr] 
\mme \biggl[ \frac{1}{1-\hrhoe(X_i)} \mme \biggl[ Y_iQ_{Y_i\mid S_i,X_i}(Y_i\mid S_i,X_i) \mid W_i = 0, X_i \biggr] (1-\hrhoe(X_i)) \biggr] \\
&= \mme \biggl[ \mme \biggl[ Y_iQ_{Y_i\mid S_i,X_i} (Y_i\mid S_i,X_i) \mid W_i = 0, X_i \biggr] \biggr].
\end{align*}
}
The first equality uses the law of total expectation and the definition of the propensity score. The second equality is algebra. Overall, we have 
\begin{align*}
&\mme \biggl[\frac{(Y_i-\mu(S_i,X_i))(\hrho(S_i,X_i) - \hrhoe(X_i))}{\hrhoe(X_i)(1-\hrhoe(X_i))} Q_{Y_i,S_i,W_i,X_i}(Y_i,S_i,W_i,X_i) \biggl] \\
=& \mme \biggl[ \mme \biggl[ Y_iQ_{Y_i\mid S_i,X_i}(Y_i\mid S_i,X_i) \mid W_i = 1, X_i \biggr] \biggr] - \mme \biggl[ \mme \biggl[ Y_iQ_{Y_i\mid S_i,X_i}(Y_i\mid S_i,X_i) \mid W_i = 0, X_i \biggr] \biggr].
\end{align*}

\item Let's consider the part of the $\psi(Y_i,S_i,W_i,X_i)$ concerning $\frac{W_i(\mu(S_i,X_i) - \mu(1,X_i))}{\hrho(X_i)}$. We have
{\footnotesize 
\begin{align*}
&\mme\biggl[ \frac{W_i(\mu(S_i,X_i) - \mu(1,X_i))}{\hrhoe(X_i)} Q_{Y_i,S_i,W_i,X_i}(Y_i,S_i,W_i,X_i)\biggr] \\
=& \mme \biggl[ \frac{W_i}{\hrho(X_i)} \mme \biggl[ \biggl\{ \mu(S_i,X_i) - \mu(1,X_i) \biggr\} \biggl\{ Q_{Y_i \mid S_i,X_i}(Y_i \mid S_i,X_i) + Q_{S_i \mid W_i,X_i}(S_i \mid W_i,X_i) + Q_{W_i \mid X_i}(W_i \mid X_i) + Q_{X_i}(X_i)\biggr\} \biggl|  W_i,X_i\biggr] \biggr] \\
=& \mme \biggl[ \frac{W_i}{\hrho(X_i)} \mme \biggl[ (\mu(S_i,X_i) - \mu(1,X_i)) Q_{Y_i \mid S_i,X_i}(Y_i \mid S_i,X_i) \biggl| W_i,X_i  \biggr] \biggr] \\
&\quad{} +  \mme \biggl[ \frac{W_i}{\hrhoe(X_i)} \mme \biggl[ (\mu(S_i,X_i) - \mu(1,X_i)) Q_{S_i \mid W_i,X_i}(S_i \mid W_i,X_i) \biggl| W_i,X_i  \biggr] \biggr] \\
&\quad{} + \mme \biggl[ \frac{W_i}{\hrhoe(X_i)} Q_{W_i \mid X_i}(W_i \mid X_i) \mme \biggl[ \mu(S_i,X_i) - \mu(1,X_i) \biggl| W_i,X_i  \biggr] \biggr] \\
&\quad{}+\mme \biggl[ \frac{W_i}{\hrhoe(X_i)} Q_{X_i}(X_i) \mme \biggl[ \mu(S_i,X_i) - \mu(1,X_i) \biggl| W_i,X_i  \biggr] \biggr] \\
=& \mme \biggl[ \frac{W_i}{\hrhoe(X_i)} \mme \biggl[ (\mu(S_i,X_i) - \mu(1,X_i)) \mme \biggl[ Q_{Y_i \mid S_i,X_i}(Y_i \mid S_i,X_i) \biggl| S_i,W_i,X_i \biggr] \bigg| W_i,X_i  \biggr] \biggr] \\
&\quad{} +  \mme \biggl[ \frac{W_i}{\hrhoe(X_i)} \mme \biggl[ (\mu(S_i,X_i) - \mu(1,X_i)) Q_{S_i\mid W_i,X_i}(S_i \mid W_i,X_i) \biggl| W_i,X_i  \biggr] \biggr] \\
=& \mme \biggl[ \frac{W_i(\mu(S_i,X_i) - \mu(1,X_i))Q_{S_i \mid W_i,X_i}(S_i \mid W_i,X_i)}{\hrhoe(X_i)} \biggr]
\end{align*}}
The first equality uses the law of total expectation.  The third equality uses the relationship $\mme [\mu(S_i,X_i)| W_i=w,X_i] = \mu(w,X_i)$. The fourth equality uses the mean-zero property of the score and the law of total expectation.

We can further simplify the above expression by noticing that
\begin{align*}
\mme \biggl[ \frac{W_i \mu(1,X_i)Q_{S_i \mid W_i,X_i}(S_i \mid W_i,X_i)}{\hrhoe(X_i)} \biggr] 
%&= E \biggl[ \frac{\mu(1,X_i)}{\hrhoe(X_i)} E\biggl[ W_iQ_{S_i \mid W_i,X_i}(S_i \mid W_i,X_i)\mid X_i \biggr] \biggr] \\
&= \mme \biggl[ \frac{W_i \mu(1,X_i)}{\hrhoe(X_i)} \mme \biggl[ Q_{S_i \mid W_i,X_i}(S_i \mid W_i,X_i)\biggl| W_i, X_i \biggr] \biggr] =0
\end{align*}
The first equality uses the law of total expectation. The second equality uses the mean-zero property of the score function. Also,
\begin{align*} \mme \biggl[ \frac{W_i \mu(S_i,X_i) Q_{S_i \mid W_i,X_i}(S_i \mid W_i,X_i)}{\hrhoe(X_i)} \biggr] &= %\mme \biggl[ \frac{1}{\hrhoe(X_i)} \mme \biggl[ W_i \mu(S_i,X_i) Q_{S_i \mid W_i,X_i}(S_i \mid W_i,X_i)\biggl| X_i \biggr] \biggr]  \\
 \mme \biggl[ \mme \biggl[\mu(S_i,X_i) Q_{S_i \mid W_i,X_i}(S_i \mid W_i=1,X_i)\biggl| W_i=1,X_i \biggr] \biggl| X_i \biggr] 
\end{align*}
The first equality uses the law of total expectation and the definition of conditional expectation with the definition $\mme [W_i \mid X_i] = \hrhoe(X_i)$. 

Overall, we end up with the following expression
\[
\mme \biggl[ \frac{W_i(\mu(S_i,X_i) - \mu(1,X_i))}{\hrhoe(X_i)} Q_{Y_i,S_i,W_i,X_i}(Y_i,S_i,W_i,X_i)\biggr] =  \mme  \biggl[  \mme \biggl[\mu(S_i,X_i) Q_{S_i \mid W_i,X_i}(S_i \mid W_i=1,X_i)\biggl| X_i,W_i=1 \biggr] \biggl| X_i \biggr]  
\]
\item Let's consider the part of the $\psi(Y_i,S_i,W_i,X_i)$ concerning $\frac{(1-W_i)(\mu(S_i,X_i) - \mu(0,X_i))}{1-\hrhoe(X_i)}$. From the above exercise, we end up with 
{\footnotesize 
\[
\mme \biggl[ \frac{(1-W_i)(\mu(S_i,X_i) - \mu(0,X_i))}{1-\hrhoe(X_i)} Q_{Y_i,S_i,W_i,X_i}(Y_i,S_i,W_i,X_i)\biggr] = \mme \biggl[ \mme \biggl[ \mu(S_i,X_i)Q_{S_i \mid W_i,X_i}(S_i \mid W_i=0,X_i) \biggl| W_i=0,X_i] \biggl| X_i\biggr]
\]}
\item Let's consider the part of the $\psi(Y_i,S_i,W_i,X_i)$ concerning $\mu(1,X_i) - \mu(0,X_i) - \tau$. We have
\begin{align*}
&\mme [ (\mu(1,X_i) - \mu(0,X_i) - \tau)Q_{Y_i,S_i,W_i,X_i}(Y_i,S_i,W_i,X_i)] \\
%=& \mme [(\mu(1,X_i) - \mu(0,X_i) - \tau) \mme [Q(Y_i,S_i,W_i,X_i) | X_i]] \\
=& \mme [(\mu(1,X_i) - \mu(0,X_i) - \tau) \mme[Q_{Y_i\mid S_i,X_i}(Y_i \mid S_i,X_i) + Q_{S_i \mid W_i,X_i}(S_i \mid W_i,X_i) + Q_{W_i \mid X_i}(W_i \mid X_i) + Q_{X_i}(X_i) \mid X_i]] \\
=&  \mme [(\mu(1,X_i) - \mu(0,X_i) - \tau) \mme [Q_{Y_i\mid S_i,X_i}(Y_i \mid S_i,X_i) + Q_{S_i \mid W_i,X_i}(S_i \mid W_i,X_i) + Q_{X_i}(X_i) \mid X_i]] \\
=& E[(\mu(1,X_i) - \mu(0,X_i) - \tau) Q_{X_i}(X_i)] \\
=& E[(\mu(1,X_i) - \mu(0,X_i)) Q_{X_i}(X_i)]
\end{align*}
The first equality uses the law of total expectation. The second equality uses the property of the score where $E[Q_{W_i \mid X_i}(W_i \mid X_i) \mid X_i] = 0$. The third equality uses both the law of total expectation and the property of the score where 
\[
\mme[Q_{Y_i \mid S_i,X_i}(Y_i \mid S_i,X_i) \mid X_i] = \mme [ \mme [Q_{Y_i \mid S_i,X_i}(Y_i \mid S_i,X_i) \mid S_i,X_i] \mid X_i] = \mme [0 \mid X_i] = 0
\]

\end{enumerate}

%\end{enumerate}

Combining the four steps (a)-(d) arrives at the desired equality between $\tau'$ and $\psi(Y_i,S_i,W_i,X_i)$.

Finally, note that the $\mathbb{V}_s$ is obtained by calculating the variance of the EIF (already written in Theorem 3):\footnote{We henceforth explicitly show the conditioning $\poin=\ppe$ to be consistent with the notation in our Theorem statement.} 

{\footnotesize \[
\psi(Y_i,S_i,W_i,X_i,\poin) = \frac{(Y_i - \mu(S_i,X_i,\ppe)) (\hrho(S_i,X_i) - \hrho(X_i))}{\hrho(X_i)(1-\hrho(X_i))}+ \frac{W_i(\mu(S_i,X_i,\ppe) - \mu(1,X_i))}{\hrho(X_i)} 
\]}
{\footnotesize \[
- \frac{(1-W_i)(\mu(S_i,X_i,\ppe) - \mu(0,X_i))}{1-\hrho(X_i)}
 + \mu(1,X_i) - \mu(0,X_i) - \tau,
\] } i.e.,
\[
\mmv_{{\rm s}} = \bigg[\psi(Y_i,S_i,W_i,X_i,\poin)^2 \bigg] =\mme \bigg[ \left( \frac{(Y_i - \mu(S_i,X_i,\ppe)) (\hrho(S_i,X_i) - \hrho(X_i))}{\hrho(X_i)(1-\hrho(X_i))} \right)^2  
+ \left( \frac{W_i(\mu(S_i,X_i,\ppe) - \mu(1,X_i))}{\hrho(X_i)} \right)^2
\]
\[
+ \left(  \frac{(1-W_i)(\mu(S_i,X_i,\ppe) - \mu(0,X_i))}{1-\hrho(X_i)} \right)^2
 + \left( \mu(1,X_i) - \mu(0,X_i) - \tau \right)^2
\bigg]
\]

\[
=\mme\left[\sigma^2(\soin,\xoin,\ppe)  \left( \frac{\hrhoe(\soin,\xoin)- \hrho(\xoin)}{\hrho(\xoin)(1-\hrho(\xoin))} \right)^2+ \left( \hkappa(1,\xoin) - \hkappa(0,\xoin) - \tau \right)^2  \right.\]
\[\left. + \frac{\woin}{\hrho(\xoin)^2} (\hmu(\soin,\xoin,\ppe) -\hkappa(1,\xoin))^2
+ \frac{1-\woin}{ (1 - \hrho(\xoin))^2} (\hmu(\soin,\xoin,\ppe) - \hkappa(0,\xoin))^2
 \right]
\]
\[
=\mme\left[\sigma^2(\soin,\xoin,\ppe)  \left( \frac{\hrhoe(\soin,\xoin)- \hrho(\xoin)}{\hrho(\xoin)(1-\hrho(\xoin))} \right)^2+ \left( \hkappa(1,\xoin) - \hkappa(0,\xoin) - \tau \right)^2  \right.\]
\[\left. + \frac{\hrhoe(\soin,\xoin)}{\hrho(\xoin)^2} (\hmu(\soin,\xoin,\ppe) -\hkappa(1,\xoin))^2
+ \frac{1-\hrhoe(\soin,\xoin)}{ (1 - \hrho(\xoin))^2} (\hmu(\soin,\xoin,\ppe) - \hkappa(0,\xoin))^2
 \right]
\]
by the law of iterated expectations, and hence we have that 

\[
\Delta=\mmv_{{\rm ns}} - \mmv_{{\rm s}}
= \mme\left[\sigma^2(\soin,\xoin,\ppe)  \left( \frac{\rho\left(S_i, X_i\right)}{\rho\left(X_i\right)^2}+\frac{1-\rho\left(S_i, X_i\right)}{\left(1-\rho\left(X_i\right)\right)^2} - \left(\frac{\hrhoe(\soin,\xoin)- \hrho(\xoin)}{\hrho(\xoin)(1-\hrho(\xoin))}\right)^2 \right)  \right]
\]
\[ = \mme\left[\sigma^2(\soin,\xoin,\ppe)  \frac{\rho\left(S_i, X_i\right)\left(1-\rho\left(S_i, X_i\right)\right)}{\rho\left(X_i\right)^2\left(1-\rho\left(X_i\right)\right)^2}  \right]
\]

\textsc{\small{}{}{}
$\square$}{\small\par}

\textit{Proof of Theorem \ref{theorem:bias}:} Consider part (i).
By the law of iterated expectations conditional on $\soin$ and $\xoin$,
we have 
\begin{align*}
\tau^{\ppe}\equiv & \mme\left[\left.\hmu(\soin,\xoin,\ppo)\cdot\frac{\woin}{\hrho(\xoin)}-\hmu(\soin,\xoin,\ppo)\cdot\frac{1-\woin}{1-\hrho(\xoin)}\right|P_{i}=\ppe\right]\\
= & \mme\left[\left.\hmu(\soin,\xoin,\ppo)\cdot\frac{\hrho(\soin,\xoin)}{\hrho(\xoin)}-\hmu(\soin,\xoin,\ppo)\cdot\frac{1-\hrho(\soin,\xoin)}{1-\hrho(\xoin)}\right|P_{i}=\ppe\right]
\end{align*}
By the proof of \eqref{peen} in Theorem \ref{theorem1} where we
don't use Surrogacy or Comparability, we get 
\begin{align*}
\tau^{\ppo}\equiv & \mme\left[\left.\yoio\cdot\frac{\hrho(\soio,\xoio)\cdot \varphi(\soio,\xoio)\cdot(1-\varphi)}{\hrhoe(\xoio)\cdot(1-\varphi(\soio,\xoio))\cdot \varphi}-\yoio\cdot\frac{(1-\hrho(\soio,\xoio))\cdot \varphi(\soio,\xoio)\cdot(1-\varphi)}{(1-\hrhoe(\xoio))\cdot(1-\varphi(\soio,\xoio))\cdot \varphi}\right|P_{i}=\ppo\right]\\
= & \mme\left[\left.\hmu(\soin,\xoin,\ppo)\cdot\frac{\hrho(\soin,\xoin)}{\hrho(\xoin)}-\hmu(\soin,\xoin,\ppo)\cdot\frac{1-\hrho(\soin,\xoin)}{1-\hrho(\xoin)}\right|P_{i}=\ppe\right]
\end{align*}
The second equality in $\tau^{\ppe}=\tau^{\ppo}=\tau^{\ppe,\ppo}$
is immediate based on only the law of iterated expectations. Finally,
by the law of iterated expectations conditional on $\xoin$, we have
\begin{align*}
 & \mme\left[\left.\hmu(\soin,\xoin,\ppo)\cdot\frac{\woin}{\hrho(\xoin)}-\hmu(\soin,\xoin,\ppo)\cdot\frac{1-\woin}{1-\hrho(\xoin)}\right|P_{i}=\ppe\right]\\
= & \mme\left[\left.\mme\left[\left.\hmu(\soin,\xoin,\ppo)\cdot\frac{\woin}{\hrho(\xoin)}-\hmu(\soin,\xoin,\ppo)\cdot\frac{1-\woin}{1-\hrho(\xoin)}\right|\xoin,P_{i}=\ppe\right]\right|P_{i}=\ppe\right]
\end{align*}
By Assumption \ref{ass:unconf} (unconfoundedness), we have %{\small{}{}{}
\begin{align*}
 & \mme\left[\left.\mme\left[\left.\hmu(\soin,\xoin,\ppo)\cdot\frac{\woin}{\hrho(\xoin)}-\hmu(\soin,\xoin,\ppo)\cdot\frac{1-\woin}{1-\hrho(\xoin)}\right|\xoin,P_{i}=\ppe\right]\right|P_{i}=\ppe\right]\\
= & \mme\left[\left.\mme\left[\left.\hmu(\soine,\xoin,\ppo)\cdot\frac{\woin}{\hrho(\xoin)}-\hmu(\soinn,\xoin,\ppo)\cdot\frac{1-\woin}{1-\hrho(\xoin)}\right|\xoin,P_{i}=\ppe\right]\right|P_{i}=\ppe\right]\\
= & \mme\left[\left.\mme\left[\hmu(\soine,\xoin,\ppo)\mid\xoin,P_{i}=\ppe\right]-\mme\left[\hmu(\soinn,\xoin,\ppo)\mid\xoin,P_{i}=\ppe\right]\right|P_{i}=\ppe\right]
\end{align*}
Undoing the law of iterated expectations give the desired result.%{\small\par}

%{\small{}{}{}
For parts (ii)-(iv), we prove (iv) first. By Assumption
\ref{ass:unconf} (unconfoundedness), we have 
\[
\tau=\mme\left[\mme\left[\yoin|\woin=1,\xoin,P_{i}=\ppe\right]\mid P_{i}=\ppe\right]-\mme\left[\mme\left[\yoin|\woin=0,\xoin,P_{i}=\ppe\right]\mid P_{i}=\ppe\right].
\]
By iterated expectations, this is equal to 
\begin{align*}
\tau & =\mme\left[\mme\left[\mme\left[\yoin|\soin,\woin=1,\xoin,P_{i}=\ppe\right]|\woin=1,\xoin,P_{i}=\ppe\right]\mid P_{i}=\ppe\right]\\
 & \quad{}-\mme\left[\mme\left[\mme\left[\yoin|\soin,\woin=0,\xoin,P_{i}=\ppe\right]|\woin=0,\xoin,P_{i}=\ppe\right]\mid P_{i}=\ppe\right]\\
 & =\mme\left[\mme\left[\mu(S_{i},1,X_{i},\ppe)|\woin=1,\xoin,P_{i}=\ppe\right]\mid P_{i}=\ppe\right]-\mme\left[\mme\left[\mu(S_{i},0,X_{i},\ppe)|\woin=0,\xoin,P_{i}=\ppe\right]\mid P_{i}=\ppe\right]
\end{align*}
Thus, we have 
\begin{align*}
 & \tau-\mme\left[\hmu(\soine,\xoin,\ppo)-\hmu(\soinn,\xoin,\ppo)\mid P_{i}=\ppe\right]\\
= & \mme\left[\mme\left[\mu(S_{i},1,X_{i},\ppe)|\woin=1,\xoin,P_{i}=\ppe\right]\mid P_{i}=\ppe\right]-\mme\left[\mme\left[\mu(S_{i},0,X_{i},\ppe)|\woin=0,\xoin,P_{i}=\ppe\right]\mid P_{i}=\ppe\right]\\
 & \quad{}-\left\{ \mme\left[\mme\left[\hmu(S_{i},\xoin,\ppo)\mid W_{i}=1,X_{i},P_{i}=\ppe\right]\mid P_{i}=\ppe\right]-\mme\left[\mme\left[\hmu(S_{i},\xoin,\ppo)\mid W_{i}=0,X_{i},P_{i}=\ppe\right]\mid P_{i}=\ppe\right]\right\} 
\end{align*}
We add and subtract 
\[
\mme\left[\mme\left[\hmu(\soin,\xoin,\ppe)|\woin=1,\xoin,P_{i}=\ppe\right]\mid P_{i}=\ppe\right]-\mme\left[\mme\left[\hmu(\soin,\xoin,\ppe)|\woin=0,\xoin,P_{i}=\ppe\right]\mid P_{i}=\ppe\right]
\]
to get 
\begin{align*}
 & \tau-\mme\left[\hmu(\soine,\xoin,\ppo)-\hmu(\soinn,\xoin,\ppo)\mid P_{i}=\ppe\right]\\
= & \mme\left[\mme\left[\mu(S_{i},1,X_{i},\ppe)|\woin=1,\xoin,P_{i}=\ppe\right]\mid P_{i}=\ppe\right]-\mme\left[\mme\left[\mu(S_{i},0,X_{i},\ppe)|\woin=0,\xoin,P_{i}=\ppe\right]\mid P_{i}=\ppe\right]\\
 & \quad{}-\mme\left[\mme\left[\hmu(\soin,\xoin,\ppe)|\woin=1,\xoin,P_{i}=\ppe\right]\mid P_{i}=\ppe\right]+\mme\left[\mme\left[\hmu(\soin,\xoin,\ppe)|\woin=0,\xoin,P_{i}=\ppe\right]\mid P_{i}=\ppe\right]\\
 & \quad{}+\mme\left[\mme\left[\hmu(\soin,\xoin,\ppe)|\woin=1,\xoin,P_{i}=\ppe\right]\mid P_{i}=\ppe\right]-\mme\left[\mme\left[\hmu(\soin,\xoin,\ppe)|\woin=0,\xoin,P_{i}=\ppe\right]\mid P_{i}=\ppe\right]\\
 & \quad{}-\left\{ \mme\left[\mme\left[\hmu(S_{i},\xoin,\ppo)\mid W_{i}=1,X_{i},P_{i}=\ppe\right]\mid P_{i}=\ppe\right]-\mme\left[\mme\left[\hmu(S_{i},\xoin,\ppo)\mid W_{i}=0,X_{i}P_{i}=\ppe\right]\mid P_{i}=\ppe\right]\right\} 
\end{align*}
Rearranging the terms, we have 
%}{\footnotesize{}{}{} 
\begin{align}
 & \tau-\mme\left[\hmu(\soine,\xoin,\ppo)-\hmu(\soinn,\xoin,\ppo)\mid P_{i}=\ppe\right]\\
= & \mme\left[\mme\left[\mu(S_{i},1,X_{i},\ppe)|\woin=1,\xoin,P_{i}=\ppe\right]\mid P_{i}=\ppe\right]-\mme\left[\mme\left[\hmu(\soin,\xoin,\ppe)|\woin=1,\xoin,P_{i}=\ppe\right]\mid P_{i}=\ppe\right]\label{feen}\\
 & -\mme\left[\mme\left[\mu(S_{i},0,X_{i},\ppe)|\woin=0,\xoin,P_{i}=\ppe\right]\mid P_{i}=\ppe\right]+\mme\left[\mme\left[\hmu(\soin,\xoin,\ppe)|\woin=0,\xoin,P_{i}=\ppe\right]\mid P_{i}=\ppe\right]\label{ftwee}\\
 & +\mme\left[\mme\left[\hmu(\soin,\xoin,\ppe)|\woin=1,\xoin,P_{i}=\ppe\right]\mid P_{i}=\ppe\right]-\mme\left[\mme\left[\hmu(\soin,\xoin,\ppo)|\woin=1,\xoin,P_{i}=\ppe\right]\mid P_{i}=\ppe\right]\label{fdrie}\\
 & +\mme\left[\mme\left[\hmu(\soin,\xoin,\ppo)|\woin=0,\xoin,P_{i}=\ppe\right]\mid P_{i}=\ppe\right]-\mme\left[\mme\left[\hmu(\soin,\xoin,\ppe)|\woin=0,\xoin,P_{i}=\ppe\right]\mid P_{i}=\ppe\right]\label{fvier}
\end{align}
%}{\small{}{}{} 
Next, by the definition of expectations, 
\begin{align*}
\hmu(s,x,\ppe)= & \mme[\yoin|\soin=s,\xoin=x,P_{i}=\ppe]\\
= & \mme[\yoin|\soin=s,\woin=1,\xoin=x,P_{i}=\ppe]\cdot{\rm pr}(\woin=1|\soin=s,\xoin=x,P_{i}=\ppe)\\
 & \quad{}+\mme[\yoin|\soin=s,\woin=0,\xoin=x,P_{i}=\ppe]\cdot{\rm pr}(\woin=0|\soin=s,\xoin=x,P_{i}=\ppe)\\
= & \mu(s,1,x,\ppe)\cdot \hrho(s,x)+\mu(s,0,x,\ppe)\cdot(1-\hrho(s,x))
\end{align*}
Use this to write (\ref{feen}) as {\footnotesize{}{}{} 
\begin{align*}
 & \mme\left[\mme\left[\mu(S_{i},1,X_{i},\ppe)|\woin=1,\xoin,P_{i}=\ppe\right]P_{i}=\ppe\right]\\
 & \quad{}-\mme\left[\mme\left[\mu(S_{i},1,X_{i},\ppe)\cdot \hrho(S_{i},X_{i})+\mu(S_{i},0,X_{i},\ppe)\cdot(1-\hrho(S_{i},X_{i}))|\woin=1,\xoin,P_{i}=\ppe\right]\mid P_{i}=\ppe\right]\\
= & \mme\left[\mme\left[\left(\mu(S_{i},1,X_{i},\ppe)-\mu(S_{i},0,X_{i},\ppe)\right)\cdot(1-\hrho(S_{i},X_{i}))|\woin=1,\xoin,P_{i}=\ppe\right]\mid P_{i}=\ppe\right]\\
= & \mme\left[\mme\left[\left(\mu(S_{i},1,X_{i},\ppe)-\mu(S_{i},0,X_{i},\ppe)\right)\cdot\frac{(1-\hrho(S_{i},X_{i}))\cdot \hrho(S_{i},X_{i})}{\hrhoe(X_{i})}|\xoin,P_{i}=\ppe\right]\mid P_{i}=\ppe\right]\\
= & \mme\left[\left(\mu(S_{i},1,X_{i},\ppe)-\mu(S_{i},0,X_{i},\ppe)\right)\cdot\frac{(1-\hrho(S_{i},X_{i}))\hrho(S_{i},X_{i})}{\hrhoe(X_{i})}\mid P_{i}=\ppe\right]
\end{align*}
%}{\small{}{}{} 
Using the same argument we can write (\ref{ftwee})
as 
\begin{align*}
 & -\mme\left[\mme\left[\mu(S_{i},0,X_{i},\ppe)|\woin=0,\xoin,P_{i}=\ppe\right]\mid P_{i}=\ppe\right]+\mme\left[\mme\left[\hmu(\soin,\xoin,\ppe)|\woin=0,\xoin,P_{i}=\ppe\right]\mid P_{i}=\ppe\right]\\
= & -\mme\left[\mme\left[\mu(S_{i},0,X_{i},\ppe)|\woin=0,\xoin,P_{i}=\ppe\right]\mid P_{i}=\ppe\right]\\
 & \quad{}+\mme\left[\mme\left[\mu(S_{i},1,X_{i},\ppe)\cdot \hrho(S_{i},X_{i})+\mu(S_{i},0,X_{i},\ppe)\cdot(1-\hrho(S_{i},X_{i}))|\woin=0,\xoin,P_{i}=\ppe\right]\mid P_{i}=\ppe\right]\\
= & \mme\left[\mme\left[\left(\mu(S_{i},1,X_{i},\ppe)-\mu(S_{i},0,X_{i},\ppe)\right)\cdot \hrho(S_{i},X_{i})|\woin=0,\xoin,P_{i}=\ppe\right]\mid P_{i}=\ppe\right]\\
= & \mme\left[\mme\left[\left(\mu(S_{i},1,X_{i},\ppe)-\mu(S_{i},0,X_{i},\ppe)\right)\cdot\frac{(1-\hrho(S_{i},X_{i}))\cdot \hrho(S_{i},X_{i})}{1-\hrhoe(X_{i})}|\xoin,P_{i}=\ppe\right]\mid P_{i}=\ppe\right]\\
= & \mme\left[\left(\mu(S_{i},1,X_{i},\ppe)-\mu(S_{i},0,X_{i},\ppe)\right)\cdot\frac{(1-\hrho(S_{i},X_{i}))\cdot \hrho(S_{i},X_{i})}{1-\hrhoe(X_{i})}\mid P_{i}=\ppe\right]
\end{align*}
Combining the results for (\ref{feen}) and (\ref{ftwee}) leads to
\[
\mme\left[\left(\mu(S_{i},1,X_{i},\ppe)-\mu(S_{i},0,X_{i},\ppe)\right)\cdot\frac{(1-\hrho(S_{i},X_{i}))\cdot \hrho(S_{i},X_{i})}{(1-\hrhoe(X_{i}))\cdot \hrhoe(X_{i})}\mid P_{i}=\ppe\right]
\]
Collecting the last two terms, (\ref{fdrie}) and (\ref{fvier}),
we have }{\footnotesize{}{}{} 
\begin{align*}
 & \mme\left[\mme\left[\hmu(\soin,\xoin,\ppe)|\woin=1,\xoin,P_{i}=\ppe\right]\mid P_{i}=\ppe\right]-\mme\left[\mme\left[\hmu(\soin,\xoin,\ppo)|\woin=1,\xoin,P_{i}=\ppe\right]\mid P_{i}=\ppe\right]\\
 & +\mme\left[\mme\left[\hmu(\soin,\xoin,\ppo)|\woin=0,\xoin,P_{i}=\ppe\right]\mid P_{i}=\ppe\right]-\mme\left[\mme\left[\hmu(\soin,\xoin,\ppe)|\woin=0,\xoin,P_{i}=\ppe\right]\mid P_{i}=\ppe\right]\\
= & \mme\left[\mme\left[\hmu(\soin,\xoin,\ppe)\cdot\frac{\hrho(S_{i},X_{i})}{\hrhoe(X_{i})}|\xoin,P_{i}=\ppe\right]\mid P_{i}=\ppe\right]-\mme\left[\mme\left[\hmu(\soin,\xoin,\ppo)\cdot\frac{\hrho(S_{i},X_{i})}{\hrhoe(X_{i})}|\xoin,P_{i}=\ppe\right]\mid P_{i}=\ppe\right]\\
 & +\mme\left[\mme\left[\hmu(\soin,\xoin,\ppo)\cdot\frac{1-\hrho(S_{i},X_{i})}{1-\hrhoe(X_{i})}|\xoin,P_{i}=\ppe\right]\mid P_{i}=\ppe\right]-\mme\left[\mme\left[\hmu(\soin,\xoin,\ppe)\cdot\frac{1-\hrho(S_{i},X_{i})}{1-\hrhoe(X_{i})}|\xoin,P_{i}=\ppe\right]\mid P_{i}=\ppe\right]\\
= & \mme\left[\mme\left[\left(\hmu(\soin,\xoin,\ppe)-\hmu(S_{i},X_{i},\ppo)\right)\cdot\frac{\hrho(S_{i},X_{i})}{\hrhoe(X_{i})}|\xoin,P_{i}=\ppe\right]\mid P_{i}=\ppe\right]\\
 & \quad{}-\mme\left[\mme\left[\left(\hmu(\soin,\xoin,\ppe)-\hmu(S_{i},X_{i},\ppo)\right)\cdot\frac{1-\hrho(S_{i},X_{i})}{1-\hrhoe(X_{i})}|\xoin,P_{i}=\ppe\right]\mid P_{i}=\ppe\right]\\
= & \mme\left[\mme\left[\left(\hmu(\soin,\xoin,\ppe)-\hmu(S_{i},X_{i},\ppo)\right)\cdot\frac{\hrho(S_{i},X_{i})-\hrhoe(X_{i})}{(1-\hrhoe(X_{i}))\cdot \hrhoe(X_{i})}|\xoin,P_{i}=\ppe\right]\mid P_{i}=\ppe\right]\\
= & \mme\left[\left(\hmu(\soin,\xoin,\ppe)-\hmu(S_{i},X_{i},\ppo)\right)\cdot\frac{\hrho(S_{i},X_{i})-\hrhoe(X_{i})}{(1-\hrhoe(X_{i}))\cdot \hrhoe(X_{i})}\mid P_{i}=\ppe\right]
\end{align*}
}{\small{}{}{} Combining the terms together, we obtain the expression
in (iv) 
\begin{align*}
 & \tau-\mme\left[\hmu(\soine,\xoin,\ppo)-\hmu(\soinn,\xoin,\ppo)\mid P_{i}=\ppe\right]\\
= & \mme\left[\left(\mu(S_{i},1,X_{i},\ppe)-\mu(S_{i},0,X_{i},\ppe)\right)\cdot\frac{(1-\hrho(S_{i},X_{i}))\cdot \hrho(S_{i},X_{i})}{(1-\hrhoe(X_{i}))\cdot \hrhoe(X_{i})}\mid P_{i}=\ppe\right]\\
 & \quad{}+\mme\left[\left(\hmu(\soin,\xoin,\ppe)-\hmu(S_{i},X_{i},\ppo)\right)\cdot\frac{\hrho(S_{i},X_{i})-\hrhoe(X_{i})}{(1-\hrhoe(X_{i}))\cdot \hrhoe(X_{i})}\mid P_{i}=\ppe\right]
\end{align*}
Finally for part (ii), under Assumption
\ref{ass:comp} (Comparability), but not Assumption \ref{ass:surro}
(Surrogacy), $\hmu(\soin,\xoin,\ppe)-\hmu(S_{i},X_{i},\ppo)=0$ and
the result is immediate from (iv). For part (iii), under Assumption \ref{ass:surro} (Surrogacy),
but not Assumption \ref{ass:comp} (Comparability), $\mu(S_{i},1,X_{i},\ppe)-\mu(S_{i},0,X_{i},\ppe)=0$
and the result is immediate from (iv).  $\square$}%{\small\par}

\noindent{\bf Proof of Lemma \ref{lemma1}}
We can identify, given overlap, the surrogate score $\hrho(s,x)$, the propensity score $\hrho(X)$,  the surrogate index $\hmu(s,x,\ppo)$, and the joint distribution of $(\soin,\xoin,\poin)$. 
This implies that to derive upper and lower bounds we just need to derive upper and lower bounds for the difference $\mu(s,1,x,\ppe)-\mu(s,0,x,\ppe)$ for each value of $(s,x)$ and then integrate these bounds. We will demonstrate the sharpness of these bounds by showing that there exist data distributions consistent with all assumptions such that these bounds are achieved.

Part $(i)$: 
By Theorem \ref{theorem:bias} the surrogacy bias can be characterized as
\[\textrm{\rm surrogacy-bias}=
\mme\left[\left.\Bigl\{\mu(\soin,1,\xoin,\ppe)-\mu(\soin,0,\xoin,\ppe)\Bigr\}\cdot\frac{\hrho(\soin,\xoin)\cdot(1-\hrho(\soin,\xoin))}{\hrho(\xoin)\cdot(1-\hrho(\xoin))}\right|P_{i}=\ppe\right].
\]
The data are not directly informative about the two conditional expectation $\mu(s,w,x,\ppe)$ (because we do not observe the outcome in the experimental sample) beyond their relation to the surrogacy index:
\[ \hmu(s,x,\ppo)=\hrho(s,x) \hmu(s,1,x,\ppe)+(1-\hrho(s,x)) \hmu(s,0,x,\ppe),\qquad \forall s,x.\]
This implies the difference $\mu(s,1,x,\ppe)-\mu(s,0,x,\ppe)$ can be written as
\[\mu(s,1,x,\ppe)-\mu(s,0,x,\ppe)=\frac{\hmu(s,x,\ppo)}{\hrho(s,x)}-\frac{\mu(s,0,x,\ppe)}{\hrho(s,x)}.\]
Fixing $\hmu(s,x,\ppo)$, $\hrho(s,x)$, and $\mu(s,0,x,\ppe)$ this places no restrictions on the difference $\mu(s,1,x,\ppe)-\mu(s,0,x,\ppe)$ and thus no restrictions on the bias, and therefore any value for the treatment effect on the whole real line is consistent with the data in the absence of surrogacy.

Part $(ii)$:
If the outcome is binary, then some values can be ruled out. Because $\hmu(s,w,x,\ppe)$ is the conditional expectation of the outcome given some conditioning variables, it obviously must be inside the interval $[0,1]$, and both $\mu(s,1,x,\ppe)$ and $\mu(s,0,x,\ppe)$ must lie inside the interval $[0,1]$. This directly implies that  
$\mu(s,1,x,\ppe)-\mu(s,0,x,\ppe)\in[-1,1]$.
However, we can sharpen these bounds exploiting the fact that $\hmu(s,x,\ppo)=\hrho(s,x)\mu(s,1,x,\ppe)+(1-\hrho(s,x))\mu(s,0,x,\ppe)$. This implies
that
\begin{equation}\mu(s,1,x,\ppe) =\frac{\hmu(s,x,\ppo)-\mu(s,0,x,\ppe)(1-\hrho(s,x))}{\hrho(s,x)}.\label{eq6}\end{equation}
First consider the upper bound on $\mu(s,1,x,\ppe)-\mu(s,0,x,\ppe)$. 
The question is what the pairs of values $(\mu(s,1,x,\ppe),\mu(s,0,x,\ppe))$ are that both lie inside $[0,1]$, such that 
$\hmu(s,x,\ppo)=\hrho(s,x)\mu(s,1,x,\ppe)+(1-\hrho(s,x))\mu(s,0,x,\ppe)$ for given $\hmu(s,x,\ppo)$ and $\hrho(s,x)$, and that maximize the difference $\mu(s,1,x,\ppe)-\mu(s,0,x,\ppe)$. There are two possibilities. Either $\hmu(s,x,\ppo)\geq \hrho(s,x)$ or
$\hmu(s,x,\ppo)< \hrho(s,x)$.

If $\hmu(s,x,\ppo)\geq \hrho(s,x)$, then
the smallest value for $\mu(s,0,x,\ppe)$ such that the  value  for $\mu(s,x,\ppe)$ implied by (\ref{eq6}) is less than or equal to one is
$\mu(s,0,x,\ppe)=(\hmu(s,x,\ppo)-\hrho(s,x))/(1-\hrho(s,x))$. This value has to be less than one by the assumption that there is a pair of values $(\mu(s,0,x,\ppe),\mu(s,1,x,\ppe))$ that satisfies (\ref{eq6}). In this case upper bound for the difference $\mu(s,1,x,\ppe)-\mu(s,0,x,\ppe)$ is equal to $(1-\hmu(s,x,\ppo))/(1-\hrho(s,x))$. 
If $\hmu(s,x,\ppo)\leq \hrho(s,x)$, then the largest value for $\mu(s,1,x,\ppe)$ such that $\mu(s,0,x,\ppe)$ is nonnegative is $\hmu(s,x,\ppo)/\hrho(s,x)$. In that case the upper bound 
for the difference $\mu(s,1,x,\ppe)-\mu(s,0,x,\ppe)$ is equal to $\hmu(s,x,\ppo)/\hrho(s,x)$.

In summary, to demonstrate sharpness, consider the following data distributions:

If \(\mu(s, x, \ppo) \geq \rho(s, x)\), set \(\mu(s, 0, x, \ppe) = \frac{\mu(s, x, \ppo) - \rho(s, x)}{1 - \rho(s, x)}\) and \(\mu(s, 1, x, \ppe) = 1\).

 If \(\mu(s, x, \ppo) < \rho(s, x)\), set \(\mu(s, 0, x, \ppe) = 0\) and \(\mu(s, 1, x, \ppe) = \frac{\mu(s, x, \ppo)}{\rho(s, x)}\)

In both cases, these distributions are admissible under our assumptions, and also achieve the bounds, demonstrating that the bounds are sharp.

Therefore,
the sharp upper bound is
\[ \Delta^U_S(s,x)=
\left\{
\begin{array}{ll}
(1-\hmu(s,x,\ppo))/(1-\hrho(s,x))\qquad & \mathrm{if }\quad  \hmu(s,x,\ppo)\geq \hrho(s,x)\\
\hmu(s,x,\ppo)/\hrho(s,x)\qquad & \mathrm{if }\quad \hmu(s,x,\ppo)\leq \hrho(s,x),\end{array}
\right.\]\[
=\min\left(\frac{\hmu(s,x,\ppo)}{\hrho(s,x)},\frac{1-\hmu(s,x,\ppo)}{1-\hrho(s,x)}\right).
\]
The proof for the lower bound follows the same argument.

Part $(iii)$: 
\[\textrm{\rm surrogacy-bias}=
\mme\left[\left.\Bigl\{\mu(\soin,1,\xoin,\ppe)-\mu(\soin,0,\xoin,\ppe)\Bigr\}\cdot\frac{\hrho(\soin,\xoin)\cdot(1-\hrho(\soin,\xoin))}{\hrho(\xoin)\cdot(1-\hrho(\xoin))}\right|P_{i}=\ppe\right]
\]
\[\leq 
\mme\left[\left.\left|\Bigl\{\mu(\soin,1,\xoin,\ppe)-\mu(\soin,0,\xoin,\ppe)\Bigr\}
\right|\cdot\left|\frac{\hrho(\soin,\xoin)\cdot(1-\hrho(\soin,\xoin))}{\hrho(\xoin)\cdot(1-\hrho(\xoin))}\right|\right|P_{i}=\ppe\right]
\]
\[\leq 
c \cdot \mme\left[\left.\frac{\hrho(\soin,\xoin)\cdot(1-\hrho(\soin,\xoin))}{\hrho(\xoin)\cdot(1-\hrho(\xoin))}\right|P_{i}=\ppe\right]
\]
The upper bound can be achieved by setting
$\mu(s,0,x,\ppe)=\hmu(s,x,\ppo)-c\cdot \hrho(s,x)$ and $\mu(s,1,x,\ppe)=\mu(s,0,x,\ppe)+c,$
These distributions are admissible under our assumptions, and hence sharpness is obtained.
We can likewise obtain the lower bound.

$\Box$

\noindent {\bf Proof of Lemma \ref{lemma2}}
We show that the derived bounds are sharp by demonstrating that there exist data distributions consistent without assumptions that achieve these bounds.
$(i)$ In the absence of Comparability the data imply no restrictions on the values for $\hmu(s,x,\ppe)$, and so as long as there is some difference between $\hrho(s,x)$ and $\hrhoe(x)$ there is no bound on the bias.
\\
$(ii)$ If the outcomes are binary, the only restrictions implied on $\hmu(s,x,\ppe)$ are that all values lie inside $[0,1]$.  The upper bound comes from imputing 1 for $\hmu(s,x,\ppe)$ if $\hrho(s,x)>\hrhoe(x)$ and $0$ if $\hrho(s,x)<\hrhoe(x)$, a choice of distribution that is admissible. This directly implies the bounds on the bias.
\\
 $(iii)$ 
\[\textrm{\rm comparability-bias}
=\mme\left[\left.\Bigl\{ \hmu(\soin,\xoin,\ppe)-\hmu(\soin,\xoin,\ppo)\Bigr\}\cdot\frac{\hrho(\soin,\xoin)-\hrho(\xoin)}{\hrho(\xoin)\cdot(1-\hrho(\xoin))}\right|P_{i}=\ppe\right].
\]
Then
\[\left|\mme\left[\left.\Bigl\{ \hmu(\soin,\xoin,\ppe)-\hmu(\soin,\xoin,\ppo)\Bigr\}\cdot\frac{\hrho(\soin,\xoin)-\hrho(\xoin)}{\hrho(\xoin)\cdot(1-\hrho(\xoin))}\right|P_{i}=\ppe\right]\right|
\]
\[\leq\mme\left[\left.\left|\Bigl\{ \hmu(\soin,\xoin,\ppe)-\hmu(\soin,\xoin,\ppo)\Bigr\}\right|\cdot\frac{\left|\hrho(\soin,\xoin)-\hrho(\xoin)\right|}{\hrho(\xoin)\cdot(1-\hrho(\xoin))}\right|P_{i}=\ppe\right] \]
\[\leq c\cdot \mme\left[\left.\frac{\left|\hrho(\soin,\xoin)-\hrho(\xoin)\right|}{\hrho(\xoin)\cdot(1-\hrho(\xoin))}\right|P_{i}=\ppe\right] .\]
The upper bound can be attained by
setting 
\[ \hmu(s,x,\ppe)=
\left\{
\begin{array}{ll}
\hmu(s,x,\ppo)+c\qquad & \mathrm{if}\quad \hrho(s,x)\geq \hrhoe(x),\\
\hmu(s,x,\ppo)-c & \mathrm{otherwise,}
\end{array}
\right.
\]
and similarly for the lower bound.
$\Box$

\noindent \textbf{\large{}{}C. Illustration of Bias Bounds Calculation}{\large\par}

We will provide a simple illustration of how the theoretical bias bounds we calculated in Section \ref{subsec:bounds_bias} look like in practice. We focus on the employment outcome to illustrate the surrogacy bias and comparability bias bounds in the binary case (Case (ii)). 

Table \ref{tab:bias_surr} and \ref{tab:bias_comp} show the bounds on the treatment effects using the Influence Function Estimator under potential violations of Surrogacy and Comparability, respectively.\footnote{If we are interested in conducting inference on the partial identification bounds, we can take the approach illustrated in, e.g., \cite{ImbensManski2004,molinari2020microeconometrics}.} This demonstrates that in the binary outcome of employment, the sign can still be credibly inferred under the latter half even under surrogacy violation. The comparability bias seems to be non-negligible, part of our design of choosing Riverside (experimental data) due to its unique "jobs first" approach, in contrast to the "human capital" approach used in LA, San Diego, and Alameda counties (observational data). Further work must be done to ensure cases where comparability bias is minimal. We can similarly compute non-binary outcomes like Earnings, with some plausible range of user-specified parameter $c$ (Case (iii) in Section \ref{subsec:bounds_bias}).

%%%the bias version 
% \begin{table}[ht]
% \centering
% \caption{Surrogacy Bias Bounds for Employment Outcome} 
% \label{tab:bias_surr}

% \begin{tabular}{rrr}
%   \hline
% t & Lower Bound & Upper Bound \\ 
%   \hline
%   1 & -0.89 & 0.33 \\ 
%     2 & -0.63 & 0.16 \\ 
%     3 & -0.44 & 0.11 \\ 
%     4 & -0.34 & 0.08 \\ 
%     5 & -0.28 & 0.06 \\ 
%     6 & -0.24 & 0.05 \\ 
%    12 & -0.12 & 0.03 \\ 
%    18 & -0.08 & 0.02 \\ 
%    24 & -0.06 & 0.01 \\ 
%    30 & -0.05 & 0.01 \\ 
%    36 & -0.04 & 0.01 \\ 
%    \hline
% \end{tabular}
% \end{table}

%the point estimate version
\begin{table}[ht]
\centering
\caption{ Bounds on the Influence Function Estimator without Surrogacy for Employment Outcome}
\label{tab:bias_surr}
\begin{tabular}{rrr}
\hline
t & Lower Bound & Upper Bound \\
\hline
1 & -0.791 & 0.369 \\
2 & -0.608 & 0.212 \\
3 & -0.417 & 0.163 \\
4 & -0.310 & 0.140 \\
5 & -0.232 & 0.128 \\
6 & -0.186 & 0.124 \\
12 & -0.058 & 0.102 \\
18 & -0.017 & 0.092 \\
24 & 0.001 & 0.081 \\
30 & 0.018 & 0.078 \\
36 & 0.026 & 0.076 \\
\hline
\end{tabular}
%\caption{ Bounds on the Influence Function Estimator without Surrogacy for Employment Outcome}
\end{table}

% \begin{table}[ht]
% \centering
% \caption{Comparability Bias Bounds for Employment Outcome} 
% \label{tab:bias_comp}

% \begin{tabular}{rrr}
%   \hline
% t & Lower Bound & Upper Bound \\ 
%   \hline
%   1 & -0.04 & 0.02 \\ 
%     2 & -0.09 & 0.07 \\ 
%     3 & -0.12 & 0.10 \\ 
%     4 & -0.13 & 0.12 \\ 
%     5 & -0.14 & 0.12 \\ 
%     6 & -0.14 & 0.13 \\ 
%    12 & -0.15 & 0.14 \\ 
%    18 & -0.15 & 0.15 \\ 
%    24 & -0.16 & 0.16 \\ 
%    30 & -0.17 & 0.17 \\ 
%    36 & -0.17 & 0.17 \\ 
%    \hline
% \end{tabular}

% \end{table}

\begin{table}[ht]
\centering
\caption{ Bounds on the Influence Function Estimator without Comparability for Employment Outcome}
\label{tab:bias_comp}

\begin{tabular}{rrr}
\hline
t & Lower Bound & Upper Bound \\
\hline
1 & -0.031 & 0.029 \\
2 & -0.058 & 0.102 \\
3 & -0.077 & 0.143 \\
4 & -0.080 & 0.170 \\
5 & -0.082 & 0.178 \\
6 & -0.076 & 0.194 \\
12 & -0.078 & 0.212 \\
18 & -0.077 & 0.223 \\
24 & -0.089 & 0.231 \\
30 & -0.102 & 0.238 \\
36 & -0.104 & 0.236 \\
\hline
\end{tabular}

\end{table}

\end{document}